\documentclass[journal]{IEEEtran}

\ifCLASSINFOpdf
  \usepackage[pdftex]{graphicx}
  \graphicspath{{../pdf/}{../bitmap/}}
  \DeclareGraphicsExtensions{.pdf,.jpeg,.png}
\else
\fi
\ifCLASSOPTIONcompsoc
  \usepackage[caption=false,font=normalsize,labelfont=sf,textfont=sf]{subfig}
\else
  \usepackage[caption=false,font=footnotesize]{subfig}
\fi
\usepackage{amsmath}
\usepackage{multirow}
\usepackage{url}

\usepackage[style=numeric-comp,backend=biber,sorting=none,sortcites=true]{biblatex}
\usepackage{lipsum,amsmath,verbatim}
\usepackage[colorlinks=false]{hyperref}
\usepackage[compress,sort]{cleveref}
\usepackage{tikz,pgfplots}
\usepackage{balance}
\usepackage[inline]{enumitem}
\usepackage[detect-all]{siunitx}
\usepackage{xcolor,colortbl}
\usepackage[normalem]{ulem}

\pgfplotsset{compat=1.18}
\usetikzlibrary{math} 
\usetikzlibrary{calc}
\usepgfplotslibrary{external}
\usetikzlibrary{arrows, decorations.markings}

\AtBeginBibliography{\footnotesize}

\hypersetup{
     colorlinks=true,linkcolor=black,anchorcolor=black,citecolor=black,filecolor=black,menucolor=black,runcolor=black,urlcolor=black
}

\DeclareSIUnit\nat{nat}

\def\cloudleak{Cloud leak}
\def\falseactivation{False activation}
\def\cloudaccess{Cloud access}
\def\WASNauthentication{{WASN} authentication}
\def\speechinterface{Speech interface leak}
\def\discussionleak{Discussion leak}
\def\shareddevice{Shared device}

\newif\ifChanges
\Changesfalse

\ifChanges
\newcommand{\changed}[1]{{\color{orange}{#1}}}%
\newcommand{\removed}[1]{\sout{#1}}%

\usepackage[%
linewidth=1pt,
middlelinecolor= orange,
linecolor=orange,
middlelinewidth=0.4pt,
roundcorner=1pt,
topline = false,
leftline = false,
rightline = true,
bottomline = false,
rightmargin=-.1cm,
innerleftmargin=0cm,
innerrightmargin=.1cm,
innertopmargin=0pt,
innerbottommargin=0pt,
]{mdframed}

\newmdenv[
middlelinecolor= green,
linecolor=green,
]{newcontent}

\else 

\newcommand{\changed}[1]{#1}%
\newcommand{\removed}[1]{}

\fi

\begin{document}
%
\title{Privacy in Speech Technology}
%
%
%

\author{Tom~Bäckström,~\IEEEmembership{Senior Member,~IEEE}        
\thanks{T. Bäckström is with the Department
Information and Communications Engineering, Aalto University, Finland e-mail:tom.backstrom@aalto.fi.}
\thanks{Manuscript received Month Day, Year; revised Month Day, Year.}}

%
%

\markboth{Proceedings of the IEEE, 
~Vol.~XX, No.~YY, Month~202Z}%
{Bäckström \MakeLowercase{\textit{et al.}}: Privacy in Speech Technology}
%



\maketitle

\begin{abstract}
Speech technology for communication, accessing information, and services has rapidly improved in quality. It is convenient and appealing because speech is the primary mode of communication for humans. Such technology, however, also presents proven threats to privacy. Speech is a tool for communication, and it will thus inherently contain private information. Importantly, it also contains a wealth of side information, including details about health, emotions, affiliations, and relationships, all of which are private. Exposing such private information can lead to serious threats such as price gouging, harassment, extortion, and stalking. This paper is a tutorial on privacy issues related to speech technology, modeling their threats, approaches for protecting users' privacy, measuring the performance of privacy-protecting methods, perception of privacy, as well as societal and legal consequences. In addition to a tutorial overview, it also presents lines for further development where improvements are most urgently needed.
\end{abstract}


%
\IEEEpeerreviewmaketitle

\section{Introduction}\label{sec:intro}
%
%
%
%

\IEEEPARstart{S}{peech} is a mode of communication and thus inherently contains a wealth of information (see \cref{tab:categories}). 
In addition, {to the desired information of the communication,} speech thus also contains a wide range of side information like the state of health and emotions, as well as physical, psychological, and social identity, most of which is private information. {Moreover}, speaking is a dynamic interaction between two or more speakers. Dialogues thus also contain information about the relationship between participants, such as their level of familiarity, affiliation, intimacy, relative hierarchy, shared interests, and history. 
From an information content perspective, we can thus expect that \emph{privacy is a multifaceted issue in all areas of speech technology}. \smallskip

\begin{table}[t]
    \centering
    \caption{\strut Private information potentially identifiable from speech signals, and the extent to which they are sustained over time and can be willfully controlled~{\cite{singh2019profiling,kroger2019privacy,kroger2022personal,singh2025humanvoiceunique}}.
    }
    \begin{tabular}{p{1.6cm}p{2.8cm}p{1.47cm}p{.9cm}}
       \textbf{Category} & \textbf{Examples} & \textbf{Permanence} & \textbf{Control} \\\hline
        Biological
        & Body characteristics & {Long-term} & No \\  \cline{2-4}     
        & State of health & {Short-term} & No\\\cline{2-4}
        &{Level of intoxication} & {{Short-term}} & {Yes}\\\hline
        Psychological 
        & Emotions & {Short-term} & Partly\\\cline{2-4}
        &{Health} & {Mixed} & {Partly}\\\cline{2-4}
        & Intelligence & {Long-term} & No \\\cline{2-4}
        & Education, skill & {Long-term} & Yes \\\cline{2-4}
        & Gender identity & {Long-term} & No \\\hline
        Message 
        & Text, emphasis, style,\newline expression & {Short-term} & Yes \\\cline{2-4}
        & Mannerisms, context & {Short-term} & Partly\\\cline{2-4}
        & {Language choice} & {Short-term} & {Yes}\\\cline{2-4}
        & {Language skills} & {Long-term} & {Partly}\\\hline
        Affiliation
        & Ethnic, national,\newline cultural, religious,\newline political, etc. & {Long-term} & {Partly} \\\hline
        Relationship\newline character
        & Hierarchy, familiarity,\newline attraction, intimacy & {Mixed} & Partly 
        \\\hline
        Physical\newline environment
        & Background sounds,\newline distance to sensor,\newline transmission distance,\newline reverberation, location & {Short-term} & Yes \\\hline
        Hardware \&\newline software used
        & Sensor and software\newline type \& manufacturer & {Short-term} & Yes \\
        \hline
    \end{tabular}
    \label{tab:categories}
\end{table}

\begin{table}[t!]
\caption{Practical examples of threats related to private and sensitive information that is conveyed by people's voices.
}\label{tab:exploits}
\begin{tabular}{p{1.6cm}p{6.41cm}
}
     \textbf{{Threat}} & \textbf{Example} 
     \\\hline
     Price gouging      
     &  Signs of depression or other health problems in users' voices could be misused to trigger an increase in their insurance premiums. Signs of users' emotions could be exploited to offer them products at higher prices.
    \\\hline
    Tracking, stalking & Voice re-identification could link users across platforms, i.e., from work-related social media to online support groups and dating apps, making it possible to follow them anywhere.
    \\\hline
    Extortion, public\newline humiliation & Private health problems, and romantic affairs could be detected in the voice and used for blackmail or made public against a user's wishes.
    \\\hline
     Algorithmic stereotyping & Recommender systems based on voice can become biased with respect to age, identity, religion, or ethnicity in ways that are nearly impossible to monitor.%
    \\\hline
     Harassment, inappropriate advances & Users in chat rooms or virtual reality could be automatically singled out by gender or opinions, making them a target for unwanted attention and harassment.
     \\\hline
     {Fear of\newline monitoring}, loss of\newline dignity & {The subjective feeling of being continuously monitored may cause psychological damage and stifle political expression, damaging democratic societies. Users may feel they lose control of personal aspects of their lives.}
    \\\hline
\end{tabular}
\end{table}

Studying and improving privacy in speech technology is important because breaches in privacy can have serious consequences. \Cref{tab:exploits} lists examples of threats and exploits. The list is incomplete, and we can expect new threats to be discovered. The generic solution template is, however, typically always the same: \emph{Minimize transmission, storage, and access to sensitive information that is irrelevant to the service that the users want.} It is then ''merely'' a question of \emph{how} such minimization is achieved, \emph{what} information is relevant, and \emph{how} to determine what the user wants (provide information and enable control of services and threats).

With respect to threats, media attention is often focused on \emph{breaches which have large economic consequences}~\cite{Lynskey2019GuardianPrivacyScandal,hern2019,brewster2021}. While such breaches are important by themselves, the media attention introduces unfortunate biases in two ways. First, a breach is a worst-case event, whereas threats that have not yet led to a breach event can already have a large impact. For example, users may choose not to use systems that threaten their privacy; known weaknesses and even a lax attitude of the service provider toward privacy can, therefore, have an effect on the adoption, retention, and sales of products and services. Importantly, users can avoid systems that are \emph{perceived} threatening, even when there is no actual threat. To maintain users' trust, it is therefore essential to both uphold the users' actual privacy but also design systems that actively and frequently communicate the level of privacy, including known threats, their current status, and measures taken to protect against them. 

Second, while a single breach in a large service can have large economic, psychological, societal, and legal consequences, small crimes are so common that their joint effect is comparable or larger in size~\cite{levy2020intimate}. With speech technology, such ''small'' crimes include stalking, extortion, harassment, humiliation, and inappropriate advances (see \cref{tab:exploits}). While the economic damage of a single such incident can be small, when combined, their joint psychological and societal effect is potentially large, and their prevalence makes them a considerable threat also economically. 


\medskip

This paper presents a tutorial overview of privacy in speech technology that covers a wide range of threats, methodologies, and algorithms. The range of applications and content types where privacy-preserving processing methods are needed is vast, and \cref{sec:content} presents a brief overview thereof, and \cref{sec:def} defines the vocabulary. Analysis of threats demonstrates that while attack surfaces have a great variety, the threat models are similar (see \cref{sec:threat}). Protections against those threats have distinct categories of approaches, isolating target information by removing private information, limiting access to private information through cryptography, reducing centralized threats through non-centralized processing, limiting access to reproduced audio, improving overall system performance, and design principles from theoretical frameworks (see \cref{sec:protections}).
To quantify the extent of protections, we further need methods for evaluation of privacy and utility (see \cref{sec:objmeasures}). While objective evaluation quantifies the actual level of privacy, users' impressions of privacy do not necessarily follow the objective level. For the best user experience, therefore, we need to quantify and understand users' perceptions, experiences, and preferences related to privacy, as well as design systems accordingly (see \cref{sec:psycho}). 
Finally, threats to and breaches in privacy have considerable consequences on both societal, economic, and individual levels, which have motivated governments to increasingly regulate the use of technology (see \cref{sec:law}).

\begin{figure}[t]
\centering
\footnotesize
    
    \begin{tikzpicture}
        \tikzsetfigurename{allthreats} 
        \node[matrix,column sep=.05cm] at (0,0) { 
            \node {Bob\strut}; & 
            \node {\parbox{1.15cm}{\centering Acoustic\\pathway\strut}}; &
            \node {\parbox{.9cm}{\centering\textbf{User\\ Alice\strut}}}; &
            \node {\parbox{1.15cm}{\centering Acoustic\\pathway\strut}}; & 
            \node{\parbox{.8cm}{\centering Local\\ device\strut}}; &
            \node {Network\strut};
            & \node{\parbox{.9cm}{\centering Remote\\service\strut}}; \\
            \node (user2) {\includegraphics[scale=0.15]{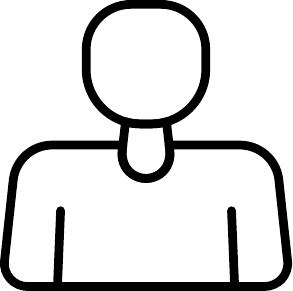} };
            & \node (speechinteraction) {}; &
            \node (user1) {\includegraphics[scale=0.15]{pdf/AdobeStock_290226498_user_extract.pdf} };
            & \node (edgeinteraction) {}; &
            \node (edge1) {\includegraphics[scale=0.15]{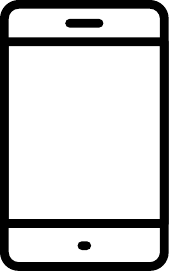} };
            & \node (cloudinteraction) {}; &
            \node (cloud1) {\includegraphics[scale=0.15]{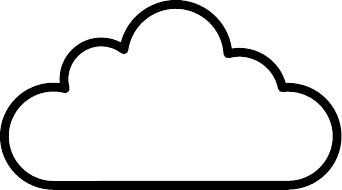} };
            \\ \node{};\\\node{};\\\node{};\\\node{};\\
            &&
            \node (eve) {\includegraphics[scale=0.15]{pdf/AdobeStock_290226498_user_extract.pdf} };
            &&
            \node (threatdevice) {\includegraphics[scale=0.15]{pdf/AdobeStock_290226498_phone_extract.pdf} };
            &&
            \node (threatcloud) {\includegraphics[scale=0.15]{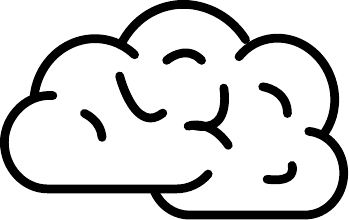} };
            \\
            &&
            \node { Eve };
            &&
            \node {\parbox{.9cm}{\centering Local\\device} };
            &&
            \node {\parbox{1.2cm}{\centering Secondary\\service}};
            \\
        };
        \usetikzlibrary {arrows.meta} 
        \draw [->,thick] (user1) to [bend left=20] (edge1);
        \draw [->,thick] (edge1) to [bend left=20] (user1);
        \draw [->,thick] (cloud1) to [bend left=20] (edge1);
        \draw [->,thick] (edge1) to [bend left=20] (cloud1);
        \draw [<-,thick] (user1) to [bend left=20] (user2);
        \draw [<-,thick] (user2) to [bend left=20] (user1);
        \draw [->,thick,dashed] (threatdevice) to  (eve);
        \draw [<->,thick,dashed] (threatdevice) to (threatcloud);

        \draw [->,red,thick,dotted] (user2) ++(.2,-.5) to [bend right=20] node [midway,below] {X} (eve);
        \draw [->,red,thick,dotted] (speechinteraction) ++(0,-.5) 
        -- node[midway,below]{X} 
        (eve);        
        \draw [red,thick] (edgeinteraction) ++(-.4,-.4) edge [->] node[near start,left]{5} (eve);        
        \draw [red,thick] (edge1) ++ (-.3,-.25) edge [->] node[very near end, below] {7} (eve);
        \draw [->,red,thick] (speechinteraction) ++(.4,-.4) -- node[near end, below] {6} (threatdevice);        
        \draw [->,red,thick] (edgeinteraction) ++(0,-.4) -- node[near end,above]{5} (threatdevice);        
        \draw [->,red,thick] (edge1) -- node[midway,right]{4} (threatdevice);        
        \draw [->,red,thick] (cloud1) -- node[near end,below]{3} (threatdevice);        
        \draw [->,red,thick] (edge1) ++(.4,-.4) -- node[near end,below]{2} (threatcloud);        
        \draw [->,red,thick,dotted] (cloudinteraction) ++(.2,-.4) -- node[near end,above]{X} (threatcloud);        
        \draw [->,red,thick] (cloud1) -- node[midway,right]{1} (threatcloud);        
    \end{tikzpicture}
\caption{Threats to the privacy of a user Alice when speaking with another person Bob or a local device, where information is leaked or shared through the acoustic pathways, the {local} device, the network, or the cloud service, to another person Eve, device or service, contrary to the preferences of Alice (red arrows). Each threat is marked with the number of the corresponding part of \cref{sec:surface}.
{Dotted red arrows marked with ``{X}'' indicate threats outside the current scope, and dashed black arrows indicate the potential retransmission of
leaked data beyond the primary attack surface.}}
    \label{fig:allthreats}
\end{figure}

{This paper focuses primarily on speech technology in human-computer interfaces (see \cref{fig:allthreats}), but will also address telecommunication where appropriate. Focus is here, however, limited to the acoustic speech signal, also known as the \emph{voice}, since natural language processing is a distinct and largely independent field that warrants treating it separately (e.g.~\cite{mahendran2021privacy}). } 
Concerning telecommunication, approaches to ethical, legal, and technological questions related to privacy in telecommunication over landlines are well-established, and open discussions are primarily related to how the existing regulation and oversight should be extended also to cover mobile telecommunication and voice-over-Internet protocols (VoIP)~\cite{penders2004privacy,wong2009privacy,arapinis2017analysis}. 

\medskip

To the author's knowledge, this is the widest tutorial overview of privacy in speech technology. Recent works in the field, however, include a technical overview~\cite{
nautsch2019preserving}, a survey on privacy issues with voice assistants~\cite{acosta2022survey},
as well as a popular-science review~\cite{backstrom2020privacy} of this area,  and several doctoral theses have their respective summaries, {e.g.~\cite{Nelus2022,das2021thesis,zarazaga2022thesis,pathak2012privacy,williams2022learning,portelo2015privacy,vestman2020phd,nespoli2025phd}}. Privacy has been extensively discussed in other areas of science, such as the neighboring field of natural language processing~\cite{mahendran2021privacy}, in statistical theory of privacy~\cite{cummings2023challenges}, in social sciences~\cite{petronio2002boundaries} and psychology~\cite{Derlega1977PrivacyAS}. There is even an excellent and thorough meta-study of all areas of science that discuss privacy~\cite{van2018handbook}. {Central contributions of this work include
\begin{enumerate*}
    \item structuring and collecting information, research results, and vocabulary into a unified format,
    \item proposing a framework for analysis of attack vectors (see \cref{fig:allthreats}), and
    \item an introduction to the psychological aspects of privacy that is not often discussed in engineering-oriented publications (see \cref{sec:psycho}).
\end{enumerate*}
}


\section{Content categories and applications}\label{sec:content}
To give a complete picture of the implications of privacy in speech technology, this section provides a brief discussion on both categories of private information, complementing \cref{tab:categories} as well as the application of such information. Observe that this is not a complete list (nor is \cref{tab:categories}) of the content nor of application categories. The purpose is merely to provide a characterization of work done and challenges in the research field, with accompanying references.

First, note that while all presented categories are (potentially) private information, all \emph{sustained} information {(in the sense that it does not change over a ``long'' time)} can potentially be used to identify a speaker. For example, while the current emotional state cannot alone identify a speaker, the tendency to display emotional states can aid in identification. However, where we want to \emph{verify} that a speaker is who they claim to be, we can use only information that cannot be willfully changed. That is, speech style is a particular example of a property that a good voice actor can freely choose, making it ``easy'' to change also for fraudulent purposes. {Using variable attributes, or their distributions, to identify or verify an identity is thus likely to be complicated.}

Recognizing the speaker's identity then becomes the natural starting point for studies in privacy. We can attempt to recognize who is speaking (speaker recognition), verify whether a speaker has the claimed identity (speaker verification), clustering audio to segments with a single speaker (speaker diarization) and we can develop methods for deceiving identity (spoofing), e.g. \cite{chouchane21_interspeech,granqvist20_interspeech,mawalim20_interspeech,mawalim20_interspeech,nautsch19b_interspeech,nelus19_interspeech,nelus19b_interspeech,noe21_interspeech,perero2022x,zewoudie2021federated,zewoudie2021use,teixeira22_interspeech,treiber2019privacy,nelus19_interspeech,nelus19b_interspeech,BAI202165,han2020voice,KINNUNEN201012,hansen2015speaker,parthasarathi2013wordless,pathak11_interspeech,parthasarathi11_interspeech,pathak2012ppsa,jimenez2017two,portelo2014bprivacy}.
Similarly, by voice conversion we can anonymize a speaker identity by replacing it with a random identity (anonymization) or a specific one (pseudonymization), e.g.~\cite{noe20_interspeech,pierre22_interspeech,prajapati21_interspeech,shamsabadi2023diff,srivastava20_interspeech,Srivastava2022taslp,mawalim20_interspeech,maouche20_interspeech,Fang2019}. Speaker characterization is the natural complement of speaker idenfication~\cite{nautsch2019preserving,nguyen2010automatic}.
Such methods related to speaker identity can be applied, for example, to recognize the user of a device, verify a customer at a bank, or as a voice avatar in online gaming. Similarly, anonymization and pseudonymization can be used to hide the identity from the public media and gaming.

Speech recognition, as in speech-to-text, is probably the largest sub-area of speech research. With respect to privacy, it presents two obvious challenges. We can try to limit side-information, such as eliminating all non-text information from the data stream, e.g.~\cite{shoemate2022sottovoce,nguyen2023federated,aloufi21_interspeech,garg20b_interspeech,shah21_interspeech,srivastava19_interspeech,tomashenko2022privacy}, or we can use natural language processing to anonymize the text content, e.g.~\cite{liu22_interspeech,maouche2022enhancing,adelani20_interspeech,novotney21_interspeech,mahendran2021privacy,habernal2021differential}. 

Privacy is even more critical in always-on applications like wake-word detection, i.e., when the interface is triggered by a specific keyword like ``Computer'' in ``Computer, lights-off''. This application is more sensitive exactly because it is always on, as users cannot choose when data is processed~\cite{koppelmann21_interspeech,koppelmann22_interspeech,stoidis22_interspeech,yang21_interspeech}.
The always-on characteristic is also prominent in assisted-living applications, which can ``monitor people’s daily exercises, consumption of calories and sleeping patterns, and to provide coaching interventions to foster positive behavior''~\cite{haider19_interspeech,salah2025privacy}. Such ambient voice interfaces are often implemented through acoustic sensor networks, which pose their own challenges~\cite{koppelmann22_interspeech,oconnor20_interspeech,malekzadeh2018replacement,zewoudie2021use,zarazaga2020acoustic,nelus2021,backstrom2025privacy}.

Speech enhancement refers to removing background noises and distortions from the desired speech signal ({target information})~\cite{benesty2008springer}. This task can have an impact on privacy in two {primary} ways. First, the recording environment can reveal private information about the speaker, and second, background noises, like a competing talker, can contain private information~\cite{zarazaga21_interspeech,rech2022multi}. Attenuating {or removing} background noises and competing speakers, as well as removing reverberation, can thus improve privacy. In addition, while using a sensor network to capture speech can improve utility, it introduces novel threats as well~\cite{nelus2021,zarazaga2022thesis,hendriks2013ppdistributed}.

Speech is typically considered to be \emph{biometric information} in its entirety, including all information categories in \cref{tab:categories} as such categories can be used to identify a person~\cite{adler2006towards,nautsch20_interspeech,nautsch2019preserving,ISObiometric,edpb2021guidelines}.
Some, however, limit the term biometric to refer only to physical and behavioral characteristics. Such properties include health~\cite{natarajan2021prioris}, emotions~\cite{feng22b_interspeech} and gender identity~\cite{nelus2018itg,stoidis21_interspeech,stoidis22_interspeech,nelus19_interspeech}, each warranting their own treatment.

\section{Definitions}\label{sec:def}

\subsection{Privacy and security}
Privacy is closely related to challenges in security, and it is often difficult to distinguish between them. Here, 
{\emph{information security} refers to enforcing rightful access to information in scenarios where the access rights are binary (granted or not granted). In contrast, \emph{information privacy} refers to the application of rules to information access and use, in scenarios where \emph{access permissions are partial and rules context-dependent} (adapted from~\cite{nissenbaum2004privacy}). The rules may distinguish between the ability to access and \emph{de facto} accessing the data, different ways and purposes of using the data, and sharing the data with third parties.
The rules may be explicit (e.g., verbalized or documented) or implicit (based on social norms and tacit expectations), and can include requirements for monitoring, reporting, and control.
} For example, a voice interface can be used to control home automation, but if the service provider shares that voice data with advertisers against the users' preferences, then it violates privacy (see \cref{fig:allthreats}).
Consequently, security concerns such as identity spoofing, deep fakes, and attacks on devices or networks are excluded here, as these threats are more related to security than privacy.

\subsection{Information categories}
\begin{figure}[t]
\centering
\footnotesize
    \begin{tikzpicture}
        \node[matrix,column sep=1mm] at (0,0) { 
            \node {Sender}; 
            & \node{\parbox{1.5cm}{\centering Channel}}; 
            & \node {Recipient}; \\
            \node  {\includegraphics[height=8mm]{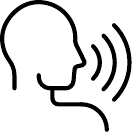} };
            \node[circle,minimum size=10mm]  (user1) {};
            &&
            \node at (-.1,.7) {\scalebox{-1}[1]{\includegraphics[height=8mm]{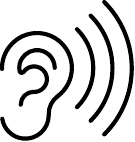} }};
            \node[circle,minimum size=10mm] at (0,.7) (cloud1) {};
            \node[right of=cloud1,xshift=1mm] {Legitimate};
            \node at (-.1,-.7)  {\includegraphics[height=6mm]{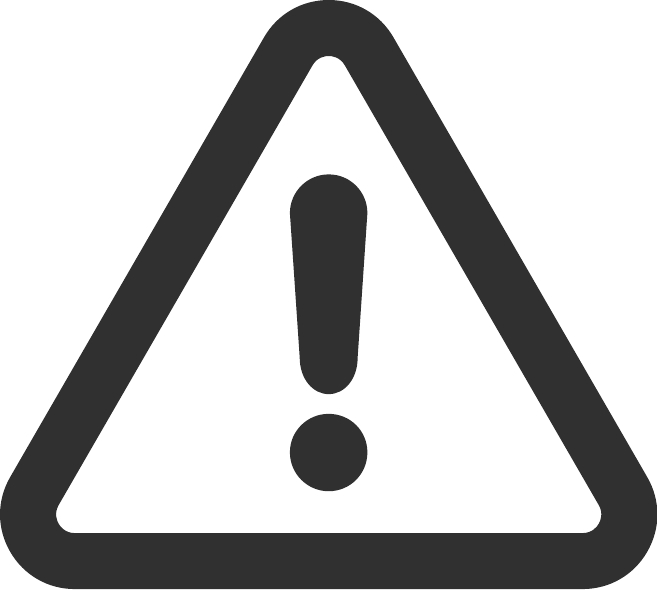} };
            \node[circle,minimum size=8mm] at (-.1,-.7) (threat1) {};
            \node[right of=threat1,xshift=1.5mm] {Undesired};
            & 
            \\
        };
        \usetikzlibrary {arrows.meta} 
        
        \draw [->,thick,red] ($(user1.east)+(0,.1)$) -- ++(1.3,0) -- node[very near end,above]{!} ($(threat1.west) + (.1,.2)$);
        \draw [->,red] ($(user1.east)+(0,-.1)$) -- ++(1.3,0) -- node[near end,below]{!} ($(threat1.west) + (0,0)$);
        \draw [->,thick] ($(user1.east)+(0,.1)$) -- node[midway,above]{\parbox{1.5cm}{\centering{Target information}}} ++(1.3,0) -- ($(cloud1.west) + (0,-.1)$);
        \draw [->] ($(user1.east)+(0,-.1)$) -- node[midway,below]{\parbox{1.5cm}{\centering Side-information}} ++(1.3,0) -- ($(cloud1.west) + (.0,-.3)$);
    \end{tikzpicture}
    \caption{ The high-level \emph{threat model} of speech interaction, where \changed{target information} is sent through a channel to the legitimate recipient, but consequential side-information is bundled to that message. It is a threat to privacy when an undesired recipient gains access to that {target information} or side information (marked by red arrows and exclamation marks).
    }
    \label{fig:threatmodel}
\end{figure}

\emph{{Target information}} is the content of \changed{a message} that is required to fulfill the purpose of that communication. \emph{Side-information} {(also known as auxiliary information, external knowledge, or background knowledge~\cite{ganta2008composition})} is any and all information in \changed{a message} other than the {target information}. See \cref{fig:threatmodel}. 

\emph{Attributes} {(or characteristics~\cite{singh2025humanvoiceunique})} of speech are properties that can be distinguished from a speech signal and describe the speaker(s) or their state, the speech content, style, or environment, the recording configuration, or any other related features. \Cref{tab:categories} lists examples of speech attributes with private information.

Observe that while the target information is often the lexical information contained in speech, in many applications, the target information can be some other attribute. For example, in medical analysis, the target information is the health state of the speaker. In that case, the lexical information may be private side information.

\subsection{Tasks and processing objectives}\label{sec:tasks}
When a speaker makes a speech command to a computer or service, this implies that the speaker is giving a task to the computer, and this is known as the \emph{trusted task}. In contrast, we use \emph{threat tasks} to model the activity of attackers trying to extract information or perform other nefarious tasks. \cite{nelus19b_interspeech,nelus2018itg}.

Voice conversion refers to any speech processing method that replaces the speaker's identity or other attributes in a speech signal with those of another.
\emph{Anonymization} of a speech signal refers to voice conversion where the \emph{purpose is to conceal} private information.
\emph{Pseudonymization} further refers to reversible anonymization, where the original signal can be recovered if the concealed attributes are known.\footnotemark{}
The anonymized speech signal then corresponds to a speech signal of an artificial person, who is known as a \emph{pseudospeaker}. 
\footnotetext{These definitions of anonymization and pseudonymization are outcomes of a community-wide discussion about personally identifying information (PII) at the Lorentz Center workshop ``Speech as PII: Ethics \& Usability in vis-à-vis with Security \& Privacy'' in 2021.}

\emph{Attribute or property inference} refers to the identification of private attributes of speakers of known identity or otherwise~\cite{gong2018attribute}. \emph{Re-identification} refers to identifying individuals from apparently anonymized information~\cite{ISObiometric,ISOIEC224745_2022}.

{In anonymization tasks,} \emph{enrollment data} is reference data, typically labeled with, for example, speakers' identities or other attribute(s), to which observations are compared with the purpose of identifying said identity or attribute(s). It is used in the context of trusted tasks, where the agent is authorized to access this information. In threat tasks, where an attacker has gained access to similar reference data, the corresponding data is known as \emph{found data}. Conversely, speech samples that are compared to a database of samples are here referred to as \emph{observations}, and in literature, they are sometimes known as \emph{trial} samples, \emph{recognition} samples, or \emph{public data}. \cite{tomashenko2022voiceprivacy}

\subsection{Attack and attacker properties}
We use the terms \emph{attacker} and unintended recipient loosely to describe any entity or agent that gains access to or uses private information in a way that conflicts with the user's preferences. A \emph{malicious} attacker is intentionally seeking to gain private information,
a \emph{targeted} attacker attempts to gain such access to a specific person, an \emph{opportunistic} attacker to any person, a \emph{motivated} attacker is seeking ways to attack, and a \emph{honest-but-curious} attacker seizes emerging opportunities~\cite{CASEY20038,le2023privacy}. 
The \emph{attack surface} {or vector} is the specific point in the system where private information leaves the authorized path to make its way to the attacker~\cite{THEISEN201894}.

\section{Threat and attack models}\label{sec:threat}
{
To better understand the decisive aspects of threats, it is useful to model threats and their components. 
The most important components of such models include 
the exposed data sources (\cref{sec:data}), 
tactics used by, and data available to the attacker (\cref{sec:tactic}),
the flow of data on an abstract level, also known as the attack model (\cref{sec:attackmodel}), 
as well as the specific point in the system where private information leaves the intended path, known as the attack surface (\cref{sec:surface}). 
}

\subsection{\texorpdfstring{{Vulnerable data sources}}{Vulnerable data sources}}\label{sec:data}

Speech {data} can be exposed in two ways (see \cref{fig:threatmodel}): First, when  \emph{{target information} is transmitted}, it is a threat to privacy when, contrary to the preferences of the user, it is used in an unexpected way or transmitted to a third party. Second, since speech contains a wide range of private information (see \cref{tab:categories}) which is bundled in the acoustic signal in complicated ways, it is challenging to extract only the desired message. Typically, speech always contains consequential side information bundled into the {target information}.
It is a threat to privacy when such \emph{private side-information is transmitted} to an undesired recipient alongside the {target information}. By definition, we here assume that the legitimate recipient can be trusted with the side information and that any undesirable use of it is labeled as an undesirable service. 

The difference between the two forms of information makes an important difference in approaches to mitigation. {Transmitting the {target information} to the recipient is the desired outcome of the interaction, and we can thus not complete the desired task without sending the message. In some cases, though, we can reduce the amount of information content in the {target information}, and more generally, we should always secure the transmission, for example, through
cryptography (see \cref{sec:isolation,sec:crypto,sec:design_of_protections}). Some cryptographic methods, such as homomorphic encryption, even keep the message secret from the receiver but allow the processing of a desired task. }
However, when side-information is exposed, as additional methods we can also use signal processing to better remove, replace, or distort such side-information (see \cref{sec:isolation}).


\subsection{\texorpdfstring{{Attack types}}{Attack types}}\label{sec:tactic}

There are three main types of attacks: motivated, opportunistic, and incidental. 
In \emph{motivated attacks}, the attacker attempts to extract information about a specific person. First, the attacker can try to extract private information from speech samples from that person. For example, an attacker could extract private health information from a public speech of a famous person. Second, an attacker can browse through a database of anonymous speakers in an attempt to identify the target person. This is known as \emph{re-identification}, as the attacker thus recovers the identity of a speaker from a database where the identities were not available. When the target speaker is re-identified, the attacker can then extract private information from the re-identified sample. Moreover, the fact that the voice of the target person \emph{is in} the database can be private or sensitive information in itself. For example, if a database contains speech samples from the members of a particular association, then the re-identification of a target person reveals that person's affiliation with that association. 

\changed{In opportunistic attacks, the attacker uses a known weakness on a large pool of potential victims. It is similar to spam email; if the attacker sends a large enough number of such emails, then, randomly, someone triggers the trap by mistake as in a \emph{birthday attack}~\cite{menezes2018handbook}. } In this form of attack, anyone could be the victim. {For example, a speech service could analyze private speaker attributes (for, e.g., health problems) or affiliations (with, e.g., minority groups), for purposes of targeted advertisement or surveillance.} Where motivated attacks typically target high-value victims such as famous persons, where both protections and rewards are high, \changed{opportunistic} attacks typically target ordinary people, where the likelihood of success is larger but rewards are smaller. 

The third category of ``attacks'' is \emph{incidental}, where an agent gains excessive access to information unintentionally or by mistake. For example, if a speech device misinterprets an utterance, it can trigger functionalities that reveal private information to unauthorized parties. Such incidents are thus neither intentional nor nefarious, but in the interest of consistency, here they are still called attacks.

\subsection{Attack model}\label{sec:attackmodel}
In the design of privacy-protecting methods, attack models simulate and quantify an attacker's performance. In other words, by putting themselves in the attacker's place, safeguards designers can better understand pertinent aspects of weaknesses and quantify risks.

\Cref{fig:anonobj} illustrates an attack model for measuring the extent of privacy~\cite{maouche20_interspeech}. We assume that there is some private data, which is anonymized such that it can be used in a trusted task. The attacker has access to the anonymized data, which is here called ``public data''. Observe that the public data is not necessarily openly available for anyone to see, but it only indicates that the data is sufficiently freely available that the attacker has access to it. The attacker has also access to some other data about speakers, found from some other source (found data), which helps in extracting private information. Again, the term ``found data'' is a loose term and denotes any data other than public data that is available to the attacker.

To complement an Attack(er) model, it can also be helpful to define a \emph{Protector model} for the use case of the system. Jointly, the two can be documented using the Scenario-of-Use scheme, as illustrated in \cref{tab:scenarioofuse}~\cite{rahman24_spsc}. The accurate definition of the protector model and use-cases is essential, since they define both the target information needed as well as how performance should be evaluated~\cite{meyer2025usecasesvoiceanonymization}. 

    \begin{table*}[t!]
        \centering
        
        \caption{{The Scenario-of-Use scheme~\cite{rahman24_spsc}.}}
        \label{tab:scenarioofuse}
        {
        \begin{tabular}{p{1.5cm}p{6cm}|p{1.5cm}p{6cm}}
             \multicolumn{2}{c|}{\textbf{Attacker Model (Attack on Privacy)}} & \multicolumn{2}{c}{\textbf{Protector Model (Protection of Privacy)}} \\ \hline
             Objective & The information about the target speakers the attacker seeks to obtain. &
             Objective & 
             \textbf{Defense objective:} The information in the spoken audio that must be protected.\newline
             \textbf{Utility objective:} The information in the spoken audio that must be retained.
             \\\hline
             Opportunity & The attacker's access to the target, including the availability of the target speakers’ audio. Also, access to (and any knowledge about) the protection. &
             Opportunity & The possibilities available for protection (i.e., counter-measures). Also, access to (and any knowledge about) the attack.
             \\\hline
             \changed{Additional\newline resources} & The knowledge, data, and compute available to the attacker to carry out the attack (beyond the access to the target specified under Opportunity). &
             \changed{Additional\newline resources} & The knowledge, data, and compute available to the protector to defend against attack (beyond the access to the attack specified under Opportunity).
             \\\hline
        \end{tabular}}
    \end{table*}

Attacks can be classified according to the amount of information available for the attacker, such as information about the speakers and about the models used. {Access to speech samples from the target user (known as \emph{enrollment data}) is particularly important for the attacker. }
For example, in the anonymization task of the VoicePrivacy 2022 challenge, the objective was to replace (pseudonymize) speaker identity but retain all other speech characteristics, such as linguistic information. The anonymized sentences are known as \emph{trial} utterances. The attack scenarios were then classified as~\cite{tomashenko2022voiceprivacy}:
\begin{enumerate}
    \item \emph{Unprotected}: no anonymization is performed; attackers have access to original trial and enrollment data. 
    \item \emph{Ignorant attacker}: Trial data is anonymized, but attackers are unaware of it, hence they use original data for enrollment.
    \item \emph{Lazy-informed}: Both trial and enrollment data are available to attackers, but anonymized with a different pseudo-speaker.
    \item \emph{Semi-informed}: Similar to lazy-informed, attackers can also train their model using the same anonymization system but with different pseudo-speakers.
\end{enumerate}
With increasing information, obviously, the attacker's task becomes easier, and the accuracy of the attacker's models improves. Further, such attack scenarios can then be devised according to the specific use case. For example, suppose the task is to anonymize emotional or health status. In that case, we can define different attack scenarios depending on the extent to which the attacker has information about categories of emotions and health status used in anonymization.    

\def\db{%
            \draw[line width=.022cm] (0,.25) ellipse (.4cm and .2cm);
            \draw[line width=.022cm] (0,-.25) ellipse (.4cm and .2cm);
            \draw[fill=white,color=white] (-.4,0) -- (.4,0) -- (.4,-.25) -- (-.4,-.25) -- (-.4,0);%
            \draw[line width=.022cm] (-.4,-.25) -- (-.4,.25);%
            \draw[line width=.022cm] (.4,-.25) -- (.4,.25)}%

\begin{figure}[t]
    \centering
    \footnotesize
    \begin{tikzpicture}
        
        
        \node[matrix,column sep=.1cm] at (0,0) {
            \node {\parbox{1.2cm}{\centering Private data}}; 
            & \node{Anonymization}; 
            & \node{\parbox{1.2cm}{\centering Public data}}; 
            & \node{\parbox{1.5cm}{\centering Downstream application}}; \\
            \db; \node[minimum width=.8cm] (private) {}; 
            & \node[draw,rounded corners=.03cm,minimum width=.8cm,minimum height=.5cm,line width=.022cm] (anon) {}; 
            & \db; \node[minimum width=.8cm,minimum height=.9cm] (public) {}; 
            & \node[draw,rounded corners=.03cm,minimum width=.8cm,minimum height=.5cm,line width=.022cm] (asr) {}; 
            \\
            \node{~};\\ \node{~};\\ 
            & \db; \node[minimum width=.8cm] (found) {};
            & \node[draw,red,rounded corners=.03cm,minimum width=.8cm,minimum height=.5cm,line width=.022cm] (attack) {!}; 
            & \node[red] (out) {\parbox{1.4cm}{\centering Private\newline information}};
            \\
            & \node{\parbox{1.6cm}{\centering Found data}}; 
            & \node{\parbox{1.2cm}{\centering Attacker}};\\
        };        

        \draw[->,thick] (private) -- (anon);
        \draw[->,dashed,thick] (anon) -- (public);
        \draw[->,thick,dashed] (public) -- (asr);
        \draw[->,dashed,red,thick] (public) -- (attack);
        \draw[->,red,thick] (found) -- (attack);
        \draw[->,red,thick] (attack) -- (out);
        
    \end{tikzpicture}
    \caption{An \emph{attack model} for the evaluation of privacy-preserving anonymization, where private data is anonymized to remove private information, and the anonymized data is shared publicly. An attacker uses any available (found) data and anonymized public data to infer private information contrary to the users' preferences. Anonymized data flow is indicated by dashed lines, and the attack is represented by red lines with an exclamation mark.}
    \label{fig:anonobj}
\end{figure}


\subsection{Attack Surfaces}\label{sec:surface}

We categorize scenarios according to the attack surface from where information is extracted (see \cref{fig:allthreats}){:} 
a channel between remote (cloud) services (\cref{sec:cloudserviceeve}),
a user interface to the local (edge) device (\cref{sec:falseactivation}) or
to the cloud service (\cref{sec:cloudeve}),
from the local network (\cref{sec:neteve}), 
the acoustic pathway (\cref{sec:acousticeve}),
or 
through the shared user interface of the local device (\cref{sec:shareddevice}). 
We consider only cases where a piece of technology is receiving or transmitting information and exclude human-to-human communication without devices. We also assume that devices and network connections are secured such that only authorized services can communicate with them. The threats we focus on are thus illustrated with numbered, solid red lines in \cref{fig:allthreats}.

\subsubsection{\cloudleak}\label{sec:cloudserviceeve}
\begin{figure}[t]
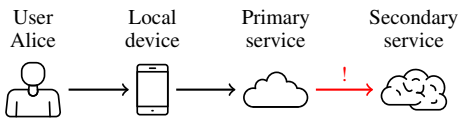

\centering
\footnotesize
    
    \begin{tikzpicture}
        \node[matrix,column sep=.6cm] at (0,0) { 
            \node {\parbox{.8cm}{\centering User\\Alice}}; & \node{\parbox{.8cm}{\centering Local\\ device}}; & \node{\parbox{.9cm}{\centering Remote\\service}}; & \node{\parbox{1.2cm}{\centering Secondary\\service}};\\
            \node (user1) {\includegraphics[scale=0.15]{pdf/AdobeStock_290226498_user_extract.pdf} };
            &
            \node (edge1) {\includegraphics[scale=0.15]{pdf/AdobeStock_290226498_phone_extract.pdf} };
            &
            \node (cloud1) {\includegraphics[scale=0.15]{pdf/AdobeStock_290226498_cloud_extract.pdf} };
            &
            \node (threat1) {\includegraphics[scale=0.15]{pdf/AdobeStock_290226498_cloud2_extract.pdf} };
            \\
        };
        \usetikzlibrary {arrows.meta} 
        \draw [->,thick] (user1) -- (edge1);
        \draw [->,thick] (edge1) -- (cloud1);
        \draw [->,red,thick] (cloud1) -- node[midway,above]{!} (threat1);        
    \end{tikzpicture}
\caption{Threat scenario ``\emph{\cloudleak}", where a user Alice accesses a (primary) remote service using a local device, but the information is shared to a secondary service contrary to preferences (red arrow and exclamation mark). }
    \label{fig:cloudthreat}
\end{figure}

Suppose a user accesses a remote server, known as a \emph{cloud service}, through a local device, known as an \emph{edge device} (see \cref{fig:cloudthreat}). The cloud service is thus assumed to have legitimate access to the \changed target information of the user. The cloud server, however, can then use that {target information} or bundled side information for some other purpose than that requested by the user, extract more information than anticipated, combine it with other information, or share information with a third party. For example, the cloud server of a voice assistant could inappropriately share information with an advertiser.

Another example is data collection for training of deep neural networks. A majority of advanced speech processing today uses machine learning, which has to be trained using large databases of speech. The best quality data correspond closely to scenarios where the services are used. Recording users' interactions with their devices is then attractive for training improved models since it corresponds precisely to the use case. This presents a considerable threat to privacy because such unrestricted recording could capture a wide range of private information, including all interactions with the device, but potentially also any and all speech in the vicinity of the device.

Observe that though \cref{fig:cloudthreat,fig:allthreats} show the secondary remote service as an impersonal cloud server, this analysis includes harms done by the people and organizations that control those servers. For the current analysis, it, however, does not make a difference who is the actor in perpetuating harm or is responsible for the harm caused to the user, but the pertinent difference is the first point where private information departs from the intended flow of information (cf. \cref{sec:cloudeve}).

\begin{figure}[t]
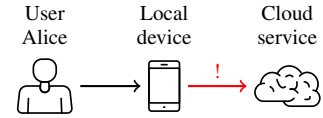

\centering
\footnotesize
    
    \begin{tikzpicture}
        \node[matrix,column sep=.6cm] at (0,0) { 
            \node {\parbox{.8cm}{\centering User\\Alice}}; & \node{\parbox{.8cm}{\centering Local\\ device}}; & \node{\parbox{.9cm}{\centering Remote\\service}}; \\
            \node (user1) {\includegraphics[scale=0.15]{pdf/AdobeStock_290226498_user_extract.pdf} };
            &
            \node (edge1) {\includegraphics[scale=0.15]{pdf/AdobeStock_290226498_phone_extract.pdf} };
            &
            \node (threat1) {\includegraphics[scale=0.15]{pdf/AdobeStock_290226498_cloud2_extract.pdf} };
            \\
        };
        \usetikzlibrary {arrows.meta} 
        \draw [->,thick] (user1) -- (edge1);
        \draw [->,thick,red] (edge1) -- node[midway,above]{!} (threat1);        
    \end{tikzpicture}
\caption{Threat scenario ``\emph{\falseactivation}", where a user Alice accesses a local device, but the information is shared to a cloud service contrary to preferences (red arrow and exclamation mark). }
    \label{fig:falseactivation}
\end{figure}

\subsubsection{\falseactivation}\label{sec:falseactivation}
Devices with speech interfaces can hear all conversations in the same acoustic space as the device and, therefore, need mechanisms to determine which utterances are intended for the speech interface (see \cref{fig:falseactivation}). A popular approach is to use a specific utterance, known as a wake word, to start all interactions with the interface~\cite{wang2021wake}. The wake word serves as a rudimentary password, preventing the interface from activating when speech is not directed to the device. Unfortunately, designing wake word detectors is non-trivial, and they will occasionally make mistakes. They might sometimes miss a wake word when it is spoken (false negative) or mistake some other unrelated sound as a wake word (false positive). While false negatives are annoying for the user when the service does not activate, false positives potentially present serious threats to privacy. 
In some famous cases, speech interfaces have been activated from sounds on the television to buy unwanted items at the users' behest, and users' private conversations have been leaked to third parties~\cite{liptak2017alexa,moscaritolo2018alexa}. 



\begin{figure}[t]
    \centering 
    \footnotesize
    \begin{tikzpicture}
        \node[matrix,column sep=.6cm] at (0,0) {
            &\node {User}; & \node{\parbox{.8cm}{\centering Local device}};  \\
            \node{Alice}; &\node (user1) {\includegraphics[scale=0.15]{pdf/AdobeStock_290226498_user_extract.pdf} };
            & \node (edge1) {\includegraphics[scale=0.15]{pdf/AdobeStock_290226498_phone_extract.pdf} }; \\
            \node{Eve}; &\node (user2) {\includegraphics[scale=0.15]{pdf/AdobeStock_290226498_user_extract.pdf} };
            & \node (edge2) {\includegraphics[scale=0.15]{pdf/AdobeStock_290226498_phone_extract.pdf} };
            \\
        };        
        \node[matrix,column sep=.6cm] at (4,.13) {
            \node{\parbox{.9cm}{\centering Remote\\service}}; & \node{\parbox{.9cm}{\centering Cloud storage}}; \\
            \node{~};\\\node{~};\\
            \node (cloud) {\includegraphics[scale=0.15]{pdf/AdobeStock_290226498_cloud_extract.pdf} };
            & \node (db) {\includegraphics[scale=0.03]{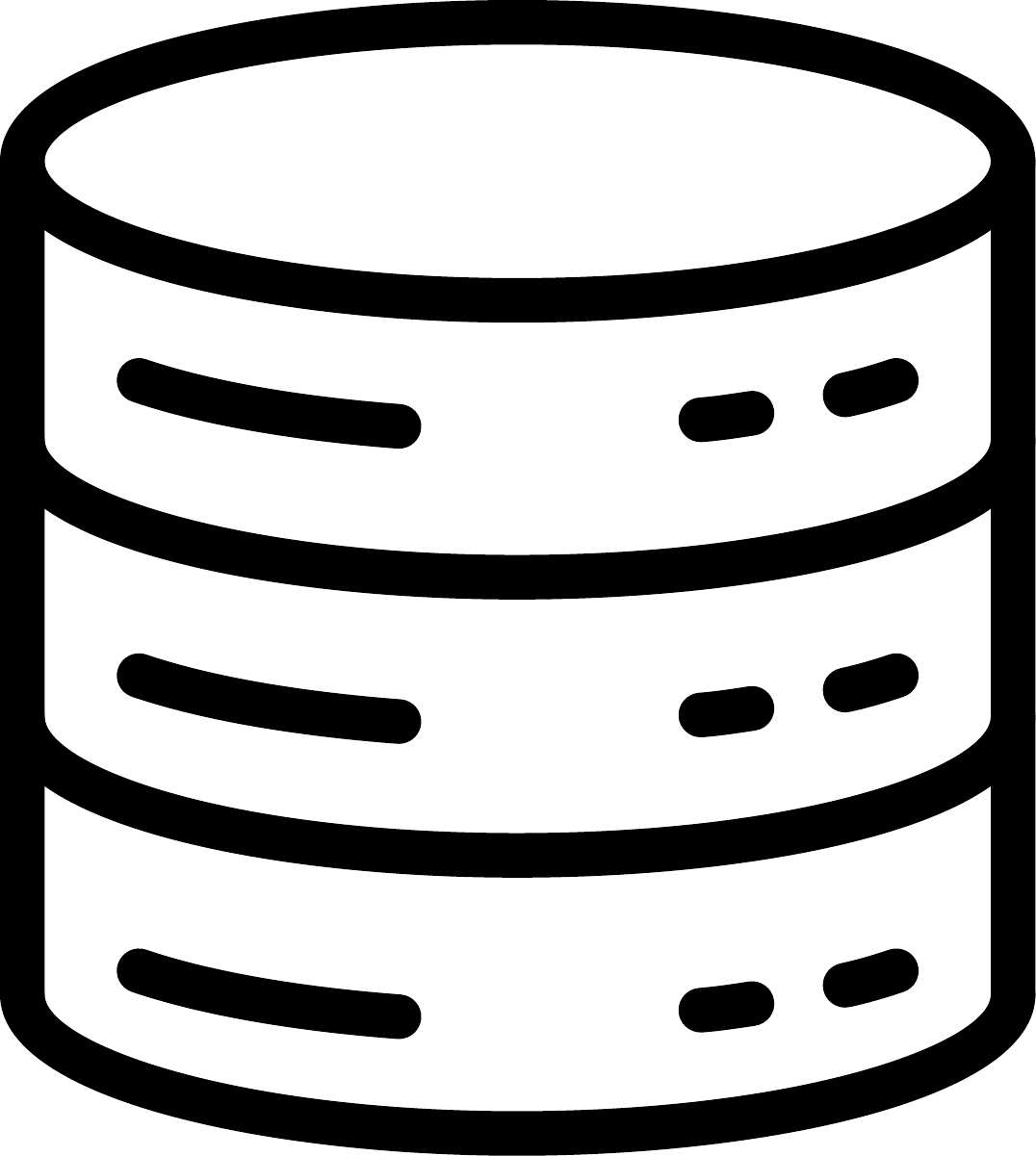} }; 
            \\
            \node{~};\\\node{~};\\
        };        
        \usetikzlibrary {arrows.meta} 
        \draw [->,thick] (user1) to (edge1);
        \draw [->,thick,red] (edge2) to node[midway,below] {!} (user2);
        \draw [->,thick] (edge1) to (cloud);
        \draw [<->,thick] (cloud) to (db);
        \draw [->,thick,red] (cloud) to node[midway,below] {!} (edge2);
    \end{tikzpicture}
    \caption{Threat scenario ``\emph{\cloudaccess}", where user Alice uses a remote service through a local device, where it is potentially stored in the cloud and shared with user Eve, contrary to Alice's preferences (red arrow and exclamation mark).}
    \label{fig:cloudapi}
\end{figure}

\subsubsection{\cloudaccess}\label{sec:cloudeve}
\Cref{fig:cloudapi} illustrates the threat where a user Alice accesses a cloud service through an edge device, and where some private information of Alice is stored. A second user, Eve, can then access the same cloud service through another device and potentially gain access to the stored private information, contrary to Alice's preferences. 
This attack surface is similar to the \cloudleak~scenario, with the main difference being the recipient, which is here a person, whereas, in the \cloudleak~scenario, it is an automated agent. 

An example of this threat is medical records, which can be collected from patients and stored in a centralized database. A researcher with authorization to access the database can then potentially extract private information beyond the expected~\cite{Teixeira2018}. 



\begin{figure}[t]
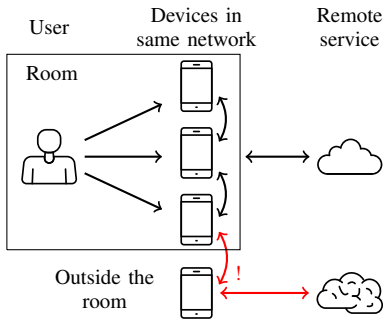

    \centering 
    \footnotesize
    \begin{tikzpicture}
        \node[matrix,column sep=.6cm] {
            \node {User}; & \node{\parbox{1.6cm}{\centering Devices in same network}}; & \node{\parbox{.9cm}{\centering Remote\\service}}; \\
            \node (user) {\includegraphics[scale=0.15]{pdf/AdobeStock_290226498_user_extract.pdf} };
            &
            \node[] (wasn) {
            \parbox{1.1cm}{\centering
                {\includegraphics[scale=0.15]{pdf/AdobeStock_290226498_phone_extract.pdf} } \\[.2cm]
                {\includegraphics[scale=0.15]{pdf/AdobeStock_290226498_phone_extract.pdf} }\\[.2cm]
                {\includegraphics[scale=0.15]{pdf/AdobeStock_290226498_phone_extract.pdf} }\\[.1cm]
                }
                };
            &
            \node (cloud) {\includegraphics[scale=0.15]{pdf/AdobeStock_290226498_cloud_extract.pdf} };
            \\            
            & \node (local2) {\includegraphics[scale=0.15]{pdf/AdobeStock_290226498_phone_extract.pdf} };
            & \node (cloud2) {\includegraphics[scale=0.15]{pdf/AdobeStock_290226498_cloud2_extract.pdf} }; \\
        };        
        \usetikzlibrary {arrows.meta} 
        \draw [->,thick] (user) to ++(1.5cm,0);
        \draw [->,thick] (user) to ++(1.5cm,.7cm);
        \draw [->,thick] (user) to ++(1.5cm,-.7cm);
        \draw [<->,thick] (wasn) to (cloud);
        \draw [<->,thick,red] (local2) to (cloud2);
        \draw [<->,thick] (wasn)++(.3cm,.2cm) to [bend right] ++(0,.6cm);
        \draw [<->,thick] (wasn)++(.3cm,-.2cm) to [bend left] ++(0,-.6cm);
        \draw [<->,thick,red] (wasn)++(.3cm,-1cm) to [bend left] node [midway,right,yshift=-.2cm] {!} ++(0,-.7cm);
        \draw (-2.6,-1.15) -- (-2.6,1.45) -- (.5,1.45) -- (.5,-1.15) -- (-2.6,-1.15);
        \node at (-1.3,-1.67) {\parbox{1.7cm}{\centering Outside the room}}; 
        \node at (-2,1.2) {Room};
    \end{tikzpicture}
    \caption{Threat scenario ``\emph{\WASNauthentication}", where a user uses a service through a wireless acoustic sensor network (WASN) inside a room, but another device, outside the room in the same network, joins the distributed sensor network and shares information contrary to preferences (red arrow and exclamation mark).}
    \label{fig:wasnthreat}
\end{figure}

\subsubsection{\WASNauthentication}\label{sec:neteve}
The audio quality of speech pickup (in terms of, for example, signal-to-noise ratio or perceptual quality), as well as the usability of voice interfaces, can be improved by using all available connected devices with microphones in the same room or acoustic space~\cite{das2021enhancement,zarazaga2020acoustic,sigg2020provable}. Such collaboration can be realized with {wireless acoustic sensor networks (WASNs)}, where several independent devices simultaneously pick up speech, and those channels are combined to obtain a high-quality signal (see \cref{fig:wasnthreat})~\cite{Nelus2022,zarazaga2022thesis}. A sensitive question, however, is authorization: Which devices are allowed to share information?

One approach is to assume that devices that are in the same room or acoustic space can hear the same signal~\cite{zarazaga2022thesis,koppelmann22_interspeech}. Presence in the same space is thus already an implicit authorization to participate in a joint signal pickup. To protect privacy, it is then necessary to determine which devices reside in the same acoustic space. Devices in a different room can belong to the same company or family and be connected to the same network, but they are still outside the sphere of the current discussion. Without proper authorization mechanisms, devices outside the room could then gain access to private speech inside the room.

\begin{figure}[t]
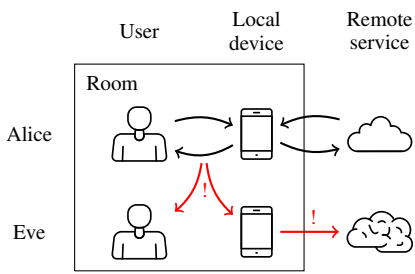

    \centering 
    \footnotesize
    \begin{tikzpicture}
        \node[matrix,column sep=.6cm] {
            &\node {User}; & \node{\parbox{.8cm}{\centering Local\\ device}}; & \node{\parbox{.9cm}{\centering Remote\\service}}; \\ \node{~}; \\ \node{~}; \\ \node{~}; \\
            \node{Alice}; &\node (user) {\includegraphics[scale=0.15]{pdf/AdobeStock_290226498_user_extract.pdf} };
            &
            \node (edge) {\includegraphics[scale=0.15]{pdf/AdobeStock_290226498_phone_extract.pdf} };
            &
            \node (cloud) {\includegraphics[scale=0.15]{pdf/AdobeStock_290226498_cloud_extract.pdf} };
            \\ \node{~}; \\ \node{~}; \\ \node{~}; 
            \node{Eve}; &\node (Eve) {\includegraphics[scale=0.15]{pdf/AdobeStock_290226498_user_extract.pdf} };
            &
            \node (edge2) {\includegraphics[scale=0.15]{pdf/AdobeStock_290226498_phone_extract.pdf} };
            &
            \node (cloud2) {\includegraphics[scale=0.15]{pdf/AdobeStock_290226498_cloud2_extract.pdf} };
            \\
        };
        \usetikzlibrary {arrows.meta} 
        \draw [->,thick] (user) to [bend left=20] (edge);
        \draw [->,thick] (edge) to [bend left=25] node [midway] (mid) {} (user);
        \draw [<-,thick] (edge) to [bend left=20] (cloud);
        \draw [<-,thick] (cloud) to [bend left=20] (edge);
        \draw [->,red,thick] (mid) to [bend left=20] node[midway,right]{!~} (Eve);        
        \draw [->,red,thick] (mid) to [bend right=20] (edge2);        
        \draw [->,thick,red] (edge2) to node[midway,above] {~!} (cloud2);
        \draw (-1.8,-1.8) -- (-1.8,.95) -- (1.25,.95) -- (1.25,-1.8) -- (-1.8,-1.8);
        \node at (-1.3,.7) {Room};
    \end{tikzpicture}
    \caption{Threat scenario ``\emph{\speechinterface}", where a user, Alice, accesses an edge device (potentially including also a remote service), but another local person, Eve, or device in the same room or acoustic space overhears the speech interaction contrary to preferences (red arrow and exclamation mark).}
    \label{fig:reproductionthreat}
\end{figure}

{
\subsubsection{\speechinterface}\label{sec:acousticeve}
\Cref{fig:reproductionthreat} illustrates a scenario where a device or person overhears a discussion between a user, Alice, and a local device. The defining property is that the leak is in the acoustic pathway, necessitating that the eavesdropper is physically present in the same acoustic space as the private communication. The difference between these leaks and False Activation (\cref{sec:falseactivation,fig:falseactivation}) is that here, the eavesdropper is collecting data inappropriately, whereas in False Activation, an incorrect functionality is activated.
}

In the case where Eve overhears Alice's speech, we assume that people have a learned awareness of which other people are present in the same room and adjust their speech accordingly. Then, either Alice does not mind that Eve overhears her speech (inconsequential information), Alice can change her speech style to a whisper, or change the content such that Eve does not hear anything private (reduced information transmission), or she can go to a different room to continue the interaction in private (modified acoustic channel). 
Whether users are similarly aware of devices present in the room, through anthropomorphization of devices or otherwise, and adjust their speech accordingly is an open question.

\begin{figure}[t]
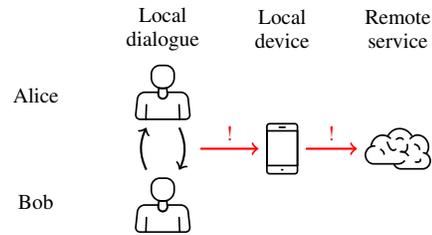

    \centering 
    \footnotesize
    \begin{tikzpicture}
        \node[matrix,column sep=.5cm,row sep=0cm] {
            &\node {\parbox{1cm}{\centering Local\\dialogue}}; & \node{\parbox{.8cm}{\centering Local\\ device}}; & \node{\parbox{.9cm}{\centering Remote\\service}}; \\
            \node{\parbox{1cm}{\centering Alice\\[1.1cm]Bob}}; & \node[yshift=.75cm] (user) {\includegraphics[scale=0.15]{pdf/AdobeStock_290226498_user_extract.pdf} };
            \node[yshift=-.75cm] (user2) {\includegraphics[scale=0.15]{pdf/AdobeStock_290226498_user_extract.pdf} };
            &
            \node (edge) {\includegraphics[scale=0.15]{pdf/AdobeStock_290226498_phone_extract.pdf} };
            &
            \node (threat) {\includegraphics[scale=0.15]{pdf/AdobeStock_290226498_cloud2_extract.pdf} };
            \\
        };
        \usetikzlibrary {arrows.meta} 
        \draw [->,thick] (user) to [bend left=25] (user2);
        \draw [->,thick] (user2) to [bend left=25] (user);
        \draw [->,thick,red] (user2) ++(.5,.75cm) -- node[midway,above]{!} (edge);
        \draw [->,red,thick] (edge) -- node[midway,above]{!} (threat);        
    \end{tikzpicture}
    \caption{Threat scenario ``\emph{\discussionleak}", where a dialogue between two users, Alice and Bob, is picked up by a local device and potentially shared with a remote service contrary to preferences (red arrow and exclamation mark).}
    \label{fig:localthreat}
\end{figure}

{
\subsubsection{\discussionleak}\label{sec:discussionleak}
\Cref{fig:localthreat} illustrates a discussion between two persons overheard by a local device and, like \falseactivation, triggers unintended functionalities in the local device. The main difference to a \speechinterface{} (\cref{sec:acousticeve,fig:reproductionthreat}) is that the type of speech is different. Interaction with a speech interface at the current level of technology is, roughly speaking, limited to speech commands, question-answer pairs, and one-sided monologues. In contrast, spoken human-to-human interaction has potentially a much more dynamic character. This is potentially beneficial in that the speaking style of human-to-human interaction can be recognized, thereby preventing unintended triggering of functionalities. 
}

\begin{figure}[t]
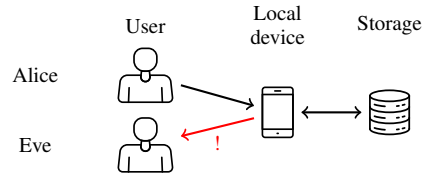

    \centering 
    \footnotesize
    \begin{tikzpicture}
        \node[matrix,column sep=.6cm] at (0,0) {
            &\node {User};  \\
            \node{Alice}; &\node (user1) {\includegraphics[scale=0.15]{pdf/AdobeStock_290226498_user_extract.pdf} };\\
            \node{Eve}; &\node (user2) {\includegraphics[scale=0.15]{pdf/AdobeStock_290226498_user_extract.pdf} };
            \\
        };        
        \node[matrix,column sep=.4cm] at (3.2,.13) {
            \node{\parbox{.9cm}{\centering Local\\ device}}; & \node{\parbox{.9cm}{\centering Storage}}; \\
            \node{~};\\\node{~};\\
            \node (edge1) {\includegraphics[scale=0.15]{pdf/AdobeStock_290226498_phone_extract.pdf} };
            & \node (db) {\includegraphics[scale=0.03]{pdf/AdobeStock_336370726_db_extract.pdf} }; 
            \\
            \node{~};\\\node{~};\\
        };        
        \usetikzlibrary {arrows.meta} 
        \draw [->,thick] (user1) to (edge1);
        \draw [->,thick,red] (edge1) to node[midway,below] {!} (user2);
        \draw [<->,thick] (edge1) to (db);
    \end{tikzpicture}
    \caption{Threat scenario ``\emph{\shareddevice}", where user Alice uses an edge device, where information is stored and shared with another user, Eve, contrary to Alice's preferences (red arrow and exclamation mark).}
    \label{fig:shareddevice}
\end{figure}

\subsubsection{\shareddevice}\label{sec:shareddevice}
Many practical scenarios include \emph{multiple users} sharing one or several {speech-operated devices}. For example, a family can share a smart speaker or television, an office meeting room can have smart devices, and customer service points can have a phone shared across duty officers. Each of these devices can collect private information about the user(s) over time and can potentially share it with other users (see \cref{fig:shareddevice}). Notably, this scenario highlights that a leak happens over a distance in \emph{time}, whereas threats occurring over a \emph{spatial} distance often receive most of the attention.

This threat model clearly demonstrates that devices and services used by multiple users need to employ access control and authorization management if they store any private information. It is also obviously related to many security threats -- unauthorized access to devices should be prevented -- but those are outside the scope of this work.

\section{Protections}\label{sec:protections}

The main avenues for protecting privacy in speech technology are:
\begin{enumerate*}
    \item \emph{Information isolation}, where private information is filtered or otherwise removed from a transmitted (or stored) channel of speech information (\cref{sec:isolation}),
    \item \emph{Secure processing}, where access to private information is prevented with methods of cryptography (\cref{sec:crypto}),
    \item \emph{Privacy-preserving architectures}, where private information is not sent to the cloud, but processed either only locally (on the edge), or in a distributed network of (nearby) devices (\cref{sec:diss}),
    \item \emph{Acoustic interventions}, where reproduction of audio is changed such that private information is not revealed to eavesdroppers (\cref{sec:reproduction}), 
    \item \emph{Improving {utility}}, where the intended functionality of a system is improved such that processing errors leak less private information (\cref{sec:improve}) {and}
    \item {\emph{Design of provable protections} by applying theoretical frameworks such as differential privacy (\cref{sec:design_of_protections}).}
\end{enumerate*}

\subsection{Information isolation}\label{sec:isolation}
By isolation of information, we refer to the design of a channel of information where the target information is isolated from other (undesired) private information. {This applies at least and in particular to the attack vectors \cloudleak{} (\cref{sec:cloudserviceeve}), \falseactivation{} (\cref{sec:falseactivation}), and \cloudaccess{} (\cref{sec:cloudeve}), where a reduction in flow of private information would reduce threats to privacy. Central concepts in information isolation include
\begin{enumerate*}
    \item \emph{information bottlenecks} (or privacy funnels)
, where information is compressed to a low-capacity channel such that only target information is passed through (see \cref{sec:bottle}) and where the neural network implementation usually takes the structure of an \emph{autoencoder}~\cite{li2023comprehensive},
    \item \emph{adversarial training}, where an information channel (such as an information bottleneck) is trained to contain target information to fulfill the trusted task and simultaneously trained such that a competing, attacking adversarial model fails to extract private information (threat task), (see \cref{sec:adversarial}) and
    \item \emph{disentanglement}, where different attributes of speech are isolated to separate channels, as a generalization of information bottlenecks (see \cref{sec:diss}). 
\end{enumerate*}
Note that these concepts are overlapping and should be seen as characterizations of algorithms and methodology, rather than alternative approaches.
}


\begin{figure}[t]
    \centering
    \footnotesize
    \begin{tikzpicture}

        \begin{scope}[yshift=.4cm]
            \draw[fill=lightgray!20,color=lightgray!20] (-2.7,.5) -- (2.9,.5) -- (2.9,-1.9) -- (-2.7,-1.9) -- (-2.7,.5);
            \draw[fill=gray!30,color=gray!30] (-2.4,.4) -- (1.3,.4) -- (1.3,-1.1) -- (-2.4,-1.1) -- (-2.4,.4);
            \node at (-1.4,.2) {Inference};
            \node at (-1.4,-1.7) {Training};        
        \end{scope}
        
        \node[matrix,column sep=.1cm] at (0,0) {
            \node {\parbox{1.2cm}{\centering User}}; 
            & \node (enclabel) {Encoder}; 
            & \node (bnlabel) {Bottleneck}; 
            & \node (declabel) {Decoder}; 
            & \node{\parbox{1.1cm}{\centering Loss}};\\
            \node{~}; \\ \node{~}; \\ \node{~}; \\
            \node (user1) {\includegraphics[scale=0.15]{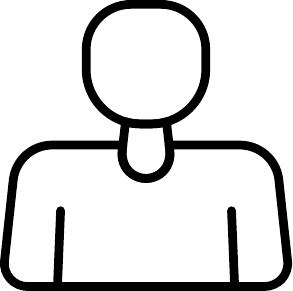} };
            & \draw node[rounded corners=.03cm,minimum size=.5cm,line width=.022cm] (enc) {}; \draw[rounded corners=.03cm, line width=.022cm] (-.25,0) -- (-.25,.4) -- (.25,.2) -- (.25,-.2) -- (-.25,-.4) -- (-.25,0); 
            & \draw node[rounded corners=.03cm,minimum size=.5cm,line width=.022cm] (bn) {}; \draw[rounded corners=.03cm, line width=.022cm] (-.25,0) -- (-.25,.2) -- (.25,.2) -- (.25,-.2) -- (-.25,-.2) -- (-.25,0); 
            & \draw node[rounded corners=.03cm,minimum size=.5cm,line width=.022cm] (dec) {}; \draw[rounded corners=.03cm, line width=.022cm] (.25,0) -- (.25,.4) -- (-.25,.2) -- (-.25,-.2) -- (.25,-.4) -- (.25,0); 
            & \draw node[rounded corners=.03cm,minimum size=.4cm,line width=.022cm,draw] (loss) {}; 
            &&& \node(out){Output};
            \\ 
            \node{~};\\\node{~};\\
            &&&& \node(label){Label};
            \\
        };        

        \usetikzlibrary{decorations.pathreplacing,angles,quotes}
        \draw[decoration={brace,raise=5pt},decorate] ($(enclabel.west) + (0,0cm)$) -- node[above=8pt] {Local device} ($(bnlabel.east) + (0,0cm)$);
        \draw[decoration={brace,raise=5pt},decorate] (declabel.west) -- node[above=8pt] {\parbox{.9cm}{\centering Remote\\service}} (declabel.east);
        
        \draw[->,thick] (user1) -> (enc);
        \draw[->,thick] (label) -> (loss);
        \draw[->,thick] (enc) -> (bn);        
        \draw[->,thick,dashed] (bn) -> (dec);
        \draw[->,thick,dashed,red] ($.5*(bn)+.5*(dec)$) -- node[midway,right]{!} ++(0,-.75);
        \draw[->,thick] ($(dec.east)+(0,0)$) -> ($(loss.west)+(0,0)$);
        \draw[->,thick] ($(dec.east)+(0,0)$) -- ++(.1,.0) -- ++(0,.5) -- ++(1.5,0) -- ++(0,-.5) -- (out);
        \draw[->,thick,dashed,red] ($(out.west)-(.24,0)$) -- node[midway,right]{!} ++(0,-.75);
    \end{tikzpicture}
    \caption{Training of a privacy-preserving speech analysis method with the \emph{information bottleneck} principle. Potential points of attack are indicated by red, dashed arrows and exclamation marks.}
    \label{fig:bottleneck}
\end{figure}

\begin{figure}[t]
    \centering
    \footnotesize
    \begin{tikzpicture}
        
        \draw[fill=lightgray!20,color=lightgray!20] (-2.7,.5) -- (2.9,.5) -- (2.9,-1.9) -- (-2.7,-1.9) -- (-2.7,.5);
        \draw[fill=gray!30,color=gray!30] (-2.4,.4) -- (1.3,.4) -- (1.3,-1.1) -- (-2.4,-1.1) -- (-2.4,.4);
        \node at (-1.4,.2) {Inference};
        \node at (-1.4,-1.7) {Training};        
        
        \node[matrix,column sep=.1cm] at (0,0) {
            \node {\parbox{1.2cm}{\centering User}}; 
            & \node{Encoder}; 
            & \node{Bottleneck}; 
            & \node{Decoder}; 
            & \node{\parbox{1.1cm}{\centering Loss}};\\
            \node{~}; \\ \node{~}; \\ \node{~}; \\
            \node (user1) {\includegraphics[scale=0.15]{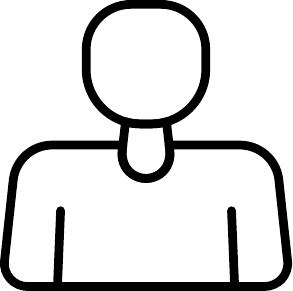} };
            & \draw node[rounded corners=.03cm,minimum size=.5cm,line width=.022cm] (enc) {}; \draw[rounded corners=.03cm, line width=.022cm] (-.25,0) -- (-.25,.4) -- (.25,.2) -- (.25,-.2) -- (-.25,-.4) -- (-.25,0); 
            & \draw node[rounded corners=.03cm,minimum size=.5cm,line width=.022cm] (bn) {}; \draw[rounded corners=.03cm, line width=.022cm] (-.25,0) -- (-.25,.2) -- (.25,.2) -- (.25,-.2) -- (-.25,-.2) -- (-.25,0); 
            & \draw node[rounded corners=.03cm,minimum size=.5cm,line width=.022cm] (dec) {}; \draw[rounded corners=.03cm, line width=.022cm] (.25,0) -- (.25,.4) -- (-.25,.2) -- (-.25,-.2) -- (.25,-.4) -- (.25,0); 
            & \draw node[rounded corners=.03cm,minimum size=.4cm,line width=.022cm,draw] (loss) {}; 
            &&& \node(out){Output};
            \\
        };        
        \draw[->,thick] (user1) -> (enc);
        \draw[->,thick] (user1) -- ++(.8,0) -- ++(0,-1) -- ($(loss.west) + (-.4,-1)$) -- ++(0,+.9) -> node[midway,below]{$-$} ($(loss.west)+(0,-0.1)$);
        \draw[->,thick] (enc) -> (bn);        
        \draw[->,thick,dashed] (bn) -> (dec);
        \draw[->,thick,dashed,red] ($.5*(bn)+.5*(dec)$) -- node[midway,right]{!} ++(0,-.75);
        \draw[->,thick] ($(dec.east)+(0,0.1)$) -> node[near end,above]{$+$} ($(loss.west)+(0,0.1)$);
        \draw[->,thick] ($(dec.east)+(0,.1)$) -- ++(.1,.0) -- ++(0,.4) -- ++(1.5,0) -- ++(0,-.5) -- (out);
        \draw[->,thick,dashed,red] ($(out.west)-(.24,0)$) -- node[midway,right]{!} ++(0,-.75);
    \end{tikzpicture}
    \caption{Training of a privacy-preserving an \emph{autoencoder} structure as an example of the information bottleneck principle. Potential points of attack are indicated by red, dashed arrows and exclamation marks.}
    \label{fig:autoencoder}
\end{figure}

\subsubsection{Information bottleneck}\label{sec:bottle}
The purpose of an information bottleneck is to use an encoder to distill the information needed to perform a downstream (trusted) task into a bottleneck, such that all other (private) information is removed. A decoder can then be used to decipher the information in the bottleneck into a format usable for the downstream task. The overall approach is generic and can be applied, for example, in cases where 
\begin{enumerate*}\item the decoder resynthesizes the signal with some attributes anonymized~\cite{wang2019vector} or 
\item information extraction scenarios such as classification~\cite{tishby2000information}, where the decoder can often be trivial.
\end{enumerate*}

A concept related to information bottlenecks is the \emph{privacy funnel}~\cite{makhdoumi2014information}. Where the original definition of the information bottleneck~\cite{tishby2000information} aims to maximize the utility of a channel with a fixed rate, the privacy funnel seeks to minimize the information rate with a fixed utility. The information bottleneck is, however, much more widely known in the community and often informally used as an umbrella term to cover both cases.

To distill only the {target information} and discard any side-information, speech information {can be} passed through a bottleneck so tight that only the legitimate message \emph{at full accuracy} can pass through~\cite{Tishby2015,makhdoumi2014information} (see \cref{fig:bottleneck}). This approach thus protects against an attacker who has access to the output of the bottleneck.
The challenges lie in designing a training methodology and model structure with a sufficiently tight bottleneck, so that the legitimate message is accurately retained while private side information is discarded. {Observe that a generic implementation of an information bottleneck does not guarantee privacy as side-information can sneak through the bottleneck alongside the target information~\cite{raj2019probing}. Instead, to be privacy preserving, a channel must be designed or optimized for privacy, warranting the term privacy funnel.
}

A frequently used approach to implementing a bottleneck is an autoencoder structure~\cite{li2023comprehensive}, consisting of an encoder, bottleneck, and a decoder, where the objective is to reconstruct the input signal from the bottleneck output~\cite{malekzadeh2018replacement,perero2022x,lu2013speech} (see \cref{fig:autoencoder}). To be privacy-preserving, the bottleneck should be sufficiently tight that the original speech signal cannot be perfectly reconstructed to resemble the original in its waveform with respect to some distance function. Designing a loss function is then challenging when the bottleneck is so tight that the decoded signal is very far from the original. One solution akin to the machine learning method known as representation learning is to feed the output of the autoencoder again to the encoder and compare the bottleneck features~\cite{neumann2019improving,chorowski2019unsupervised}.

Information content at the bottleneck can be reduced by either reducing the signal rank (viz. dimensionality reduction, e.g., \cite{perero2022x}) or by quantizing and coding the signal at a low bitrate (e.g., \cite{wang2019vector,williams2021revisiting}). The trade-off between the accuracy of the trusted task and bottleneck entropy then determines the extent of privacy (see \cref{sec:objmeasures}). Observe that dimensionality reduction is, in general, not alone sufficient because if the channel's floating- (or fixed)-point representation has $K$ dimensions each with $B$ bits, then a sufficiently advanced encoder can pack $KB$ bits of information into that channel. Since typical representations have $B=32$ or $B=64$, which can encode up to $2^B$ unique speaker identities, we see that the bottleneck is not particularly tight. As a general rule, therefore, \emph{bottlenecks should always be quantized} to the lowest bitrate sufficient to represent the target information. 

A central advantage of the information bottleneck approach is that it can give a provable level of privacy. If the bitrate of the bottleneck is $B_b$ and that of the target information is $B_t$, then the privacy leak is at most $B_{leak}\leq B_t - B_b$. This applies to single vectors of target information passed through a bottleneck with full accuracy. Streaming data through the bottleneck, however, over time, leaks a flow of private information that increases in size without bound~\cite{kairouz2015composition,backstrom2025privacydisclosure}. This approach addresses only the average disclosure of private information, without guaranteeing worst-case disclosure~\cite{nautsch20_interspeech,vali2024porcupine}.

Information bottlenecks, or more accurately, privacy funnels, are thus theoretically well-suited for privacy-preserving applications. They have also been used in other areas, such as privacy-aware transmission of images~\cite{sun2024privacyaware} and face recognition~\cite{razeghi2024deep}. However, the above-mentioned applications in speech either do not specifically apply it \emph{for privacy} or use information bottlenecks optimized for privacy with additional components.

\begin{figure}[t]
    \centering
    \footnotesize
    \begin{tikzpicture}
        
        \draw[fill=lightgray!20,color=lightgray!20] (-2.4,1.3) -- (2.3,1.3) -- (2.3,-1.4) -- (-2.4,-1.4) -- (-2.4,1.3);
        \draw[fill=gray!30,color=gray!30] (-2.2,1.2) -- (.6,1.2) -- (.6,-.3) -- (-2.2,-.3) -- (-2.2,1.2);
        \node at (-1.5,1) {Inference};
        \node at (-1.5,-1.1) {Training};        

        \node[matrix,column sep=.4cm] at (0,0) {
            \node {\parbox{.8cm}{\centering User}}; 
            & \node{Encoder}; 
            & \node{\parbox{.9cm}{\centering Trusted Task}};
            & \node{\parbox{.9cm}{\centering Trusted Loss}}; \\
            \node{~}; \\ \node{~}; \\ \node{~}; \\
            \node (user1) {\includegraphics[scale=0.15]{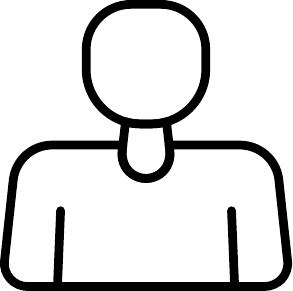} };
            & \draw node[rounded corners=.03cm,minimum size=.5cm,line width=.022cm] (enc) {}; \draw[rounded corners=.03cm, line width=.022cm] (-.25,0) -- (-.25,.4) -- (.25,.2) -- (.25,-.2) -- (-.25,-.4) -- (-.25,0); 
            & \draw node[rounded corners=.03cm,minimum width=.5cm,minimum height=.6cm,line width=.022cm,draw] (trust) {}; 
            & \draw node[rounded corners=.03cm,minimum size=.4cm,line width=.022cm,draw] (loss) {}; 
            & \node(out){Output};
            \\\node{~}; \\ \node{~}; \\
            && \draw node[rounded corners=.03cm,minimum width=.5cm,minimum height=.6cm,line width=.022cm,draw,red] (threat) {}; 
            & \draw node[rounded corners=.03cm,minimum size=.4cm,line width=.022cm,draw,red] (threatloss) {}; 
            \\ \node{~}; \\ \node{~}; \\ 
            && \node{\parbox{1cm}{\centering Threat Task}}; 
            & \node{\parbox{1cm}{\centering Threat Loss}}; \\
        };        
        \draw[->,thick] (user1) -> (enc);
        \draw[->,thick,dashed] (enc) -> (trust);
        \draw[->,thick,red,dashed] (enc)++(.85,0) -- ++(0,-1.15cm) -> node[midway, above]{!} (threat);
        \draw[->,thick] ($(trust.east)+(0,.1)$) -> node[near end, above]{$+$} ($(loss.west) + (0,.1)$);
        \draw[->,thick] ($(trust.east)+(0,.1)$) -- ++(.1,.0) -- ++(0,.4) -- ++(1.7,0) -- ++(0,-.5) -- (out);
        \draw[->,thick,red] (threat.east) -- ++(.6,0) -- ++(0,1.05) -> node[midway,below]{$-$} ($(loss.west) + (0,-.1)$);
        \draw[->,thick,red] (threat.east) -- ++(.5,0) -> node[midway,above]{} (threatloss.west);
        \draw[->,thick,red] (threatloss.south) -- ++(0,-.3) -- ($(threat.south) + (0,-.2)$) -> (threat.south);
        \draw[->,thick] (loss.south) -- ++(0,-.3) -- ++(-.42,0) -- ++(-.2,0) arc  [x radius=.1, y radius=.1, start angle=0, end angle=180] -- ($(trust.south) + (0,-.2)$) -> (trust.south);
    \end{tikzpicture}
    \caption{Training of a privacy-preserving speech analysis method with an \emph{adversarial} approach. The trusted task is authorized to extract some information, while the threat task (drawn in red) is extracting some other private information. During training, the trusted task competes with the adversarial threat task, such that in the encoder block, all private information is removed. During inference, the threat task can be ignored.}
    \label{fig:adversarial}
\end{figure}

\subsubsection{\texorpdfstring{Adversarial {training}}{Adversarial training}}\label{sec:adversarial}
{Privacy-preserving neural networks can be trained using adversarial methods}, such that side-information in the bottleneck is minimized~\cite{noe21_interspeech,srivastava19_interspeech,nelus2018itg}. \Cref{fig:adversarial} illustrates the model structure, where the trusted and threat tasks correspond respectively to utility and privacy, which in turn respectively correspond to the extraction of the {target information} and side-information. The threat task is independently optimized to extract private side information from the bottleneck output. While keeping the threat task fixed, the encoder and the trusted tasks are jointly optimized to maximize the accuracy\footnote{Accuracy refers here to some similarity measure between the message and a corrupt version of the message.}  of the {target information} \emph{and} to minimize the accuracy of private side information. 

Adversarial training thus combines maximizing and minimizing the accuracy of private side information. This forces the encoder to minimize private side information passed through the bottleneck, as the threat task aims to maximize the extraction of such information. {Similar effects can be achieved by a variety of other formulations, for example, by Siamese training, where the analysis model is specifically optimized to minimize and maximize feature distances between speakers within and across target classes~\cite{nelus19b_interspeech}. This approach should cluster extracted information into groups corresponding to the trusted task while minimizing private information.}

An advantage of an adversarial configuration is that the designer can specifically choose which type of threat and information category is removed from the data. The system can also be optimized end-to-end, such that all components are jointly optimized for the best performance. This can lead to an efficient model in terms of the trade-off between computational cost and output quality. 

The principal issue with adversarial training is that {theoretical frameworks for certifying performance (e.g.~\cite{NEURIPS2023_e8b0c97b}) are not yet widely accepted and there are thus no} formal assurances of privacy (cf. \cref{sec:objprivacy}). The best it can do is to demonstrate the extent to which the chosen adversarial model was unable to extract private side information from the selected category of private information. We can thus expect that some other attacker with a better model could still extract private side-information \emph{and} that other categories of private information are potentially present in the bottleneck output.


\begin{figure}[t]
    \centering
    \footnotesize
    \begin{tikzpicture}
        
        
        \node[matrix,column sep=.1cm] at (0,0) {
            \node {\parbox{1.5cm}{\centering User}}; 
            & \node{Encoder}; 
            & \node{Bottleneck}; 
            &&& \node{Decoder}; \\
            \node (user1) {\includegraphics[scale=0.15]{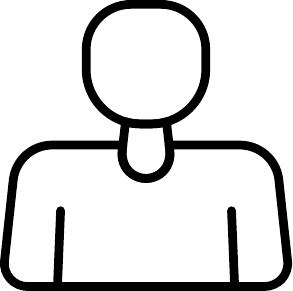} };
            & \draw node[rounded corners=.03cm,minimum size=.5cm,line width=.022cm] (enc) {}; \draw[rounded corners=.03cm, line width=.022cm] (-.25,0) -- (-.25,.6) -- (.25,.4) -- (.25,-.4) -- (-.25,-.6) -- (-.25,0); 
            & \draw node[rounded corners=.03cm,minimum size=.2cm,line width=.022cm,draw,xshift=.2cm,yshift=.25cm] (bn1) {}; 
              \draw node[rounded corners=.03cm,minimum size=.2cm,line width=.022cm,draw] (bn2) {}; 
              \draw node[rounded corners=.03cm,minimum size=.2cm,line width=.022cm,draw,xshift=-.2cm,yshift=-.25cm] (bn3) {}; 
            &&& \draw node[rounded corners=.03cm,minimum size=.5cm,line width=.022cm] (dec) {}; \draw[rounded corners=.03cm, line width=.022cm] (.25,0) -- (.25,.6) -- (-.25,.4) -- (-.25,-.4) -- (.25,-.6) -- (.25,0); 
            &&&&& \node(out){Output};
            \\\node{~};\\
            && \draw node[rounded corners=.03cm,minimum height=.2cm, minimum width=1.2cm,line width=.022cm,draw] (anon) {}; \\
            && \node {Anonymization}; \\
        };        
        \draw[->,thick] (user1) -> (enc);
        \draw[->,thick] (enc) -> (bn2);
        \draw[->,thick,dashed] (bn2) -> (dec);
        \draw[->,thick] ($(enc.east)+(0,.25cm)$) -> (bn1);
        \draw[->,thick,dashed] (bn1) -> ($(dec.west)+(0,.25cm)$);
        \draw[->,thick] ($(enc.east)+(0,-.25cm)$) -> (bn3);
        \draw[->,thick,dashed] (bn3) -> ($(dec.west)+(0,-.25cm)$);
        \draw[->,thick] (dec) -- (out);
        \draw[->,thick] ($(anon.north)+(.2cm,0)$) -- (bn1);
        \draw[->,thick] (anon) -- (bn2);
        \draw[->,thick] ($(anon.north)-(.2cm,0)$) -- (bn3);

        \draw[->,thick,dashed,red] ($(dec.west)+(-.3,.25cm)$) -- node[near end,right]{!} ++(0,-1.2);
        \draw[->,thick,dashed,red] ($.5*(dec.east)+.5*(out.west)$) -- node[midway,right]{!} ++(0,-.96);
        
    \end{tikzpicture}
    \caption{Privacy-preserving speech processing through \emph{disentanglement}, where speech is decomposed into independent streams of information, each representing a distinct category of information, and where the level of anonymization can be individually chosen for each category. Potential points of attack are indicated by red, dashed arrows and exclamation marks.}
    \label{fig:disentanglement}
\end{figure}

\subsubsection{Disentanglement}\label{sec:diss}
With disentanglement, we refer to methods that split a speech signal into channels corresponding to distinct categories of information. For example, selecting from \cref{tab:categories}, we can choose to decompose speech into linguistic information, the speaker's (voice) identity, and emotions~\cite{zhou2021vaw}.  
Such disentanglement can be achieved by either or both through adversarial training or by generalizing
information bottlenecks to encompass multiple bottleneck channels, where each bottleneck represents a \emph{disentangled} representation (see \cref{fig:disentanglement}). Note that provable privacy requires that such channels be statistically independent of each other. That is, each channel represents a distinct and statistically independent attribute of the signal such that the extent of anonymization can be cherry-picked per channel as per use-case e.g.,~\cite{aloufi2020privacy,stoidis21_interspeech,zhou2021vaw,williams2022learning}. 

{The privacy-preserving performance of a disentangling system can be readily formulated, provided that channels $k$ are uncorrelated. If the bitrates for private information of respective channels are $B_k$, then using the composition theorem, the joint disclosure is $\sum_kB_k$~\cite{kairouz2015composition}. Per-channel disclosures are thus additive such that the design and performance analysis of the privacy-preserving systems is straightforward.}

An important challenge with disentanglement is to define the constraints with which information is funneled to the corresponding channel and the methods by which they are made independent of each other. For example, a model can be trained to match specific channels with labels in the dataset (e.g.~\cite{stoidis21_interspeech,zhou2021vaw}). Alternatively, using representation learning, we can constrain channels based on objective criteria such as time-scope, i.e., the length of time over which data is integrated (e.g.~\cite{Niekerk2020,neumann2019improving,lotfidereshgi2022cognitive}). Specifically, attributes such as the speaker's identity are constant over an utterance, while phonemes, intensity, and pitch change over a time scale of tens of milliseconds. By modeling the signal in terms of such time scales, we can disentangle attributes that appear on the corresponding time scales.

\begin{figure}[t]
\centering
\footnotesize
    
    \begin{tikzpicture}

        \draw[fill=lightgray!20,color=lightgray!20] (-.6,.25) -- (2.7,.25) -- (2.7,-1.1) -- (-.6,-1.1) -- (-.6,.25);
    
        \node[matrix,column sep=.6cm] at (0,0) { 
            \node {\parbox{.8cm}{\centering User\\Alice}}; & \node{\parbox{.8cm}{\centering Local\\ device}}; && \node{\parbox{.8cm}{Remote\\service}};\\ \node{~};\\ \node{~};\\
            \node (user1) {\includegraphics[scale=0.15]{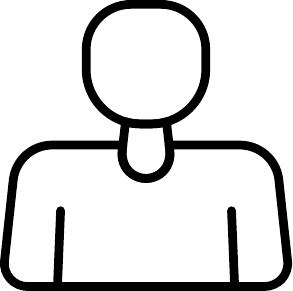} };
            &
            \node (edge1) {\includegraphics[scale=0.15]{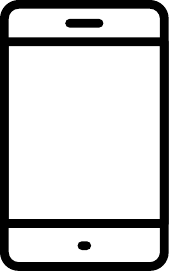} };
            &
            \node[minimum width=.5cm] {};
            &
            \node (cloud) {\includegraphics[scale=0.15]{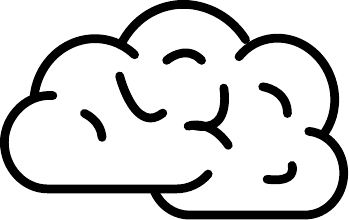} };
            \\
        };
        \usetikzlibrary {arrows.meta} 
        \draw [->,thick] (user1) to [bend left=20] (edge1);
        \draw [->,thick] (edge1) to [bend left=20] (user1);
        \draw [->,double,red] (edge1) to [bend left=20] node[midway,above]{Encrypted}  (cloud);
        \draw [->,double,red] (cloud) to [bend left=20] node[midway,above,yshift=.15cm]{!} (edge1);        
    \end{tikzpicture}
\caption{ Protecting privacy by \emph{encrypting} communication and processing. The gray area indicates the domain that is encrypted, and the red, double arrows marked with exclamation marks indicate the protected stream of data.}
    \label{fig:crypto}
\end{figure}

\subsection{Secure processing}\label{sec:crypto}
If the transmitted information is encrypted, then it cannot be used for malicious purposes without access to the key. Surprisingly, however, it is possible to process encrypted information when using \emph{secure multi-party computation} (SMPC), such that the encrypted result of the computation can be returned and opened~\cite{yao1986generate,portelo2015logsum,portelo2014privacy,Teixeira2018,cramer2015mpc} (see \cref{fig:crypto}). {This applies largely to the same attack vectors as Information isolation, including  \cloudleak{} (\cref{sec:cloudserviceeve}), \falseactivation{} (\cref{sec:falseactivation}), and \cloudaccess{} (\cref{sec:cloudeve}), but also \WASNauthentication{} (\cref{sec:neteve}) and \shareddevice{} (\cref{sec:shareddevice}).} 

The SMPC umbrella contains several categories of methods such as homomorphic encryption, oblivious transfer, {secret sharing,} and garbled circuits. The general principle is that in an interaction between two agents, Alice and Bob, Alice can send encrypted data for processing or a request to Bob, such that Bob cannot interpret the input but can perform the requested task. Furthermore, when Alice receives Bob's response, she cannot deduce the operations that Bob has performed, such that the desired task is completed with mutual privacy preserved.
Another category of approaches is \emph{distance-preserving hashing techniques}, where {data is mapped into a space where distances between points are approximately maintained, allowing some operations like clustering, without revealing the actual data~\cite{portelo2015privacy,jimenez2015secure}.}

Such secure techniques can be applied, for example, in the extraction of spectral features of speech or speech recognition in the cloud, without plain text access to the speech signal~\cite{thaine19_interspeech,Zhang2019}. It can also be applied 
in privacy-preserving speaker recognition, authentication or verification~\cite{teixeira22_interspeech,pathak11_interspeech,pathak2012ppsa,portelo2014bprivacy}, speech search~\cite{portelo2015bprivacy}, or
for secure sound classification~\cite{shashanka2006secure}.
Finally, we can compare the similarity of audio signals without revealing them~\cite{rane2013privacy,perez2020,zarazaga2022thesis}.

A central problem with such cryptographic protocols is that {the floating-point or fixed-point} operations that can be performed in the encrypted domain are limited, typically to only additions and multiplications~\cite{smaragdis2007framework,pathak2012ppsv,pathak2011privacy,pathak2013privacy}. {The limitation does not apply to the boolean operations offered by fast fully homomorphic encryption over the torus scheme, garbled circuits, and boolean secret sharing~\cite{chillotti2020tfhe}.} In contrast, typical speech-processing algorithms also require other {floating- or fixed-point} operations, such as non-linearities, conditionals, and recursions~\cite{chialva2018conditionals}, making the algorithms impossible to implement. While many operations have standard workarounds~\cite{teixeira2019privacy},  such as approximation with additions and multiplications, such approximations are often computationally prohibitively complex, see e.g.~\cite{thaine19_interspeech}. In addition, the communication overhead, in terms of the amount of data to send in comparison to the size of the secret message, is very large.
For a broader overview of cryptographic methods in speech, see \cite{nautsch2019preserving}.

\begin{figure}[t]
\centering
\footnotesize
    
    \begin{tikzpicture}
        \node[matrix,column sep=.6cm] at (0,0) { &
            \node {User}; & \node{\parbox{.8cm}{\centering Local\\ device}}; & \node{\parbox{.9cm}{\centering Remote\\service}}; & \node{\parbox{1.2cm}{\centering Secondary\\service}};\\
            \node{(a)}; &          
            \node (user2) {\includegraphics[scale=0.15]{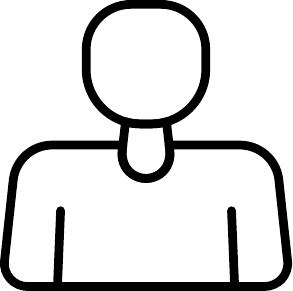} };
            &
            \node (edge2) {\includegraphics[scale=0.15]{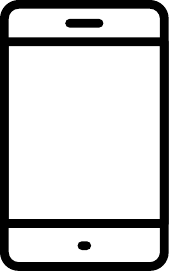} };
            &
            \node (cloud2) {\includegraphics[scale=0.15]{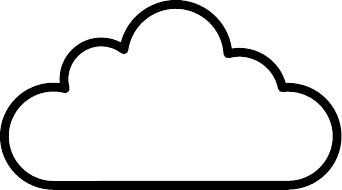} };
            &
            \node (threat2) {\includegraphics[scale=0.15]{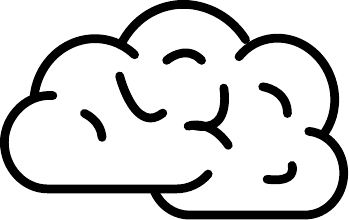} };
            \\ \node{};\\ \node{};\\ \node{};\\ \node{};\\
            \\  \node {(b)}; &
            \node (user3) {\includegraphics[scale=0.15]{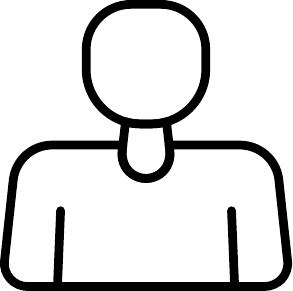} };
            &
            \node (edge3) {\includegraphics[scale=0.15]{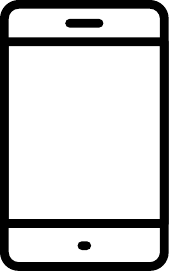} };
            &
            \\
        };
        \usetikzlibrary {arrows.meta} 
        \draw [->,thick] (user2) -- (edge2);
        \draw [->,dashed,thick] (edge2) -- node [midway,below,yshift=-.2cm,xshift=.1cm] {\parbox{1.55cm}{\centering Reduced information}} (cloud2);
        \draw [->,dashed,red,thick] (cloud2) -- node[midway,above]{!} (threat2);        
        \usetikzlibrary {arrows.meta} 
        \draw [->,thick] (user3) -- (edge3);
        \draw [->,thick] (edge3) to [out=-30,in=30,looseness=3] node[midway,right] {\parbox{1.3cm}{\centering Local processing}} (edge3);
    \end{tikzpicture}
\caption{Two alternative protections against the ``\emph{\cloudleak}'' scenario, where the transmission of private information to a remote service is either (a) reduced (by some means and to some extent) or (b)~prevented. }
    \label{fig:fnleaksolution}
\end{figure}

\begin{figure}[t]
    \centering 
    \footnotesize
    \begin{tikzpicture}
        \node[matrix,column sep=.3cm] {
            \node {Users}; & \node{\parbox{.9cm}{\centering Edge\\ devices}}; & &\node{\parbox{.9cm}{\centering Remote\\service}}; \\
            \node (user1) {\includegraphics[scale=0.15]{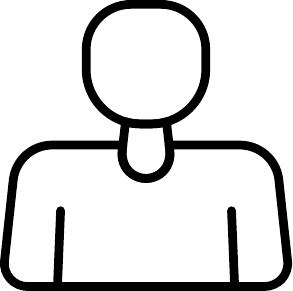} };
            &
            \node[] (edge1) {\includegraphics[scale=0.15]{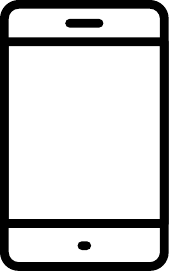}} ;
            \\ \node{~};\\\node{~};\\
            \node (user2) {\includegraphics[scale=0.15]{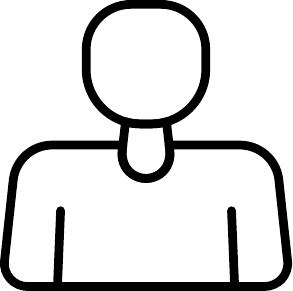} };
            &
            \node[] (edge2) {\includegraphics[scale=0.15]{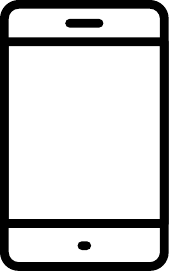}} ;            
            & \node[minimum width=.5cm]{~}; &
            \node (cloud) {\includegraphics[scale=0.15]{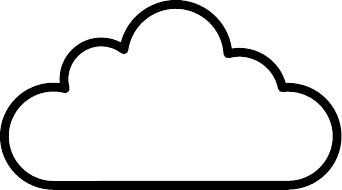} };
            \\
            \node{~};\\\node{~};\\
            \node (user3) {\includegraphics[scale=0.15]{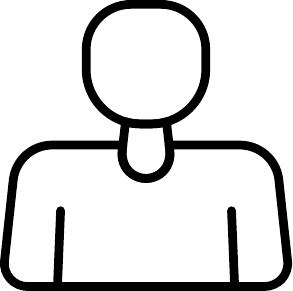} };
            &
            \node[] (edge3) {\includegraphics[scale=0.15]{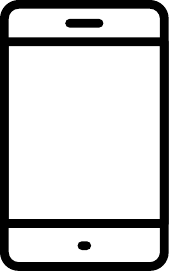} };
            \\
        };        
        \usetikzlibrary {arrows.meta} 
        \draw[->,thick] (user1) -- (edge1);
        \draw[->,thick] (user2) -- (edge2);
        \draw[->,thick] (user3) -- (edge3);
        \draw[->,thick] (edge1) to [out=-75,in=-105,looseness=4] (edge1);
        \draw[->,thick] (edge2) to [out=-75,in=-105,looseness=4] (edge2);
        \draw[->,thick] (edge3) to [out=-75,in=-105,looseness=4] (edge3);
        \draw[<->,thick,dashed,red] (edge1) to [bend left=0] node[midway,above,yshift=.1cm,xshift=.2cm] {\parbox{1cm}{\centering Model\\updates}} (cloud);
        \draw[<->,thick,dashed,red] (edge2) to [bend left=0] node[near end,above]{!} (cloud);
        \draw[<->,thick,dashed,red] (edge3) to [bend left=0] node[near end,above]{!} (cloud);
        
    \end{tikzpicture}
    \caption{\emph{Federated learning} as an example of distributed learning, where devices train models locally and transmit only model updates to the cloud. The red dashed arrows marked with exclamation marks indicate the reduced flow of information.}
    \label{fig:federated}
\end{figure}

\subsection{Privacy-preserving architectures}\label{sec:noncloud}
Perhaps the most obvious solution to protecting privacy, for example, in the \cloudleak--scenario, is to limit information flow (to some extent) from the local edge device to the cloud server (see \cref{fig:fnleaksolution}). Such limitation or reduction in information flow can be implemented by removing, replacing, or distorting private content as in \cref{fig:fnleaksolution}(a) (cf.~\cref{sec:isolation}). At the extreme, the edge device can be entirely disconnected from the cloud or network, and process information only locally, as in \cref{fig:fnleaksolution}(b)~\cite{shi2016edge}. In doing so, we then need to assume that the local device has sufficient capacity and software to complete the requested tasks, that it is not compromised, and that the restriction of information flow is sufficient to ensure that private information is not leaked. Observe, for protecting privacy in scenarios 1 to 4 in \cref{fig:allthreats}, local processing is the best protection as nothing is transmitted to a cloud service.

Distributed learning is an approach to training models without the need for centralized data collection~\cite{liu2021machine}. Models are trained on local edge devices and only model updates are shared between the nodes and/or with a central server (see \cref{fig:federated}). 
{The model updates are, however, calculated \emph{from} the raw signal and thus describe the original signal in some convoluted way. 
Though the raw data thus never leaves the edge device in distributed learning, this corresponds to a privacy protection approach where information flow is restricted (cf. \cref{sec:isolation}).}

Federated learning is one of the flavors of distributed learning, where a central, cloud server collects, merges, and redistributes model updates. It has been used, for example, in speaker and emotion recognition, language modeling, as well as for unsupervised estimation of microphone clusters in sensor networks~\cite{feng22b_interspeech,granqvist20_interspeech,zewoudie2021federated,nelus2021,shoemate2022sottovoce,ro21_interspeech,nguyen2023federated,feng2021attribute}.

Overall, distributed learning is a promising approach, even if it has several challenges. First, constructing, training, and testing system architectures is much more complicated than regular machine learning. Second, model updates are model-specific and cannot easily be reused if the model structure is updated. The learning accumulated during training is then effectively lost every time the model is updated, jeopardizing the fair comparison of competing approaches. Third, even if distributed learning does improve privacy, it is not a guarantee for privacy since model updates can also contain private information~\cite{Orekondy2018gradientleaks,liu2021machine}. {To be specific, except for federated learning, distributed learning protects efficiently against threats 1 to 3 in \cref{fig:allthreats}, but it may potentially leak private information to other devices, and federated learning may leak information to the cloud.}


\subsection{Acoustic interventions}\label{sec:reproduction}

\begin{figure}[t]
    \centering 
    \footnotesize
    \begin{tikzpicture}
        \node[matrix,column sep=.6cm,ampersand replacement=\&] at (0,0) { 
            \&\node {User};
            \&\node {Loudspeaker};
            \&\node{\parbox{.9cm}{\centering Local\\ device}}; \\
            \node{Alice}; 
            \&\node (user1) {\includegraphics[scale=0.15]{pdf/AdobeStock_290226498_user_extract.pdf} }; 
            \&\node (speaker1) {\includegraphics[scale=0.25]{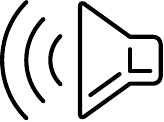} }; 
            \&\node (edge1) {\includegraphics[scale=0.15]{pdf/AdobeStock_290226498_phone_extract.pdf} };\\
            \node{~};\\\node{~};\\
            \node{Eve}; 
            \&\node (user2) {\includegraphics[scale=0.15]{pdf/AdobeStock_290226498_user_extract.pdf} };
            \&\node (microphone2) {\includegraphics[scale=0.25]{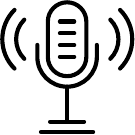} }; 
            \\
            \&\&\node{Microphone};\\
        };        
        \usetikzlibrary {arrows.meta} 
        \draw [->,thick] (edge1) to (speaker1);
        \draw [<-,thick] (edge1) to (microphone2);
        \draw [->,thick,dashed] (speaker1) to (user1);        
        \draw [->,thick,dashed,red] (speaker1) to node [near end,below] {!} (user2);        
        \draw [<-,thick] (microphone2) to (user1);        
        \draw [<-,thick] (microphone2) to (user2);              
    \end{tikzpicture}
    \caption{Protection against a reproduction leak by \emph{authorization tracking} of users, where the device keeps track of people present in the room such that private information is not shared with unauthorized listeners (red dashed line and exclamation mark).}
    \label{fig:usertracking}
\end{figure}

\begin{figure}[t]
    \centering 
    \footnotesize
    \begin{tikzpicture}
        \node[matrix,column sep=.6cm] at (0,0) {
            &\node {User};
            &\node {Loudspeakers};
            &\node{\parbox{.9cm}{\centering Local\\ device}}; \\
            \node{Alice}; 
            &\node (user1) {\includegraphics[scale=0.15]{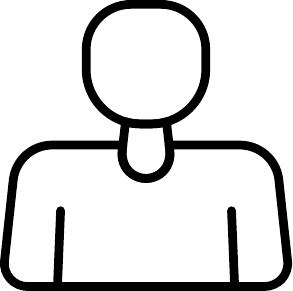} }; 
            &\node (speaker1) {\includegraphics[scale=0.25]{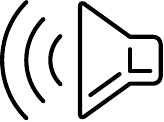} }; 
            &\node (edge1) {\includegraphics[scale=0.15]{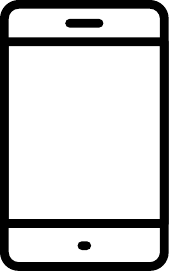} };\\
            \node{~};\\\node{~};\\
            \node{Eve}; 
            &\node (user2) {\includegraphics[scale=0.15]{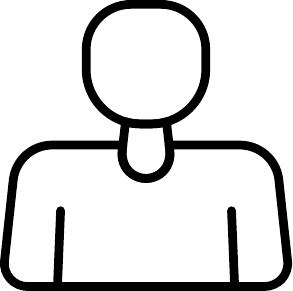} };
            &\node (speaker2) {\includegraphics[scale=0.25]{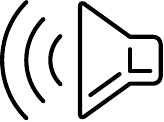} }; 
            \\
        };        
        \usetikzlibrary {arrows.meta} 
        \draw [->,thick] (edge1) to (speaker1);
        \draw [->,thick] (edge1) to (speaker2);
        \draw [->,thick] (speaker1) to (user1);        
        \draw [->,thick] (speaker2) to (user1);        
        \draw [->,thick,dashed,red] (speaker1) to node[near end,below] {!} (user2);        
        \draw [->,thick,dashed,red] (speaker2) to (user2);        
    \end{tikzpicture}
    \caption{Protection against a reproduction leak by using \emph{sound zones}, where constructive interference between loudspeakers is used to retain intelligibility for user Alice, and destructive interference distorts it for user Eve (red dashed lines and exclamation mark).}
    \label{fig:reproduction}
\end{figure}

When a user interacts with a speech interface, the spoken answer of the speech interface can contain private information. Other users can overhear this private information in the same acoustic space, and that presents a threat to privacy (see \cref{fig:reproduction,sec:acousticeve}), {as in attack vectors \falseactivation{} (\cref{sec:falseactivation}), \WASNauthentication{} (\cref{sec:neteve}), \speechinterface{} (\cref{sec:acousticeve}), \discussionleak{} (\cref{sec:discussionleak}), and \shareddevice{} (\cref{sec:shareddevice})}. In theory, it would be possible to identify and track users in the same acoustic space and communicate private information only when it does not pose a threat to privacy (see \cref{fig:usertracking}). {However, tracking itself also poses a threat to privacy, as it is yet another form of private information extracted from the users.}
This places great trust in the local device's ability to track and identify the authorization levels of the people present in the acoustic space.
Such systems have, however, not yet been widely published.

\Cref{fig:reproduction} presents another solution, which uses acoustic interventions to create physical spaces 
where the private information is, respectively, intelligible or inaudible~\cite{gardner1998_3daudio,pulkki2018parametric}. This acoustic effect can be achieved either as a \emph{sound zone} through constructive and destructive interference between loudspeakers (see e.g.~\cite{donley2016,nielsen2018,WALLACE2022101285}) or as an \emph{audio spotlight} through non-linear interference effects between ultrasound sources~\cite{yoneyama1983audio}. By choosing the spatial location where speech is intelligible and assuming we know where the target user is located, we can thus limit access to private information only to the target speaker. 

Note that sound zones with destructive interference do receive a partial observation of the {target information}. Still, it is (hopefully) low-energy and distorted to the extent that it is unintelligible. The central challenges of this approach are to make the constructive sound zone large enough that it allows for small head movements and to make the destructive interference uniform everywhere else, such that there are no isolated points with constructive interference outside the desired sound zone. A benefit of this approach is, however, that it requires tracking of only the target listener, which is, while difficult, still much simpler than tracking all the people in the room.

\subsection{\texorpdfstring{Improving {utility}}{Improving utility}}\label{sec:improve}
In many use cases, like \falseactivation{} (\cref{sec:falseactivation}), \speechinterface{} (\cref{sec:neteve}), and \discussionleak{} (\cref{sec:discussionleak}), the actual culprit is the inadequate performance of the service. For example, if a wake word detector is incorrectly triggered (false positive), the system will start to listen to a conversation when it was not supposed to, thus breaching privacy. Since the inadequate wake word detector is thus causing a privacy breach, the best solution is to improve the wake word detector. The treatment thus addresses the cause rather than the symptoms thereof. {Preserving privacy thus becomes an issue of systems design and implies using the privacy-by-design approach~\cite{Kalloniatis2009,van2011privacy}.}

This approach has, however, two principal challenges. First, improving {utility} often requires an increase in computational power and other resource consumption. This is financially costly and has an environmental penalty~\cite{patterson2021carbon}. In particular, balancing the computational load is easier on a cloud server, and thus, existing resources can be more efficiently used. Second, speech interfaces will always have occasional errors, even with improved {utility}. That is, the design of privacy-preserving speech technology must include multiple layers of protection, especially when requested operations have large consequences~\cite{hong2004privacy,backstrom2021edpbcomments}. Say, when the wake word detector is activated, then before any actions which would potentially breach privacy, the system could require that the speaker is in the same room, or the speaker's identity is verified, or an extra confirmation step ``\emph{Are you sure?}'' or similar. The intrusiveness of such additional protections should then reflect the severity of the potential breach, such that the protections are not perceived as overly obtrusive and decrease the utility of the service.

Improving performance can also mean that the system requires additional functionalities.
For example, in the case of a second local device overhearing Alice's interaction with Bob (\discussionleak) or the primary device (\speechinterface), the required privacy protection is that the secondary device either 
\begin{enumerate*}
    \item is aware that it is not the intended recipient of speech (speech analysis), or
    \item notifies the user of its presence (user-interface design).
\end{enumerate*}


\begin{figure}[t]
\centering
\footnotesize
    
    \begin{tikzpicture}
        \node[matrix,column sep=.6cm] at (0,0) { 
            \node {\parbox{.8cm}{\centering User\\Alice}}; & \node{\parbox{.8cm}{\centering Local\\ device}}; & \node (noise) {Noise}; & \node{\parbox{.8cm}{Remote\\service}};\\ \node{~};\\
            \node (user1) {\includegraphics[scale=0.15]{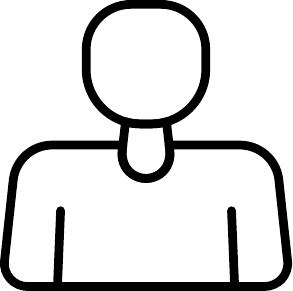} };
            &
            \node (edge1) {\includegraphics[scale=0.15]{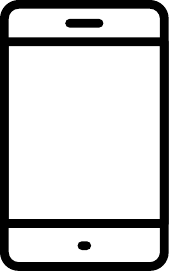} };
            &
            \node[draw,circle,minimum size=.45cm]  { ~};
            \node[circle] (add) { $+$ };
            &
            \node (threat1) {\includegraphics[scale=0.15]{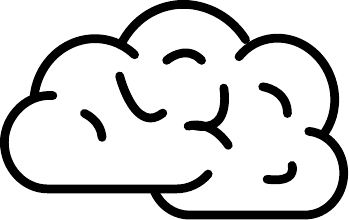} };
            \\
        };
        \usetikzlibrary {arrows.meta} 
        \draw [->,thick] (user1) -- (edge1);
        \draw [->,thick] (edge1) -- (add);
        \draw [->,thick,dashed,red] (add) -- node[midway,above]{!} (threat1);    
        \draw [->,thick] (noise) -- (add);
    \end{tikzpicture}
\caption{ Protecting privacy by adding noise to private data, following the idea of \emph{differential privacy}. }
    \label{fig:noise}
\end{figure}

\subsection{\texorpdfstring{{Privacy-by-design and provable protections}}{Privacy-by-design and provable protections}}\label{sec:design_of_protections}
{Privacy-by-design is a philosophy or approach to design, where privacy is taken into account from the start of the design process and not as an afterthought~\cite{schaar2010privacy}. This increases the chances that the system does not violate privacy unnecessarily, since the privacy-preserving performance is considered at every step of the design process. }

{The privacy-preserving performance of a system can be designed to satisfy design requirements provably, using theoretical frameworks, and \emph{differential privacy} in particular~\cite{dwork2006differential,dwork2008differential,nautsch2019preserving}. Privacy is generally enforced by reducing the accuracy of private information sufficiently, for example, through the addition of noise or by quantization (see~\cref{fig:noise} and cf.~\cref{sec:isolation}). To determine how much accuracy must be reduced,} differential privacy quantifies how the amount of information available about a speaker's affiliation to a particular group changes due to an observation. Specifically, suppose an algorithm $M()$ is applied on a set $D$ to answer $M(D)=S$. {The limit $\epsilon$ for leakage of information (in nits) is then the amount of} new information gained through a slightly different dataset $D'$ as
\begin{equation}\label{eq:diff}
    \log\left|\frac{p\left(M(D)=S\right)}{p\left(M(D')=S\right)}\right| \leq \epsilon.
\end{equation}

Leaks can also be formulated as a function of the groups' sizes, where the speaker is known to belong before and after an observation. Suppose we know that a speaker $x$ belongs to group $x\in A=\{a_1,\dotsc,\,a_M\}$ with $M$ members. {Without prior knowledge, }the speaker can be any one of the group members with equal likelihood $p\left(x=a_k\right)=\frac 1M$. The identity within this group, which is unknown to the attacker, then has an information content of $\log M$ nits. If an attacker somehow manages to identify that the speaker belongs to a subset $B\subset A$ of size $N$, then the remaining information content unknown to the attacker is $\log N$ nits. The reduction in uncertainty is thus the leak magnitude
\begin{equation}
    \epsilon = \log M - \log N = \log\frac MN.
\end{equation}
A leak is thus a function of the \emph{relative} group size before and after an observation.

It is arguably more informative for users to state the \emph{absolute} group size within which they remain anonymous. If, through the processing of an observation, an attacker can reduce the group size within which the user remains anonymous to $k$, then we say that the process is \emph{$k$-anonymous}~\cite{sweeney2002k}.
\smallskip

Even if individual modules of a system can assure privacy, their interaction in a system can expose privacy.
Therefore, the privacy-preservation performance of processes involving speech or other biometric signals must be evaluated on a systems level. Such evaluation uses a concept of \emph{biometric reference}, which is a data record that stores an attribute of the user, such as the speaker's identity. New samples are compared to such biometric references to find matching attributes.
The three key concepts in such evaluation are~\cite{ISOIEC224745_2022,nautsch2019preserving}
\begin{itemize}
    \item \emph{Unlinkability} -- comparing two or more biometric references will not reveal whether they are from the same or different persons. 
    \item \emph{Irreversibility} -- information about the biometric sample cannot be derived from the biometric reference.
    \item \emph{Renewability} -- multiple, unlinkable biometric references can be created from the same biometric samples and used to identify the person without revealing information about the biometric samples. 
\end{itemize}
Together, unlinkability and irreversibility thus minimize the amount of information leaked if biometric references are revealed. At the same time, renewability ensures that leaked references can be revoked and new ones created to replace them. Such concepts are central to the legal definitions of privacy, and first proposals for their mathematical characterization have emerged~\cite{vauquier25_interspeech,cohen2020towards}.

{These theoretical design tools and concepts can be used in the design of, say, information bottlenecks (\cref{sec:isolation}), to quantify the maximal leak of private information. However, care must be taken to ensure that prior assumptions, such as the equal likelihood of all labels, are met (or enforced) in practical scenarios (cf.~\cite{vali2024porcupine}).}
Theoretical methodology for privacy has been applied in, among others, differentially private speech recognition~\cite{shoemate2022sottovoce}, emotion recognition~\cite{feng22b_interspeech}, speaker anonymization~\cite{shamsabadi2023diff}, and
federated learning in an acoustic sensor network~\cite{el2022differential}. As theoretical proofs of privacy have obvious advantages, the adoption of this methodology will likely increase.

\begin{table}[t]
    \caption{{Dimensions of performance in privacy-preserving speech processing.}}
    \label{tab:performance}
    \centering
    \begin{tabular}{l|p{2.9cm}|p{3.4cm}}
        & \multicolumn{2}{l}{\textbf{Performance}} \\
        &\textbf{Objective} & \textbf{Subjective} \\\hline
        \textbf{Privacy} & The consequences and\newline likelihood of attacks & Users' perception and\newline experience of privacy \\\hline
        \textbf{Utility} & Performance in the\newline downstream, trusted task & User experience (UX) and\newline quality of experience (QoE) \\[.2cm]
    \end{tabular}
\end{table}

\section{Evaluating the performance of privacy-preserving methods}\label{sec:objmeasures}
To evaluate the performance of any privacy-preserving methods, we need performance measures for both utility and privacy, corresponding respectively to the trusted and threat tasks (see \cref{fig:threatmodel}). With utility, we refer here to the usefulness of the service in its intended task.
While the objective performance in both utility and privacy is a prerequisite for any good speech application, to the user, it is central that their \emph{experience} with the service or application is good. Correspondingly, performance can always be measured on two dimensions, privacy and utility, as well as objective and subjective (see \cref{tab:performance}). Performance with respect to objective and subjective privacy and utility is presented in the following subsections (\cref{sec:objprivacy,sec:subjprivacy,sec:objtrusted,sec:subjtrusted}), but the subjective \emph{experience} of privacy is a sufficiently large and independent topic that it deserves its own section (\cref{sec:psycho}). We further need to define how the objective measures are applied in systems design to ensure privacy (see \cref{sec:design_of_protections}).

The evaluation protocol of the VoicePrivacy challenge is a state-of-the-art, practical example of evaluation of both privacy and utility~\cite{tomashenko2022voiceprivacy}. {It is based on direct measurement of identifiability using signal processing and machine learning methods. The evaluation objective in such empirical privacy is to determine the extent to which an attacker can re-identify an anonymized speaker or anonymized attribute(s) of a speaker (see \cref{fig:anonobj}). }

\subsection{Objective measures of privacy}\label{sec:objprivacy}
An information-theoretic approach to the objective measurement is to quantify the new information about a speaker gained through an observed speech signal in terms of entropy and units of nits or bits. In other words, suppose an attacker has prior assumptions about a speaker, quantified as a probability distribution $p(X)$ of the possible states $X$ of the speaker. The information content of that model is the negative logarithm $I(X)=-\log p(X)$ and its entropy is the expectation thereof $H(X)=E\left[I(X)\right]$. The attacker then gains access to or makes an observation~$Y$. This new information modifies the attacker's beliefs about the speaker, expressed as the conditional model of the speaker $X$ given the information $Y$, or $X|Y$. Correspondingly, the posterior, conditional entropy $H(X|Y)$ quantifies the amount of information the attacker has about the speaker after (posterior to) the observation. 

The leak of private information $\epsilon$ is then defined as the change in entropy
\begin{equation}\label{eq:leak}
    \epsilon = H\left(X\right) - H\left(X|Y\right).
\end{equation}
The leak is thus equal to the mutual information of $X$ and $Y$.

A central issue with these metrics is that they are based on mathematical models of probability distributions, whereas speech is a natural signal that does not submit to exact formulas~\cite{nautsch20_interspeech}. To use these formulas, we must, therefore, first choose models to describe the distributions of observations and then estimate their parameters. Both steps lead to inaccuracies that dilute the evidentiary power of the proofs of privacy.

\medskip

Measures commonly used to analyze privacy in speaker recognition, verification, and biometrics include~\cite{natgunanathan2016protection}:
\begin{enumerate}
    \item The \emph{false rejection rate (FJR)} (or false negative rate) is the proportion of times a sample from a speaker with a true attribute is not accepted.
    \item The \emph{false acceptance rate (FAR)} (or false positive rate) is the proportion of times a sample from a speaker without an attribute is accepted,
    \item The \emph{equal error rate (EER)} is the proportion of misclassified samples when the threshold of acceptance is adjusted to give equal FJR and FAR~\cite{maouche20_interspeech,ISObiometric}. An increase in EER means that the attacker has made more errors, and privacy has improved.
    \item {The \emph{log-likelihood ratio (LLR)}, defined as the logarithm of the ratio of likelihoods $p(Y|H_k)$ of the observation $Y$ conditioned with the hypotheses that either the sample is $H_1$, or is not $H_0$, the target speaker~\cite{nist2021SRE}
    \begin{equation}\label{eq:llr}
    LLR(s) = \log\left(\frac{p\left(Y|H_0\right)}{p\left(Y|H_1\right)}\right) = I(Y|H_0) - I(Y|H_1).
    \end{equation}
    A large $LLR$ indicates that $H_0$ is more likely than $H_1$. Comparing to \cref{eq:leak}, we can conclude that where \cref{eq:leak} is the expectation of information content, \cref{eq:llr} quantifies information content explicitly. }
    \item The \emph{application-independent log-likelihood-ratio cost function} $C^{\min}_{llr}$ generalizes the EER by considering optimal thresholds over all possible prior probabilities and all possible error cost functions~\cite{maouche20_interspeech,brummer2006application}.
    \item The \emph{expected privacy disclosure} from the Privacy ZEBRA framework~\cite{nautsch20_interspeech} quantifies the empirical cross-entropy (the leak magnitude) integrated over all possible prior likelihoods. 
    \item The \emph{worst-case disclosure} also from the Privacy ZEBRA framework~\cite{nautsch20_interspeech} is the largest disclosure observed in a dataset.

\end{enumerate}

{Private information can be potentially extracted (and privacy-preserving performance evaluated) from either 
\begin{enumerate*}
    \item the waveform of a speech signal or 
    \item from any intermediate representation of speech or its attributes within a speech processing system, such as an information bottleneck.
\end{enumerate*}
In the analysis of waveforms, we can use generic speech analysis tools to determine the magnitude of disclosure. In contrast, intermediate representations depend on the configuration and design of the speech processing system. Analysis and evaluation tools thus have to be purpose-built and/or optimized for every different speech processing system, which is usually a time and labour-intensive effort. The benefit is, however, that when analyzing the intermediate representations, such as an information bottleneck, we analyze a compact representation of the potential leak, without the potentially confounding effects introduced by later processing steps, such as the decoder of an autoencoder. Analyzing waveforms can thus be easier, while analysis of intermediate representations is more effort but potentially more informative.}

A central downside of empirical privacy evaluation, {such as the analysis of the accuracy of a speaker recognition model}, is that it is based on a current state-of-the-art model, which can be surpassed by future technology. In other words, weaknesses in anonymization that are not (yet) known to the evaluator can compromise privacy. Moreover, it is widely known that the performance of machine learning models can be improved by making the model larger or using more data to train such models. A speech analysis model can thus be (likely) improved, not only by future developments but also with this currently known methodology that only requires additional effort. Empirical evaluation can thus never give proofs, but just \emph{characterizations} of the level of privacy. In contrast, provable privacy can be provided by theoretical frameworks such as differential privacy (cf.~\cref{sec:design_of_protections}).

As mentioned above, empirical evaluation of privacy using EER has been applied, for example, in the VoicePrivacy challenge~\cite{tomashenko2022voiceprivacy}. Both EER and $C^{\min}_{llr}$ are widely used in any speaker recognition~\cite{KINNUNEN201012}. The worst-case disclosure in a quantized bottleneck in speaker recognition has been studied in~\cite{vali2024porcupine}.
More comprehensive treatments of objective privacy metrics are provided in~\cite{nautsch2019preserving,noe2022towards}.

\subsection{Subjective evaluation of privacy}\label{sec:subjprivacy}
In speaker recognition, computers are now on par or better than humans~\cite{hansen2015speaker} and are not fooled by voice mimicry (i.e., voice actors) even when humans struggle to do so~\cite{vestman2020methods,vestman2020voice}. We can thus expect that computers are better at evaluating privacy in \emph{well-defined tasks}. However, there are at least two caveats. First, in identifying private information, human listeners can potentially use properties or features of speech signals other than those computers use. Subjective listeners can thus potentially discover new weaknesses that computers miss (though the opposite is also true). 
Second, in scenarios where humans are the potential attackers extracting information (as in threats 3, 5, and 7 in \cref{fig:allthreats}), the true measure of privacy is quantified by subjective listening. Thus, it is often prudent to complement objective privacy metrics with subjective tests.

In the subjective evaluation of privacy, the listening task should be defined to match the threat. For example, if the threat is re-identification, subjects should be tasked with identifying subjects from anonymized speech signals. Similarly, if the threat is speakers' attribute inference, the subjects should evaluate the extent to which they can identify such attributes. 

{As an example, in the VoicePrivacy 2020 challenge, subjective speaker verifiability was measured by asking subjects to rate, on a scale of \numrange{1}{10}, the similarity between an anonymized target and a reference speaker, where the reference was either the same or a different speaker~\cite{tomashenko2022voiceprivacy}. If the distributions of similarity scores have significant differences, then the anonymization has failed. Alternatives to the \numrange{1}{10} Likert-scale include, for example, AB (are these samples the same person?) and ABX (which of AB is equal to X?) tests. For a comprehensive exposition of subjective evaluation methods, see~\cite{zacharov2018sensory}.}

\subsection{Objective performance in downstream tasks}\label{sec:objtrusted}
The utility of a privacy-performing system can be quantified by measuring the performance of the downstream, trusted task. By the downstream task, we here refer to the intended functionality of the speech service. The downstream task is, in the functional flow diagram, generally implemented \emph{after} privacy-protection elements, which is why it is known as a \emph{downstream} task. In some cases, we might \emph{not} trust the provider of the downstream service and need to apply privacy protections specifically because we have no trust. The concept of a ``trusted task'' is, in such cases, misleading, and the concept of a ``downstream task'' can be preferred.

Metrics used for quantifying utility depend on the application. In applications where a computer analyzes the speech signal, we quantify utility by measuring the accuracy of analysis. For example, in automatic speech recognition, where the computer extracts the text content of a speech signal, a commonly used measure is word error rate (WER), which is the proportion of correctly identified words~\cite{rohit2024asr}. By measuring the word error rate with and without privacy-preserving processing, we can determine how much privacy protections impede utility.

\subsection{Subjective performance in downstream tasks}\label{sec:subjtrusted}
If a human user is to listen to the output signal, then that user is the ultimate measure of performance~\cite{zacharov2018sensory}. For example, if we implement speech enhancement to reduce the perceived noise in a signal, then a user can subjectively evaluate how close the enhanced signal is to the noiseless original~\cite{vincent2018audio}. Such perceptual evaluation is best implemented using subjective listening test standards such as P.800 or MUSHRA~\cite{P800,MUSHRA}. Similarly, if privacy protections degrade a speech signal, we can readily quantify such degradations by listening tests.

Subjective listening tests are, however, expensive, time-consuming, and noisy. Individual subjects, especially those without training (known as naïve subjects), are typically unable to repeat a sequence of tests such that results would match. Subjects are also slow because they must listen to samples one at a time, at real-time speeds, and require time to answer between each sample. Additionally, a multitude of subjects need to complete the same test to gain statistical confidence in the results, and tests require oversight by trained personnel. Automated tests, known as objective measures of perceptual quality, are an alternative to such subjective tests. They estimate the outcome of subjective listening tests by computational means. For example, a classic objective measure known as PESQ estimates the outcome of a P.800 test~\cite{rix2001perceptual}. Later improvements include POLQA and ViSQOL~\cite{beerends2013perceptual,chinen2020visqol}. Such objective measures can be used to automate quality testing at a lower price and faster speed, so that repeated estimates with the same input give the same result. Observe that researchers should carefully review the specifications of such algorithms to determine whether they apply to the problem at hand to ensure reliable results.

\section{Perception and psychology of privacy}\label{sec:psycho}
{Much of the work of the speech community is focused on technical aspects of privacy-preserving algorithms. Yet speech devices are used by and designed for people, and they use people's voice data. It is therefore essential to understand people's perceptions of speech devices, including how they use and want to use them, as well as their preferences regarding privacy. Much of this work falls under the fields of psychology, user-interface design, and social sciences. This section is an introduction to these topics.\medskip
}

\subsection{\texorpdfstring{{Psychological privacy and ownership}}{Psychological privacy and ownership}}
People tend to have a deep sense of \emph{ownership} of some things, both material and immaterial. In particular, people have a \emph{feeling of ownership} toward information about themselves~\cite{cichy2014extending,dawkins2017psychological}. Such feelings are, to some extent, detached from the material consequences of breaches in privacy. For example, even if publishing an audio recording revealing a secret intimate relationship would not have direct economic consequences, it can cause psychological damage and harm personal relationships. The effect of threats to privacy then necessarily becomes a question of psychology and social sciences. To qualitatively or quantitatively measure such effects, we also need subjective user tests.

Concurrently, people have well-established social rules regarding human-to-human privacy~\cite{petronio2002boundaries}. Moreover, users also tend to anthropomorphize technology, that is, treat devices and services as if they were human~\cite{nass1993anthropomorphism,cornelius2021acceptance}, such that they will likely \emph{assume} that devices apply human-like social rules to privacy. Observe that such social rules are related to human-like behavior and performance. That does not reveal whether they apply to the super-human performance that computers can possess, such as the ability to integrate information over massive databases and to retain accurate records of events long past. How machines should then behave, especially with respect to super-human capabilities, is then not only a question of user-interface design but of moral psychology~\cite{thompson2021machine}: We need discussions on a societal level of how automated services \emph{should behave} and how they are allowed to behave (see also \cref{sec:law}).

\subsection{Perception and experience of privacy}\label{sec:perception}
Irrespective of whether a system \emph{actually} preserves privacy or not, some users can \emph{perceive} the system as threatening, and others can be oblivious to the threats it poses~\cite{pepino2020distrust}. Clearly, users do not like using services they perceive as threatening. Understanding how people perceive privacy is interesting in its own right, but such understanding is essential in the design of effective user interfaces.

Perception and experience of privacy can be approached in two alternative ways. We can make user studies where people interact with either machines or each other~\cite{leschanowsky2024review}. The distinction is important in the sense that human-to-human interaction relies on social rules that are well-established and developed over a long time. We can thus expect them to be stable over time, whereas human-computer interaction is continuously evolving as people learn more. Moreover, by using studies of human-to-human interaction, we can learn effects related to human-like performance, but can probably not rely on observing effects related to super-human performance. Studies of human-to-human interaction are, however, always a proxy if the actual target is to design human-to-computer interfaces~\cite{leschanowsky2024review}. 

Studies of human-computer interaction thus characterize the desired phenomenon directly. The compromise is, however, that people's understanding of privacy with devices might not reflect the true level of privacy, and conclusions made based on the users' opinions might then not reflect their true preferences \emph{and} those preferences are moving targets.  As people gain more experience with and as they learn more about technology, their attitudes and perceptions of it change~\cite{leschanowsky2024review}. 

Despite these shortcomings, both types of experiments are essential for improving our understanding of privacy and for improving technology. For example, human-to-human studies have revealed that people experience privacy differently in different acoustic environments; a noisy cafeteria can be better at masking sounds and defending against eavesdropping than a reverberant hallway~\cite{zarazaga19_interspeech,das2021thesis,zarazaga2022thesis,backstrom2021intuitive,Leschanowsky2020}, and multiple-bed patient rooms in hospitals have privacy-concerns~\cite{reinten2017speech}. Similarly, human-machine studies have revealed that when a chatbot actively communicates choices related to privacy it improves users' experience of privacy~\cite{bruggemeier2022perceptions}, voice interfaces with unknown features cause fear in users, and reduces retention of services~\cite{yeasmin2020privacy,Lau2018,lutz2021privacy} and breaches in privacy are highly detrimental to the trust in services and reduce users' willingness to use the services, but such trust can be rebuilt~\cite{Kurz2021}.

\subsection{User interface design}\label{sec:ui}
For usable privacy, information about privacy must be readily provided, changes in the extent of privacy are promptly notified, and that users have control of the level of privacy~{\cite{bruggemeier2022perceptions,so2021voice,lewis1998designing,Lau2018,Kurz2021,dhiya2023privacy,iachello2007end,sivaraman2018smart}. As a parallel to speech interfaces}, visual interfaces can use lights or icons for monitoring and tactile interfaces for controlling the level of privacy. While sound can be used to monitor system status with \emph{sonification}~\cite{hildebrandt2016continuous,iber2021auditory}, this is not in widespread use within signaling of privacy. It is a compromise between filling the acoustic space with information and the tendency of the human auditory system to block out monotonous sounds~\cite{pulkki2015communication}. To be effective, the sound should be perceivable when the user consciously wants to check the privacy level, and changes in privacy level should evoke correspondingly and appropriately large changes in the soundscape so that users can consciously register the changes. Furthermore, user interfaces should provide accurate information about and enable control over the privacy level~\cite{Lau2018,backstrom2020privacy}.

In any case, it is imperative that services are designed such that they \emph{reflect the true extent of privacy}. 
Observe that it is likely possible to design systems that communicate an advanced level of privacy even when the system does not respect user privacy. In fact, service providers may have short-term incentives to follow such approaches to design as long as they improve overall user satisfaction. However, such approaches to design are known as \emph{dark patterns} or \emph{deceptive design patterns}, and they are considered unethical~\cite{willis2020deception}. Through deception, such dark patterns lull users into believing they are safe when, in fact, they are being abused. Therefore, good design of privacy should actively communicate and enable control of the true extent of privacy. Such design practice is not only ethical but also rewards service providers in the long term by improving retention of those services~\cite{Kurz2021}.


\subsection{Privacy and trust}\label{sec:trust}

The level of privacy provided by a speech service is intimately linked to the level of trust users experience in that service~\cite{leschanowsky2024review}. In particular, if users experience that a service is detrimental to their privacy or, more generally, to their well-being, they will obviously not trust it. We thus discuss trust mainly from a psychological perspective, unlike in security research and cryptography, where trust has a specific and unambiguous mathematical definition~\cite{balsa2022cryptography}.

While the meaning of trust may seem intuitively clear, it is, however, difficult to define exactly. A classic definition is~\cite{deutch1973resolution}
\begin{quote}
    Trust: confidence that [one] finds what is desired rather than feared. 
\end{quote}
This definition highlights the psychological nature of trust~\cite{simpson2007psychological}. From the current engineering perspective, this translates to the \emph{statistical confidence that the outcome of a process is useful}. {Here, usefulness refers to what the user subjectively perceives as useful.} Implicitly, this furthermore requires that we predict the outcome of that process in the form of a statistical distribution of possible outcomes and that the expectation over the possible outcomes is that the process is useful. Another way of expressing this is to say that trust requires dimensions of competence, integrity, and benevolence~\cite{Chen2003,Xie2009}. 

Observe the semantic difference between a system that is ``benevolent'' and one that is ``useful''. The casual use of the concept of trust seems to require that the actor is indeed benevolent. However, it is possible to imagine a system with malicious objectives concerning Alice, but that produces side effects that consistently benefit Bob. Bob thus could potentially trust the system to act beneficially even if it is not benevolent towards him. It is furthermore questionable whether we can assign an ``intent'' to a computational system due to its close association with psychology. The concept of benevolence is thus ambiguous in the context of speech technology.\smallskip

It is useful and ethical to design speech services that are \emph{trustworthy}. This means that systems should be such that it would be logical for users to trust them{~\cite{hardin1996trustworthiness}}. Speech technology should thus be designed to have useful outcomes with statistically significant confidence, especially with respect to privacy. 

A practical approach for reducing the likelihood of adverse outcomes is to build multiple layers of checks and verifications, similar to multi-factor authentication~\cite{ometov2018multi}. In \cref{fig:cloudthreat}, this can be implemented by applying layers of privacy protection in each successive connection and device, as well as in their internal modules and components, as described in \cref{sec:improve}. Activation of an undesired functionality of the system or leaking private information would thus require that every subsequent module fails simultaneously, which is unlikely in most scenarios. 

Conversely, users likely experience most trust in devices and services that are ``close'' to them and over which they exert control. Thus, it seems obvious that users will tend to trust more those devices they have in their possession or homes, rather than remote cloud services.

\section{Legal and societal landscape}\label{sec:law}
Privacy in speech technology has a great impact on both the individual and societal levels, as already discussed in \cref{sec:intro}. The magnitude of societal impact can be appreciated by recalling the Cambridge Analytica privacy scandal~\cite{schneble2018cambridge,heikkila2023cambridge_ai} where private information was extracted from social media and used for targeted political advertising. In both speech technology and social media, services operate on massive user bases and involve interaction between multiple users. This exposes both areas to the same magnitude of risks. In the case of Cambridge Analytica, the most significant consequence was its influence on election results in a major democratic country. Another prominent scandal from biometrics is the case where supposedly anonymized patient data for \SI{10}{\percent} of Australians was released to the public, only to be later shown that individuals were readily identifiable~\cite{culnane2017health}. All health data of those patients was thus made public contrary to user preferences and with unknown long-term consequences. 

While we have not (yet) seen breaches related to speech technology with consequences of comparable magnitude, these parallels highlight the \emph{potential} effect of breaches. Scandals directly related to speech technology include eavesdropping on private persons by employees and contractors of service providers~\cite{Lynskey2019GuardianPrivacyScandal,hern2019,day2019}. 

\Cref{sec:ui} also makes the point that service and technology providers have clear short-term incentives that conflict with users' preferences for privacy. We can easily find examples where users do not have real choices in protecting their privacy. For example, suppose all friends and family of a user use a particular platform for social and voice interaction, even if its privacy configuration is inadequate~\cite{backstrom2025beyonduser}. The individual user then faces the choice of either disconnecting from their social network or compromising privacy. This applies to all users with a sufficiently large portion of their communities participating in that platform. This is a variation of the prisoner's dilemma, where no single user has sufficient incentive to transition to a solution that would be best for everyone.

On the individual level, the examples of potential exploits in \cref{tab:exploits} cause clear damage to individuals, including psychological harm in particular (e.g.~\cite{natarajan2021prioris}). Additional dangers include stalkers~\cite{lu2019abusers,hulsen2022stalkers,avast2021stalkerware} {and oppression~\cite{backstrom2025beyonduser}}.
While such individual damage is ``small'' on the societal level, their prevalence makes their joint impact significant~\cite{levy2020intimate}.

These examples demonstrate the inherent need for society-level regulation of speech technology with respect to privacy. Governments have already responded to this need, with the European Union spearheading the process with the General Data Protection Regulation (GDPR)~\cite{eu2016gdpr,nautsch19c_interspeech}, with the State of California following soon thereafter with the California Consumer Privacy Act (CCPA) of 2018~\cite{california2018ccpa}. While these laws cover only a small percentage of the global population, as cloud services typically operate globally, they need the capability to follow local laws. In many cases, it can be easiest to apply the strictest laws on all users, such that the most strict laws benefit the privacy of all users. Service providers have, therefore, widely adopted the requirements of GDPR and CCPA, and that has likely had a significant impact also on users outside the scope of these regulations. {More recently, other countries have followed suit, for example, through the ``Personal Information Protection Law of the People's Republic of China''~\cite{china2021pip} and ``Lei Geral de Proteção de Dados'' (Brazilian General Data Protection Law, LGPD)~\cite{brazil2020lgpd}.}
\smallskip

With respect to regulation, an important consequence of the objective measures of privacy in \cref{sec:objmeasures} is that our tools and measurements will give as an output only \emph{statistical characterizations} of {the accuracy of estimated attributes}, but they can never give absolute confidence. This is in stark contrast with the concept of \emph{unique} identifiability used in legal documents, such as the General Data Protection Regulation (GDPR) by the European Union~\cite{eu2016gdpr}, which does not explicitly leave room for statistical uncertainty. This is reflected, for example, in the Guidelines for virtual voice assistants by the European Data Protection Board, which states that:~\cite[page 13, \textsection31]{edpb2021guidelines}
\begin{quote}
    ... voice data is inherently
biometric personal data. As a result, when such data is processed for the purpose of uniquely
identifying a natural person ... the processing must have a valid legal basis  ...
\end{quote}
This leaves the interpretation open. It is possible to argue that it is never possible to obtain absolute confidence in speaker identification such that the GDPR is never triggered. It is also possible to argue that all voice data contains personal information that can be used to uniquely identify a person, such that all processing must have a valid legal basis. Both interpretations lead to absurdity, which suggests that the truth must lie somewhere in the middle. In fact, the GDPR in practice requires (see~\cite[page 4]{edpb2021guidelines}) that the design process of voice assistants includes a \emph{data protection impact assessment}, where the risks and consequences are evaluated such that the designer can take appropriate precautions to preserve privacy. Authors of the GDPR are thus clearly aware that it is impossible to give absolute guarantees of privacy but that the impact assessment (i.e., objective measures of privacy) must necessarily be based on statistical measures, even if such measures have not been defined. Fortunately, the first mathematical characterizations of the legal criteria have started to appear~\cite{vauquier25_interspeech,cohen2020towards}.

However, even in the absence of legal requirements, the current author would encourage all researchers, developers, data analysts, project managers, and policymakers to \emph{perform impact assessments} in data projects, data management, and data policies, anytime when there is or could be the slightest concern that the dataset imposes risks to privacy. A practical tool to aid impact assessment is the ``Data Ethics Decision Aid (DEDA)'' that formalizes and enumerates all the questions that a representative group of stakeholders should review, discuss, and consider~\cite{franzke2021data}.

\medskip

While governments are in the process of regulating privacy, corporations and non-governmental organizations have also realized that proper privacy is an opportunity. For example, the Open Voice Network seeks to develop and standardize open technical standards and ethical guidelines for voice assistance~\cite{ovn2023} and the MyData Global seeks to help people and organizations to benefit from personal data in a human-centric way~\cite{mydata2023,poikola2020mydata}. Within the research community, the author of this paper has been involved in establishing a special interest group within the International Speech Communication Association (ISCA) devoted to ``Security and Privacy in Speech Communication"~\cite{spsc2023}. It is, as far as we know, the world's largest community focused on this topic.

\section{Discussion and conclusions}
The quality and use of speech interfaces have increased rapidly in recent years. As with any new technology, the rapid progress has also revealed dangers, particularly the threats to privacy it demonstrably poses. Unprotected users are exposed to threats like stalking, algorithmic stereotyping, harassment, and price gouging. Researchers, service providers, and governments thus have the impetus to protect the users, not only because it is ethical but also because it makes for better products and long-term business. 

This paper is a tutorial on privacy for speech technology. Its most notable novelty is an exhaustive categorization of threats (see \cref{fig:allthreats,sec:threat}). Protections against those threats are further categorized according to whether they relate to the {target information} or side information. The pertinent difference is that transmitting {target information} is the whole purpose of communication, and there is not much we can do to protect it other than encryption. With side information, that is, all the other information that is bundled into a speech, like health status and gender identity, we have a much larger arsenal of protections. The primary approach is, however, to remove as much of the side information as possible as early as possible. As the {target information} is all the communication we need, all side information should be removed to the extent possible. Such removal, however, rapidly demonstrates that paralinguistic information like speech style is often instrumental in conveying the intended message. It is thus not always clear what constitutes the legitimate, {target information}.

The first conclusion from this paper is that the range of possible privacy threats is vast. Each agent -- be it human or device -- participating in an interaction, as well as the acoustic pathways and network connection through which they are connected, is a potential attack surface. Any actor that can interact with the other agent or listen to the connection is a potential eavesdropper. Since we define privacy as a scenario where an agent is authorized with some access but over-exceeds that authorization (intentionally or inadvertently), we cannot just cut connections but need more refined designs and methodologies. We thus need to dynamically adjust access according to need. Conversely, systems must actively monitor the privacy status to determine appropriate actions.

Second, we find that privacy and ethics are largely overlapping challenges. Our ethical values govern our preferences for privacy. Most potential breaches of ethics in speech technology are based on breaching privacy. That means that we need a society-wide ethical discussion about what is allowed with respect to user privacy. Such discussions are necessary to prevent a Cambridge Analytica-style scandal for speech technology~\cite{heikkila2023cambridge_ai}.

A third implication of this paper is that, while research in this field has started to accelerate only recently, there is already a substantial body of research available. The research is not, however, mature but in a phase of rapid development, and there are important sub-areas that have not yet seen much work. This makes it a fruitful area for research, as we can expect important understanding to be discovered in the coming years. 

Particular research questions where the author sees an urgent need for and expects to see new results include:
\paragraph{Consent} While management of acquiring informed consent has established traditions and best practices for most interface types~\cite{friedman2000informed}, speech, audio, and ambient systems are notably unique. Namely, acoustic information is a time-varying stream. Reading out a pages-long consent form before an interaction can start is clearly much too obtrusive and unnecessarily detailed. Privacy requirements also vary over time. Consent should thus be acquired on an actual need basis. In addition to being more usable, it would also make choices better connected to the actual needs, since consent is acquired only once it is needed.
\paragraph{Metrics for streaming}
The available theoretical metrics reflect privacy with respect to a finite dataset, whereas speech is an open-ended stream of data. The consequence is that, in theory, we can resolve any private attribute or identity, provided that it has a unique probability distribution and we have a sufficiently long observation. We would thus need methodologies for characterizing the effect of the length of observation on privacy. The current author has recently made first attempts in this direction~\cite{backstrom2025privacydisclosure}.
\paragraph{Metrics for out-of-category information} The metrics discussed in \cref{sec:objmeasures} are all related to specific categories of private information, and in particular, we can provide protection only to identified threats. For example, we can measure the threat to privacy related to health information, but that does not say anything about the threat related to information about ethnic background. We thus need methods for evaluating privacy jointly with respect to \emph{all categories} of private information except the {target information}. 
\paragraph{Future-proof metrics} {Empirical} metrics are generally based on a model of the signal or the attacker. The metrics are thus subject to change when those models are improved in the future, and it will likely expose new threats. Though it is likely difficult, it would be extremely useful if we could characterize, for example, privacy threats as a function of computational complexity, similarly to how cryptography provides assurances over security based on the difficulty of solving certain difficult computational problems. Where cryptographic security is often seen as a threshold of computational complexity, however, for privacy in speech applications, it would be beneficial to also assess the leak magnitude (e.g., in bits) as a function of computational complexity.  Over time, efficiency improvements reduce algorithmic complexity, and as hardware with better capabilities becomes available, this would allow us to characterize how protections will survive the test of time.
{\paragraph{Attack and attacker characterization} The biometrics community has a long tradition of characterizing attacks as, for example, opportunistic and motivated attacks, as well as in terms of the potential rewards of such attacks~\cite{roberts2007biometric}. Characterization of speech-specific attacks has received much less attention, though \cref{fig:allthreats,sec:threat} in this paper are attempts in that direction. Further work is needed to gain privacy protections on a system-level.
}
\paragraph{Multi-user interaction} Privacy research is categorically focused on \emph{personal} and \emph{user-centric} privacy. However, speech is, by definition, communication between multiple agents, and when exposed, threatens \emph{all participants simultaneously}. This is not an issue from a legal point of view, because privacy protections apply to all individual users equally. However, from an authorization and consent management perspective, this is an underappreciated issue. If user A records a discussion with user B, then both clearly have some level of ownership and privacy requirements on that recording. Another case is smart technology with multiple users, like smart TVs; even if one user has consented to data collection, that does not mean that others would agree~\cite{backstrom2025beyonduser}. We do not yet have any widely accepted standard approaches for handling privacy, ownership, and consent in such multi-user scenarios. 
\paragraph{Disentangelement} If we could disentangle all categories of speech information as in \cref{fig:disentanglement}, it would be easy to anonymize each category to an appropriate degree. This approach thus seemingly solves all our problems. The issue is that we do not yet have sufficiently sophisticated methods for such disentanglement. The difficulty in developing disentanglement algorithms is that information categories in \cref{tab:categories} are vaguely and heuristically defined, and there is significant overlap between them. We cannot even demonstrate that this would be a complete list of information categories. Without exact definitions of those categories, we have no hope of developing methods for them. An alternative approach is to use representation learning methods to create unsupervised clustering of information categories, thus circumventing the need to define categories explicitly. The compromise is that we cannot guarantee that the learned representations correspond to heuristically meaningful categories. Still, since disentanglement is \emph{the ideal} solution, that should continue to be a central focus of research.
\paragraph{Perception, experience, and design of privacy}
Most of the speech-specific research on privacy has focused on privacy-preserving processing and system structures. This is useful because it is the mandatory prerequisite for privacy-preserving technology. However, as discussed in \cref{sec:ui}, users' experience of services is to some extent independent of the objective level of privacy. We need many more user studies on how, for example, voice characteristics and word choices influence trust, how the privacy level can be monitored during interactions and how changes are notified, how the environment and content of interaction influence user experiences, etc. By improving the user experience with respect to privacy, we are likely to improve user satisfaction and retention of the overall service, while also improving the service objectively.
\medskip

In conclusion, threats and breaches of privacy have significant negative consequences on individual, societal, ethical, and economic levels. While further improvements in smart technology are expected to improve the utility of the technology, it will likely also introduce new threats. The protection of privacy in speech technology has thus been important already for a long time, and its importance is increasing. Fortunately, research in the area has picked up speed, and this tutorial presents the most important concepts, approaches, and methodologies. It is, however, likely that fundamental results and new technologies will be introduced in the near future. This is thus an exciting time for researchers in the area.

\appendices


\section*{Acknowledgments}
A majority of the content in this paper has risen directly from discussions with a long list of colleagues, who were far too many to include as co-authors. These colleagues include, but are not limited to, in alphabetical order,
Chris Brzuska, 
Mads Græsbøll Christensen, 
Sneha Das, 
Nicholas Evans, 
Johannes Fischer, 
Francisco Florez,
Florin Ghido,
Meeri Haataja, 
Ivan Habernal,
Aki Härmä,
Dorothea Kolossa, 
Michael Laakasuo, 
Martha Larson, 
Anna Leschanowsky,
Rainer Martin, 
Joachim Meyer, 
Sebastian Möller, 
Andreas Nautsch, 
Francesco Nespoli,
Birgit Popp, 
Silas Rech,
Nitin Sawhney, 
Ingo Siegert, 
Stephan Sigg, 
Francisco Teixeira,
Isabel Trancoso, 
Sanna Toropainen,
Ville Vestman,
Emmanuel Vincent, 
Jennifer Williams, 
and 
Pablo Pérez Zarazaga. 
I will remain forever grateful. 

Feedback from the anonymous reviewers was outstanding in both quality and extent. It pains me that I can never adequately acknowledge their exceptional work.

A part of this work was supported by ‘‘Designing Inclusive \&
Trustworthy Digital Public Services for Migrants in Finland (Trust-M)’’
(Grant Number: 353529) funded by the Strategic Research Council of
Finland.


\ifCLASSOPTIONcaptionsoff
  \newpage
\fi

\IEEEtriggeratref{199}


%



\printbibliography

@inproceedings{nguyen2010automatic,
  title={Automatic classification of speaker characteristics},
  author={Nguyen, Phuoc and Tran, Dat and Huang, Xu and Sharma, Dharmendra},
  booktitle={International Conference on Communications and Electronics 2010},
  pages={147--152},
  year={2010},
  organization={IEEE},
  url={https://doi.org/10.1109/ICCE.2010.5670700},
}

@inproceedings{maouche2022enhancing,
  title={Enhancing speech privacy with slicing},
  author={Maouche, Mohamed and Srivastava, Brij Mohan Lal and Vauquier, Nathalie and Bellet, Aur{\'e}lien and Tommasi, Marc and Vincent, Emmanuel},
  year=2022,
  booktitle={Proc. Interspeech},
  url={https://hal.inria.fr/hal-03369137/},
}

@book{benesty2008springer,
  title={Springer handbook of speech processing},
  author={Benesty, Jacob and Sondhi, M Mohan and Huang, Yiteng and others},
  volume={1},
  year={2008},
  publisher={Springer},
  url={https://doi.org/10.1007/978-3-540-49127-9},
}

@article{lewis1998designing,
  title={Designing for human-agent interaction},
  author={Lewis, Michael},
  journal={AI Magazine},
  volume={19},
  number={2},
  pages={67--67},
  year={1998},
  url={https://doi.org/10.1609/aimag.v19i2.1369},
}

@book{zacharov2018sensory,
  title={Sensory evaluation of sound},
  author={Zacharov, Nick},
  year={2018},
  publisher={CRC Press},
  isbn={9780367656744},
  url={https://www.routledge.com/Sensory-Evaluation-of-Sound/Zacharov/p/book/9780367656744},
}

@book{van2018handbook,
  title={The handbook of privacy studies},
  author={Van Der Sloot, Bart and De Groot, Aviva},
  year={2018},
  publisher={Amsterdam University Press},
  url={https://doi.org/10.1515/9789048540136},
}

@manual{eu2016gdpr,
  author = {{European Parliament}},
  title = {Directive {95/46/EC} General Data Protection Regulation},
  year = 2016,
  url={http://data.europa.eu/eli/reg/2016/679},
}

@manual{china2021pip,
  title = {Personal Information Protection Law of the People's Republic of China},
  author = {The National People's Congress of the People's Republic of China},
  year = 2021,
  url = {http://en.npc.gov.cn.cdurl.cn/2021-12/29/c_694559.htm},
}

@manual{brazil2020lgpd,
   title = {Brazilian General Data Protection Law (LGPD, English translation)},
   year = 2020,
   author = {Ronaldo Lemos and Natalia Langenegger and Juliana Pacetta Ruiz and Sofia Lima Franco and Andréa Guimarães Gobbato and Daniel Douek and Ramon Alberto dos Santos and Rafael A. Ferreira Zanatta},
   url={https://iapp.org/resources/article/brazilian-data-protection-law-lgpd-english-translation/},
}

@article{wang2019vector,
  title={A Vector Quantized Variational Autoencoder ({VQ-VAE}) Autoregressive Neural $ F\_0 $ Model for Statistical Parametric Speech Synthesis},
  author={Wang, Xin and Takaki, Shinji and Yamagishi, Junichi and King, Simon and Tokuda, Keiichi},
  journal={IEEE/ACM Transactions on Audio, Speech, and Language Processing},
  volume={28},
  pages={157--170},
  year={2019},
  publisher={IEEE},
  url={https://doi.org/10.1109/TASLP.2019.2950099},
}

@inproceedings{dwork2008differential,
  title={Differential privacy: A survey of results},
  author={Dwork, Cynthia},
  booktitle={International conference on theory and applications of models of computation},
  pages={1--19},
  year={2008},
  organization={Springer},
  url={https://doi.org/10.1007/978-3-540-79228-4_1},
}

@article{tishby2000information,
  title={The information bottleneck method},
  author={Tishby, Naftali and Pereira, Fernando C and Bialek, William},
  journal={arXiv preprint physics/0004057},
  year={2000},
  url={https://arxiv.org/abs/physics/0004057},
}

@inproceedings{malekzadeh2018replacement,
  title={Replacement autoencoder: A privacy-preserving algorithm for sensory data analysis},
  author={Malekzadeh, Mohammad and Clegg, Richard G and Haddadi, Hamed},
  booktitle={2018 IEEE/ACM Third International Conference on Internet-of-Things Design and Implementation (IoTDI)},
  pages={165--176},
  year={2018},
  url={https://doi.org/10.1109/IoTDI.2018.00025},
}

@article{perero2022x,
  title={X-vector anonymization using autoencoders and adversarial training for preserving speech privacy},
  author={Perero-Codosero, Juan M and Espinoza-Cuadros, Fernando M and Hern{\'a}ndez-G{\'o}mez, Luis A},
  journal={Computer Speech \& Language},
  pages={101351},
  year={2022},
  publisher={Elsevier},
  url={https://doi.org/10.1016/j.csl.2022.101351},
}

@inproceedings{lu2013speech,
  title={Speech enhancement based on deep denoising autoencoder.},
  author={Lu, Xugang and Tsao, Yu and Matsuda, Shigeki and Hori, Chiori},
  booktitle={Interspeech},
  volume={2013},
  pages={436--440},
  year={2013},
  organization={ISCA},
  url={https://doi.org/10.21437/Interspeech.2013-130},
}

@article{das2021enhancement,
  title={Enhancement by postfiltering for speech and audio coding in ad hoc sensor networks},
  author={Das, Sneha and B{\"a}ckstr{\"o}m, Tom},
  journal={JASA Express Letters},
  volume={1},
  number={1},
  pages={015206},
  year={2021},
  publisher={Acoustical Society of America},
  url={https://doi.org/10.1121/10.0003208},
}

@inproceedings{yeasmin2020privacy,
  title={Privacy Analysis of Voice User Interfaces},
  author={Yeasmin, Farida and Das, Sneha and Bäckström, Tom},
  booktitle={Proc. 1st International Workshop on the Internet of Sound},
  year=2020,
  doi={10.5281/zenodo.4026514},
}

@article{tomashenko2022voiceprivacy,
  title={The VoicePrivacy 2020 Challenge: Results and findings},
  author={Tomashenko, Natalia and Wang, Xin and Vincent, Emmanuel and Patino, Jose and Srivastava, Brij Mohan Lal and No{\'e}, Paul-Gauthier and Nautsch, Andreas and Evans, Nicholas and Yamagishi, Junichi and O’Brien, Benjamin and others},
  journal={Computer Speech \& Language},
  volume={74},
  pages={101362},
  year={2022},
  publisher={Elsevier},
  url={https://doi.org/10.1016/j.csl.2022.101362},
}

@inproceedings{aloufi2020privacy,
  title={Privacy-preserving voice analysis via disentangled representations},
  author={Aloufi, Ranya and Haddadi, Hamed and Boyle, David},
  booktitle={Proceedings of the 2020 ACM SIGSAC Conference on Cloud Computing Security Workshop},
  pages={1--14},
  year={2020},
  url={https://doi.org/10.1145/3411495.3421355},
}

@incollection{natarajan2021prioris,
  title={PRIORIS: Enabling Secure Detection of Suicidal Ideation from Speech Using Homomorphic Encryption},
  author={Natarajan, Deepika and Dalskov, Anders and Kales, Daniel and Khanna, Shabnam},
  booktitle={Protecting Privacy through Homomorphic Encryption},
  pages={133--146},
  year={2021},
  publisher={Springer},
  url={https://doi.org/10.1007/978-3-030-77287-1_10},
}

@manual{california2018ccpa,
  author = {State of California},
  title = {California Consumer Privacy Act of 2018 (CCPA)},
  year = 2018,
  url = {https://leginfo.legislature.ca.gov/faces/codes_displayText.xhtml?division=3.&part=4.&lawCode=CIV&title=1.81.5},
  }

@article{zewoudie2021federated,
  title={Federated Learning for Privacy-Preserving Speaker Recognition},
  author={Zewoudie, Abraham and B{\"a}ckstr{\"o}m, Tom},
  journal={IEEE Access},
  volume={9},
  pages={149477--149485},
  year={2021},
  publisher={IEEE},
  url={https://doi.org/10.1109/ACCESS.2021.3124029},
}

@article{Lau2018,
author = {Lau, Josephine and Zimmerman, Benjamin and Schaub, Florian},
title = {Alexa, Are You Listening? Privacy Perceptions, Concerns and Privacy-Seeking Behaviors with Smart Speakers},
year = {2018},
issue_date = {November 2018},
publisher = {Association for Computing Machinery},
address = {New York, NY, USA},
volume = {2},
number = {CSCW},
url = {https://doi.org/10.1145/3274371},
doi = {10.1145/3274371},
journal = {Proc. ACM Hum.-Comput. Interact.},
month = nov,
articleno = {Article 102},
numpages = {31}
}

@article{shi2016edge,
  title={Edge computing: Vision and challenges},
  author={Shi, Weisong and Cao, Jie and Zhang, Quan and Li, Youhuizi and Xu, Lanyu},
  journal={IEEE internet of things journal},
  volume={3},
  number={5},
  pages={637--646},
  year={2016},
  publisher={IEEE},
  url={https://doi.org/10.1109/JIOT.2016.2579198},
}

@book{petronio2002boundaries,
  title={Boundaries of privacy: Dialectics of disclosure},
  author={Petronio, Sandra},
  year={2002},
  publisher={Suny Press}
}

@misc{poikola2020mydata,
  title={MyData -- an introduction to human-centric use of personal data},
  editor = {Viivi Lähteenoja},
  author={Antti Poikola and Kai Kuikkaniemi and Ossi Kuittinen and Harri Honko and Aleksi Knuutila and Viivi Lähteenoja},
  year={2020},
  edition = "3rd",
  month=jul,
  publisher={Ministry of Transport and Communications},
  url={http://urn.fi/URN:ISBN:978-952-243-617-7},
}

@article{Chen2003,
author = {Chen, S.C. and Dhillon, G.S.},
title = {Interpreting Dimensions of Consumer Trust in E-Commerce},
journal = {Information Technology and Management},
volume = {4},
pages = {303--318},
doi = {https://doi.org/10.1023/A:1022962631249},
year = {2003}
}

@article{Xie2009,
author = {Xie, Yi and Peng, Siqing},
title = {How to repair customer trust after negative publicity: The roles of competence, integrity, benevolence, and forgiveness},
journal = {Psychology \& Marketing},
volume = {26},
number = {7},
pages = {572-589},
url = {https://doi.org/10.1002/mar.20289},
year = {2009}
}

@article{levy2020intimate,
  title={Privacy threats in intimate relationships},
  author={Levy, Karen and Schneier, Bruce},
  journal={Journal of Cybersecurity},
  volume={6},
  number={1},
  year={2020},
  publisher={Oxford University Press},
  url = {https://doi.org/10.1093/cybsec/tyaa006},
}

@article{Lynskey2019GuardianPrivacyScandal,
author = {Dorian Lynskey},
title = {Alexa, are you invading my privacy? -- the dark side of our voice assistants},
url={https://www.theguardian.com/technology/2019/oct/09/alexa-are-you-invading-my-privacy-the-dark-side-of-our-voice-assistants},
year = 2019,
journal = "The Guardian",
}

@article{moscaritolo2018alexa,
url={https://www.pcmag.com/news/amazon-alexa-sends-familys-private-conversation-to-contact},
year = 2018,
title={Amazon Alexa Sends Family's Private Conversation to Contact},
author={Angela Moscaritolo},
journal = {PCMag},
}

@article{liptak2017alexa,
title = {Amazon’s Alexa started ordering people dollhouses after hearing its name on TV},
author = {Andrew Liptak},
journal="The Verge",
year = 2017,
url = {https://www.theverge.com/2017/1/7/14200210/amazon-alexa-tech-news-anchor-order-dollhouse},
}

@article{hern2019,
author = {Alex Hern},
title = {Apple contractors regularly hear confidential details‘ on {Siri} recordings},
journal = {The Guardian},
year = 2019,
url = {https://www.theguardian.com/technology/2019/jul/26/apple-contractors-regularly-hear-confidential-details-on-siri-recordings},
}

@article{day2019,
author = {Matt Day and Giles Turner and Natalia Drozdiak}, title = {Amazon Workers Are Listening to What You Tell {Alexa}}, 
journal = {Bloomberg.com}, 
year = 2019,
url = {https://www.bloomberg.com/news/articles/2019-04-10/is-anyone-listening-to-you-on-alexa-a-global-team-reviews-audio},
}

@article{brewster2021,
  author = {Thomas Brewster},
  title = {Fraudsters Cloned Company Director’s Voice In \$35 Million Bank Heist, Police Find},
  journal = {Forbes},
  year = 2021,
  url = {https://www.forbes.com/sites/thomasbrewster/2021/10/14/huge-bank-fraud-uses-deep-fake-voice-tech-to-steal-millions/},
  }

@article{perez2020,
author = {S. Perez},
title = {Smart speaker sales reached new record of 146.9M in 2019, up 70\% from 2018},
journal = {techcrunch.com},
url={https://techcrunch.com/2020/02/17/smart-speaker-sales-reached-new-record-of-146-9m-in-2019-up-70-from-2018/},
year = 2020,
}

@inproceedings{lotfidereshgi2022cognitive,
  title={Cognitive coding of speech},
  author={Lotfidereshgi, Reza and Gournay, Philippe},
  booktitle={IEEE International Conference on Acoustics, Speech and Signal Processing (ICASSP)},
  pages={7772--7776},
  year={2022},
  organization={IEEE},
  url={https://doi.org/10.1109/ICASSP43922.2022.9747914},
}

@MastersThesis{rech2022multi,
  title={Multi-Device Speech Enhancement for Privacy and Quality},
  author={Rech, Silas},
  school={Aalto University},
  year={2022}
}

@article{patterson2021carbon,
  publtype={informal},
  author={David A. Patterson and Joseph Gonzalez and Quoc V. Le and Chen Liang and Lluis-Miquel Munguia and Daniel Rothchild and David R. So and Maud Texier and Jeff Dean},
  title={Carbon Emissions and Large Neural Network Training},
  year={2021},
  cdate={1609459200000},
  journal={CoRR},
  volume={abs/2104.10350},
  url={https://arxiv.org/abs/2104.10350}
}

@inproceedings{backstrom2021intuitive,
  title={Intuitive Privacy from Acoustic Reach: A Case for Networked Voice User-Interfaces},
  author={B{\"a}ckstr{\"o}m, Tom and Das, Sneha and Perez Zarazaga, Pablo and Fischer, Johannes and Findling, Rainhard and Sigg, Stephan and Nguyen, Le},
  year={2021},
  booktitle={Proc. 2021 ISCA Symposium on Security and Privacy in Speech},
  url={https://research.aalto.fi/files/67799021/SIG_Symposium2021_Privacy_2_.pdf},
}

@inproceedings{zewoudie2021use,
  title={The Use of Audio Fingerprints for Authentication of Speakers on Speech Operated Interfaces},
  author={Zewoudie, Abraham Woubie and B{\"a}ckstr{\"o}m, Tom and Zarazaga, Pablo P{\'e}rez},
  year={2021},
  booktitle={Proc. 2021 ISCA Symposium on Security and Privacy in Speech},
  url = {https://research.aalto.fi/files/75674953/Woubie_Use_of_Audio_Fingerprints_isca.pdf},
}

@article{zarazaga2020acoustic,
  title={Acoustic fingerprints for access management in ad-hoc sensor networks},
  author={Zarazaga, Pablo P{\'e}rez and B{\"a}ckstr{\"o}m, Tom and Sigg, Stephan},
  journal={IEEE Access},
  volume={8},
  pages={166083--166094},
  year={2020},
  publisher={IEEE},
  doi = {10.1109/ACCESS.2020.3022618},
}

@inproceedings{sigg2020provable,
  title={Provable consent for voice user interfaces},
  author={Sigg, Stephan and Zarazaga, Pablo Perez and Bäckström, Tom},
  booktitle={2020 IEEE International Conference on Pervasive Computing and Communications Workshops (PerCom Workshops)},
  pages={1--4},
  year={2020},
  doi={10.1109/PerComWorkshops48775.2020.9156182},
}

@inproceedings{Leschanowsky2020,
  author={Anna Leschanowsky and Sneha Das and Tom Bäckström and Pablo Pérez Zarazaga},
  title={{Perception of Privacy Measured in the Crowd — Paired Comparison on the Effect of Background Noises}},
  year=2020,
  booktitle={Proc. Interspeech},
  pages={4651--4655},
  doi={10.21437/Interspeech.2020-2299},
  url={http://dx.doi.org/10.21437/Interspeech.2020-2299}
}

@inproceedings{zarazaga19_interspeech,
  author={Pablo Pérez Zarazaga and Sneha Das and Tom Bäckström and V. V. Vidyadhara Raju and Anil Kumar Vuppala},
  title={{Sound Privacy: A Conversational Speech Corpus for Quantifying the Experience of Privacy}},
  year=2019,
  booktitle={Proc. Interspeech},
  pages={3720--3724},
  doi={10.21437/Interspeech.2019-1172}
}

@phdthesis{zarazaga2022thesis,
  author = {Pablo Pérez Zarazaga},
  title   = "Preserving Speech Privacy in Interactions with Ad Hoc Sensor Networks",
  school  = "Aalto University",
  year    = "2022",
  url={http://urn.fi/URN:ISBN:978-952-64-0972-6},
  }

@phdthesis{das2021thesis,
  author = {Sneha Das},
  title   = "Robust and Efficient Methods for Distributed Speech Processing - Perspectives on Coding, Enhancement and Privacy",
  school  = "Aalto University",
  year    = "2021",
  url={http://urn.fi/URN:ISBN:978-952-64-0576-6},
  }

@inproceedings{zarazaga21_interspeech,
  author={Pablo Pérez Zarazaga and Mariem Bouafif Mansali and Tom Bäckström and Zied Lachiri},
  title={{Cancellation of Local Competing Speaker with Near-Field Localization for Distributed ad-hoc Sensor Network}},
  year=2021,
  booktitle={Proc. Interspeech},
  pages={676--680},
  doi={10.21437/Interspeech.2021-1329}
}

@inproceedings{chouchane21_interspeech,
  author={Oubaïda Chouchane and Baptiste Brossier and Jorge Esteban Gamboa Gamboa and Thomas Lardy and Hemlata Tak and Orhan Ermis and Madhu R. Kamble and Jose Patino and Nicholas Evans and Melek Önen and Massimiliano Todisco},
  title={{Privacy-Preserving Voice Anti-Spoofing Using Secure Multi-Party Computation}},
  year=2021,
  booktitle={Proc. Interspeech},
  pages={856--860},
  doi={10.21437/Interspeech.2021-983}
}

@inproceedings{shah21_interspeech,
  author={Muhammad A. Shah and Joseph Szurley and Markus Mueller and Athanasios Mouchtaris and Jasha Droppo},
  title={{Evaluating the Vulnerability of End-to-End Automatic Speech Recognition Models to Membership Inference Attacks}},
  year=2021,
  booktitle={Proc. Interspeech},
  pages={891--895},
  doi={10.21437/Interspeech.2021-1188}
}

@inproceedings{yang21_interspeech,
  author={Chao-Han Huck Yang and Sabato Marco Siniscalchi and Chin-Hui Lee},
  title={{PATE-AAE: Incorporating Adversarial Autoencoder into Private Aggregation of Teacher Ensembles for Spoken Command Classification}},
  year=2021,
  booktitle={Proc. Interspeech},
  pages={881--885},
  doi={10.21437/Interspeech.2021-640}
}

@inproceedings{koppelmann21_interspeech,
  author={Timm Koppelmann and Alexandru Nelus and Lea Schönherr and Dorothea Kolossa and Rainer Martin},
  title={{Privacy-Preserving Feature Extraction for Cloud-Based Wake Word Verification}},
  year=2021,
  booktitle={Proc. Interspeech},
  pages={876--880},
  doi={10.21437/Interspeech.2021-262}
}

@inproceedings{ro21_interspeech,
  author={Jae Ro and Mingqing Chen and Rajiv Mathews and Mehryar Mohri and Ananda Theertha Suresh},
  title={{Communication-Efficient Agnostic Federated Averaging}},
  year=2021,
  booktitle={Proc. Interspeech},
  pages={871--875},
  doi={10.21437/Interspeech.2021-153}
}

@inproceedings{novotney21_interspeech,
  author={Scott Novotney and Yile Gu and Ivan Bulyko},
  title={{Adjunct-Emeritus Distillation for Semi-Supervised Language Model Adaptation}},
  year=2021,
  booktitle={Proc. Interspeech},
  pages={866--870},
  doi={10.21437/Interspeech.2021-27}
}

@inproceedings{aloufi21_interspeech,
  author={Ranya Aloufi and Hamed Haddadi and David Boyle},
  title={{Configurable Privacy-Preserving Automatic Speech Recognition}},
  year=2021,
  booktitle={Proc. Interspeech},
  pages={861--865},
  doi={10.21437/Interspeech.2021-1783}
}

@inproceedings{stoidis21_interspeech,
  author={Dimitrios Stoidis and Andrea Cavallaro},
  title={{Protecting Gender and Identity with Disentangled Speech Representations}},
  year=2021,
  booktitle={Proc. Interspeech},
  pages={1699--1703},
  doi={10.21437/Interspeech.2021-2163}
}

@inproceedings{prajapati21_interspeech,
  author={Gauri P. Prajapati and Dipesh K. Singh and Preet P. Amin and Hemant A. Patil},
  title={{Voice Privacy Through x-Vector and CycleGAN-Based Anonymization}},
  year=2021,
  booktitle={Proc. Interspeech},
  pages={1684--1688},
  doi={10.21437/Interspeech.2021-1573}
}

@inproceedings{noe21_interspeech,
  author={Paul-Gauthier Noé and Mohammad Mohammadamini and Driss Matrouf and Titouan Parcollet and Andreas Nautsch and Jean-François Bonastre},
  title={{Adversarial Disentanglement of Speaker Representation for Attribute-Driven Privacy Preservation}},
  year=2021,
  booktitle={Proc. Interspeech},
  pages={1902--1906},
  doi={10.21437/Interspeech.2021-1712}
}

@inproceedings{koppelmann22_interspeech,
  author={Timm Koppelmann and Luca Becker and Alexandru Nelus and Rene Glitza and Lea Schönherr and Rainer Martin},
  title={{Clustering-based Wake Word Detection in Privacy-aware Acoustic Sensor Networks}},
  year=2022,
  booktitle={Proc. Interspeech},
  pages={719--723},
  doi={10.21437/Interspeech.2022-842}
}

@inproceedings{feng22b_interspeech,
  author={Tiantian Feng and Raghuveer Peri and Shrikanth Narayanan},
  title={{User-Level Differential Privacy against Attribute Inference Attack of Speech Emotion Recognition on Federated Learning}},
  year=2022,
  booktitle={Proc. Interspeech},
  pages={5055--5059},
  doi={10.21437/Interspeech.2022-10060}
}

@inproceedings{stoidis22_interspeech,
  author={Dimitrios Stoidis and Andrea Cavallaro},
  title={{Generating gender-ambiguous voices for privacy-preserving speech recognition}},
  year=2022,
  booktitle={Proc. Interspeech},
  pages={4237--4241},
  doi={10.21437/Interspeech.2022-11322}
}

@inproceedings{liu22_interspeech,
  author={Yuchen Liu and Apu Kapadia and Donald Williamson},
  title={{Preventing sensitive-word recognition using self-supervised learning to preserve user-privacy for automatic speech recognition}},
  year=2022,
  booktitle={Proc. Interspeech},
  pages={4207--4211},
  doi={10.21437/Interspeech.2022-85}
}

@inproceedings{teixeira22_interspeech,
  author={Francisco Teixeira and Alberto Abad and Bhiksha Raj and Isabel Trancoso},
  title={{Towards End-to-End Private Automatic Speaker Recognition}},
  year=2022,
  booktitle={Proc. Interspeech},
  pages={2798--2802},
  doi={10.21437/Interspeech.2022-10672}
}

@inproceedings{pierre22_interspeech,
  author={Champion Pierre and Anthony Larcher and Denis Jouvet},
  title={{Are disentangled representations all you need to build speaker anonymization systems?}},
  year=2022,
  booktitle={Proc. Interspeech},
  pages={2793--2797},
  doi={10.21437/Interspeech.2022-10586}
}

@inproceedings{adelani20_interspeech,
  author={David Ifeoluwa Adelani and Ali Davody and Thomas Kleinbauer and Dietrich Klakow},
  title={{Privacy Guarantees for De-Identifying Text Transformations}},
  year=2020,
  booktitle={Proc. Interspeech},
  pages={4666--4670},
  doi={10.21437/Interspeech.2020-2208}
}

@inproceedings{oconnor20_interspeech,
  author={Matt O’Connor and W. Bastiaan Kleijn},
  title={{Distributed Summation Privacy for Speech Enhancement}},
  year=2020,
  booktitle={Proc. Interspeech},
  pages={4646--4650},
  doi={10.21437/Interspeech.2020-1977}
}

@inproceedings{granqvist20_interspeech,
  author={Filip Granqvist and Matt Seigel and Rogier van Dalen and Áine Cahill and Stephen Shum and Matthias Paulik},
  title={{Improving On-Device Speaker Verification Using Federated Learning with Privacy}},
  year=2020,
  booktitle={Proc. Interspeech},
  pages={4328--4332},
  doi={10.21437/Interspeech.2020-2944}
}

@inproceedings{garg20b_interspeech,
  author={Abhinav Garg and Gowtham P. Vadisetti and Dhananjaya Gowda and Sichen Jin and Aditya Jayasimha and Youngho Han and Jiyeon Kim and Junmo Park and Kwangyoun Kim and Sooyeon Kim and Young-yoon Lee and Kyungbo Min and Chanwoo Kim},
  title={{Streaming On-Device End-to-End ASR System for Privacy-Sensitive Voice-Typing}},
  year=2020,
  booktitle={Proc. Interspeech},
  pages={3371--3375},
  doi={10.21437/Interspeech.2020-3172}
}

@inproceedings{noe20_interspeech,
  author={Paul-Gauthier Noé and Jean-François Bonastre and Driss Matrouf and N. Tomashenko and Andreas Nautsch and Nicholas Evans},
  title={{Speech Pseudonymisation Assessment Using Voice Similarity Matrices}},
  year=2020,
  booktitle={Proc. Interspeech},
  pages={1718--1722},
  doi={10.21437/Interspeech.2020-2720}
}

@inproceedings{srivastava20_interspeech,
  author={Brij Mohan Lal Srivastava and N. Tomashenko and Xin Wang and Emmanuel Vincent and Junichi Yamagishi and Mohamed Maouche and Aurélien Bellet and Marc Tommasi},
  title={{Design Choices for X-Vector Based Speaker Anonymization}},
  year=2020,
  booktitle={Proc. Interspeech},
  pages={1713--1717},
  doi={10.21437/Interspeech.2020-2692}
}

@inproceedings{maouche20_interspeech,
  author={Mohamed Maouche and Brij Mohan Lal Srivastava and Nathalie Vauquier and Aurélien Bellet and Marc Tommasi and Emmanuel Vincent},
  title={{A Comparative Study of Speech Anonymization Metrics}},
  year=2020,
  booktitle={Proc. Interspeech},
  pages={1708--1712},
  doi={10.21437/Interspeech.2020-2248}
}

@inproceedings{mawalim20_interspeech,
  author={Candy Olivia Mawalim and Kasorn Galajit and Jessada Karnjana and Masashi Unoki},
  title={{X-Vector Singular Value Modification and Statistical-Based Decomposition with Ensemble Regression Modeling for Speaker Anonymization System}},
  year=2020,
  booktitle={Proc. Interspeech},
  pages={1703--1707},
  doi={10.21437/Interspeech.2020-1887}
}

@inproceedings{nautsch20_interspeech,
  author={Andreas Nautsch and Jose Patino and N. Tomashenko and Junichi Yamagishi and Paul-Gauthier Noé and Jean-François Bonastre and Massimiliano Todisco and Nicholas Evans},
  title={{The Privacy ZEBRA: Zero Evidence Biometric Recognition Assessment}},
  year=2020,
  booktitle={Proc. Interspeech},
  pages={1698--1702},
  doi={10.21437/Interspeech.2020-1815}
}

@inproceedings{thaine19_interspeech,
  author={Patricia Thaine and Gerald Penn},
  title={{Extracting Mel-Frequency and Bark-Frequency Cepstral Coefficients from Encrypted Signals}},
  year=2019,
  booktitle={Proc. Interspeech},
  pages={3715--3719},
  doi={10.21437/Interspeech.2019-1136}
}

@inproceedings{nelus19b_interspeech,
  author={Alexandru Nelus and Janek Ebbers and Reinhold Haeb-Umbach and Rainer Martin},
  title={{Privacy-Preserving Variational Information Feature Extraction for Domestic Activity Monitoring versus Speaker Identification}},
  year=2019,
  booktitle={Proc. Interspeech},
  pages={3710--3714},
  doi={10.21437/Interspeech.2019-1703}
}

@inproceedings{nelus19_interspeech,
  author={Alexandru Nelus and Silas Rech and Timm Koppelmann and Henrik Biermann and Rainer Martin},
  title={{Privacy-Preserving Siamese Feature Extraction for Gender Recognition versus Speaker Identification}},
  year=2019,
  booktitle={Proc. Interspeech},
  pages={3705--3709},
  doi={10.21437/Interspeech.2019-1148}
}

@inproceedings{srivastava19_interspeech,
  author={Brij Mohan Lal Srivastava and Aurélien Bellet and Marc Tommasi and Emmanuel Vincent},
  title={{Privacy-Preserving Adversarial Representation Learning in ASR: Reality or Illusion?}},
  year=2019,
  booktitle={Proc. Interspeech},
  pages={3700--3704},
  doi={10.21437/Interspeech.2019-2415}
}

@inproceedings{nautsch19c_interspeech,
  author={Andreas Nautsch and Catherine Jasserand and Els Kindt and Massimiliano Todisco and Isabel Trancoso and Nicholas Evans},
  title={{The GDPR \& Speech Data: Reflections of Legal and Technology Communities, First Steps Towards a Common Understanding}},
  year=2019,
  booktitle={Proc. Interspeech},
  pages={3695--3699},
  doi={10.21437/Interspeech.2019-2647}
}

@inproceedings{nautsch19b_interspeech,
  author={Andreas Nautsch and Jose Patino and Amos Treiber and Themos Stafylakis and Petr Mizera and Massimiliano Todisco and Thomas Schneider and Nicholas Evans},
  title={{Privacy-Preserving Speaker Recognition with Cohort Score Normalisation}},
  year=2019,
  booktitle={Proc. Interspeech},
  pages={2868--2872},
  doi={10.21437/Interspeech.2019-2638}
}

@inproceedings{haider19_interspeech,
  author={Fasih Haider and Saturnino Luz},
  title={{A System for Real-Time Privacy Preserving Data Collection for Ambient Assisted Living}},
  year=2019,
  booktitle={Proc. Interspeech},
  pages={2374--2375},
  url={https://isca-speech.org/archive/pdfs/interspeech_2019/haider19_interspeech.pdf},
}

@article{nautsch2019preserving,
  title={Preserving privacy in speaker and speech characterisation},
  author={Nautsch, Andreas and Jim{\'e}nez, Abelino and Treiber, Amos and Kolberg, Jascha and Jasserand, Catherine and Kindt, Els and Delgado, H{\'e}ctor and Todisco, Massimiliano and Hmani, Mohamed Amine and Mtibaa, Aymen and others},
  journal={Computer Speech \& Language},
  volume={58},
  pages={441--480},
  year={2019},
  publisher={Elsevier},
  url={https://doi.org/10.1016/j.csl.2019.06.001},
}

@article{treiber2019privacy,
  title={Privacy-preserving PLDA speaker verification using outsourced secure computation},
  author={Treiber, Amos and Nautsch, Andreas and Kolberg, Jascha and Schneider, Thomas and Busch, Christoph},
  journal={Speech Communication},
  volume={114},
  pages={60--71},
  year={2019},
  publisher={Elsevier},
  url={https://doi.org/10.1016/j.specom.2019.09.004},
}

@article{bruggemeier2022perceptions,
  title={Perceptions and reactions to conversational privacy initiated by a conversational user interface},
  author={Br{\"u}ggemeier, Birgit and Lalone, Philip},
  journal={Computer Speech \& Language},
  volume={71},
  pages={101269},
  year={2022},
  publisher={Elsevier},
  url={https://doi.org/10.1016/j.csl.2021.101269},
}

@misc{pepino2020distrust,
  doi = {10.48550/ARXIV.2007.15711},
  
  url = {https://arxiv.org/abs/2007.15711},
  
  author = {Pepino, Leonardo and Riera, Pablo and Gauder, Lara and Gravano, Agustín and Ferrer, Luciana},
  
  title = {Detecting Distrust Towards the Skills of a Virtual Assistant Using Speech},
  
  publisher = {arXiv},
  
  year = {2020},
  
  copyright = {arXiv.org perpetual, non-exclusive license}
}

@article{lutz2021privacy,
  title={Privacy and smart speakers: A multi-dimensional approach},
  author={Lutz, Christoph and Newlands, Gemma},
  journal={The Information Society},
  volume={37},
  number={3},
  pages={147--162},
  year={2021},
  publisher={Taylor \& Francis},
  url={https://doi.org/10.1080/01972243.2021.1897914},
}

@inproceedings{reinten2017speech,
  title={Speech privacy in multiple-bed patient rooms},
  author={Reinten, Jikke and Braat, Ella and Zuydervliet, Ronald and Valk, Marten and De Mast, Quirijn},
  booktitle={Healthy Buildings Europe 2017, HB 2017},
  pages={Paper--ID},
  year={2017},
  organization={International Society of Indoor Air Quality and Climate (ISIAQ)},
  url={https://research.tue.nl/en/publications/speech-privacy-in-multiple-bed-patient-rooms},
}

@misc{shoemate2022sottovoce,
  doi = {10.48550/ARXIV.2207.07816},
  
  url = {https://arxiv.org/abs/2207.07816},
  
  author = {Shoemate, Michael and Jett, Kevin and Cowan, Ethan and Colbath, Sean and Honaker, James and Muthukumar, Prasanna},
  
  title = {Sotto Voce: Federated Speech Recognition with Differential Privacy Guarantees},
  
  publisher = {arXiv},
  
  year = {2022},
  
  copyright = {Creative Commons Attribution Share Alike 4.0 International}
}

@article{shamsabadi2023diff,
  title={Differentially Private Speaker Anonymization},
  author={Shamsabadi, Ali Shahin and Srivastava, Brij Mohan Lal and Bellet, Aur{\'e}lien and Vauquier, Nathalie and Vincent, Emmanuel and Maouche, Mohamed and Tommasi, Marc and Papernot, Nicolas},
  journal={Proceedings on Privacy Enhancing Technologies},
  volume={1},
  pages={98--114},
  year={2023},
  url={https://doi.org/10.56553/popets-2023-0007},
}

@INPROCEEDINGS{nielsen2018,  
author={Nielsen, Jesper Kjær and Lee, Taewoong and Jensen, Jesper Rindom and Christensen, Mads Græsbøll},  
booktitle={2018 52nd Asilomar Conference on Signals, Systems, and Computers},   
title={Sound Zones As An Optimal Filtering Problem},   
year={2018},  
volume={},  
number={},  
pages={1075-1079},  
doi={10.1109/ACSSC.2018.8645268}}

@article{WALLACE2022101285,
title = {Combining background noise and artificial masking to achieve privacy in sound zones},
journal = {Computer Speech \& Language},
volume = {72},
pages = {101285},
year = {2022},
issn = {0885-2308},
doi = {https://doi.org/10.1016/j.csl.2021.101285},
url = {https://www.sciencedirect.com/science/article/pii/S0885230821000875},
author = {Daniel Wallace and Jordan Cheer}
}

@INPROCEEDINGS{donley2016,  
author={Donley, Jacob and Ritz, Christian and Kleijn, W. Bastiaan},  
booktitle={2016 IEEE International Conference on Acoustics, Speech and Signal Processing (ICASSP)},   
title={Improving speech privacy in personal sound zones},   
year={2016},  
pages={311-315},  
doi={10.1109/ICASSP.2016.7471687}}

@INPROCEEDINGS{nelus2021,  
author={Nelus, Alexandru and Glitza, Rene and Martin, Rainer},  
booktitle={ICASSP 2021 - 2021 IEEE International Conference on Acoustics, Speech and Signal Processing (ICASSP)},   
title={Estimation of Microphone Clusters in Acoustic Sensor Networks Using Unsupervised Federated Learning},   
year={2021},  
pages={761-765},  
doi={10.1109/ICASSP39728.2021.9414186}}

@INPROCEEDINGS{Zhang2019,  
author={Zhang, Shi-Xiong and Gong, Yifan and Yu, Dong},  
booktitle={ICASSP 2019 - 2019 IEEE International Conference on Acoustics, Speech and Signal Processing (ICASSP)},   
title={Encrypted Speech Recognition Using Deep Polynomial Networks},   
year={2019},  
pages={5691-5695},  
doi={10.1109/ICASSP.2019.8683721}}

@INPROCEEDINGS{Tishby2015,  
author={Tishby, Naftali and Zaslavsky, Noga},  
booktitle={2015 IEEE Information Theory Workshop (ITW)},   
title={Deep learning and the information bottleneck principle},   
year={2015},  
pages={1-5},  
doi={10.1109/ITW.2015.7133169}}

@inproceedings{williams2021revisiting,
  title={Revisiting Speech Content Privacy},
  author={Williams, Jennifer and Yamagishi, Junichi and No{\'e}, Paul-Gauthier and Valentini-Botinhao, Cassia and Bonastre, Jean-Fran{\c{c}}ois},
  booktitle={1st ISCA Symposium of the Security \& Privacy in Speech Communication},
  year={2021},
  url={https://arxiv.org/pdf/2110.06760},
}

@INPROCEEDINGS{nelus2018itg,  
author={Nelus, Alexandru and Martin, Rainer},  
booktitle={Speech Communication; 13th ITG-Symposium},   
title={Gender Discrimination Versus Speaker Identification Through Privacy-Aware Adversarial Feature Extraction},   
year={2018},  
pages={1-5},  
url={https://ieeexplore.ieee.org/abstract/document/8578003}}

@misc{Orekondy2018gradientleaks,
  doi = {10.48550/ARXIV.1805.05838},
  
  url = {https://arxiv.org/abs/1805.05838},
  
  author = {Orekondy, Tribhuvanesh and Oh, Seong Joon and Zhang, Yang and Schiele, Bernt and Fritz, Mario},
  
  title = {Gradient-Leaks: Understanding and Controlling Deanonymization in Federated Learning},
  
  publisher = {arXiv},
  
  year = {2018},
  
  copyright = {Creative Commons Attribution Non Commercial Share Alike 4.0 International}
}

@ARTICLE{Srivastava2022taslp,  
author={Srivastava, Brij Mohan Lal and Maouche, Mohamed and Sahidullah, Md and Vincent, Emmanuel and Bellet, Aurélien and Tommasi, Marc and Tomashenko, Natalia and Wang, Xin and Yamagishi, Junichi},  
journal={IEEE/ACM Transactions on Audio, Speech, and Language Processing},   
title={Privacy and Utility of X-Vector Based Speaker Anonymization},   
year={2022},  
volume={30}, 
number={},  
pages={2383-2395},  
doi={10.1109/TASLP.2022.3190741}}

@phdthesis{Nelus2022,
  author      = {Alexandru Nelus},
  title       = {Privacy-preserving audio features for classification and clustering in acoustic sensor networks},
  school      = {Ruhr-Universit{\"a}t Bochum, Universit{\"a}tsbibliothek},
  doi       = {10.13154/294-9064},
  year        = {2022},
}

@InProceedings{Kurz2021,
author="Kurz, Miriam
and Br{\"u}ggemeier, Birgit
and Breiter, Michael",
editor="Kurosu, Masaaki",
title="Success is not Final; Failure is not Fatal -- Task Success and User Experience in Interactions with Alexa, Google Assistant and Siri",
booktitle="Human-Computer Interaction. Design and User Experience Case Studies",
year="2021",
publisher="Springer International Publishing",
address="Cham",
pages="351--369",
isbn="978-3-030-78468-3"
}

@article{backstrom2020privacy,
  title={Privacy in speech interfaces},
  author={B{\"a}ckstr{\"o}m, Tom and Br{\"u}ggemeier, Birgit and Fischer, Johannes},
  journal={VDE ITG News},
  publisher = {VDE},
  year={2020},
  url={https://www.vde.com/resource/blob/1991012/07662bec66907573ab254c3d99394ec7/itg-news-juli-oktober-2020-data.pdf},
}

@article{dawkins2017psychological,
  title={Psychological ownership: A review and research agenda},
  author={Dawkins, Sarah and Tian, Amy Wei and Newman, Alexander and Martin, Angela},
  journal={Journal of Organizational Behavior},
  volume={38},
  number={2},
  pages={163--183},
  year={2017},
  publisher={Wiley Online Library},
  url={https://doi.org/10.1002/job.2057},
}

@inproceedings{cichy2014extending,
  title={Extending the privacy calculus: the role of psychological ownership},
  author={Cichy, Patrick and Salge, Torsten-Oliver and Kohli, Rajiv},
  year={2014},
  booktitle={ICIS 2014 Proceedings},
  volume = 30,
  url={https://aisel.aisnet.org/icis2014/proceedings/ISSecurity/30},
}

@article{hulsen2022stalkers,
  title = {Stalkers intimideren ex-partners via slimme camera's, lampen en alarmen},
  author = {Stan Hulsen},
  year = 2022,
  journal = {RTL Nieuws},
  url = {https://www.rtlnieuws.nl/nieuws/nederland/artikel/5345441/stalkers-intimideren-ex-partners-slimme-cameras-lampen-en-alarmen},
  }

@article{lu2019abusers,
    url={https://www.vice.com/en/article/d3akpk/smart-home-technology-stalking-harassment},
    title = {How Abusers Are Exploiting Smart Home Devices},
    author = {Donna Lu},
    year = 2019,
    journal = {Vice},
    }

@misc{avast2021stalkerware,
   url={https://press.avast.com/use-of-stalkerware-and-spyware-apps-increase-by-93-since-lockdown-began-in-the-uk},
   title = {Use of Stalkerware and Spyware Apps Increase by 93\% since Lockdown Began in the {UK}},
   author = {Avast~Software~s.r.o.},
   year = 2021,
   }

@article{Derlega1977PrivacyAS,
  title={Privacy and self-disclosure in social relationships},
  author={Valerian J. Derlega and Alan L. Chaikin},
  journal={Journal of Social Issues},
  year={1977},
  volume={33},
  pages={102-115},
  url={https://doi.org/10.1111/j.1540-4560.1977.tb01885.x},
}

@article{penders2004privacy,
  title={Privacy in (mobile) telecommunications services},
  author={Penders, Jacques},
  journal={Ethics and Information Technology},
  volume={6},
  pages={247--260},
  year={2004},
  publisher={Springer},
  url={https://doi.org/10.1007/s10676-005-5605-9},
}

@article{wong2009privacy,
  title={Privacy in Electronic Communications: The regulation of VoIP in the EU and the United States},
  author={Wong, Rebecca and Garrie, Daniel},
  journal={Computer Telecommunications Law Review},
  pages={139--146},
  year={2009},
  url={https://dx.doi.org/10.2139/ssrn.1466153},
}

@article{arapinis2017analysis,
  title={Analysis of privacy in mobile telephony systems},
  author={Arapinis, Myrto and Mancini, Loretta Ilaria and Ritter, Eike and Ryan, Mark Dermot},
  journal={International Journal of Information Security},
  volume={16},
  pages={491--523},
  year={2017},
  publisher={Springer},
  url={https://doi.org/10.1007/s10207-016-0338-9},
  }

@book{pathak2012privacy,
  title={Privacy-preserving machine learning for speech processing},
  author={Pathak, Manas A},
  year={2012},
  publisher={Springer Science \& Business Media},
  url={https://doi.org/10.1007/978-1-4614-4639-2},
}

@inproceedings{habernal2021differential,
  title={When differential privacy meets {NLP}: The devil is in the detail},
  author={Habernal, Ivan},
  booktitle={Proceedings of the 2021 Conference on Empirical Methods in Natural Language Processing},
  pages={1522--1528},
  year={2021},
  url={http://dx.doi.org/10.18653/v1/2021.emnlp-main.114},
}

@article{culnane2017health,
  title={Health data in an open world -- A report on re-identifying patients in the {MBS/PBS} dataset and the implications for future releases of a {Australian} gogernment data},
  author={Culnane, Chris and Rubinstein, Benjamin IP and Teague, Vanessa},
  journal={arXiv:1712.05627},
  year={2017},
  url={https://doi.org/10.48550/arXiv.1712.05627},
}

@misc{nguyen2023federated,
      title={Federated Learning for {ASR} based on {Wav2vec 2.0}}, 
      author={Tuan Nguyen and Salima Mdhaffar and Natalia Tomashenko and Jean-François Bonastre and Yannick Estève},
      year={2023},
      eprint={2302.10790},
      archivePrefix={arXiv},
      primaryClass={eess.AS},
      url={https://doi.org/10.48550/arXiv.2302.10790},
}

@article{BAI202165,
title = {Speaker recognition based on deep learning: An overview},
journal = {Neural Networks},
volume = {140},
pages = {65-99},
year = {2021},
issn = {0893-6080},
doi = {https://doi.org/10.1016/j.neunet.2021.03.004},
author = {Zhongxin Bai and Xiao-Lei Zhang}
}

@report{edpb2021guidelines,
   title={Guidelines 02/2021 on virtual voice assistants},
   author={European Data Protection Board},
   year = 2021,
   url={https://www.edpb.europa.eu/system/files/2021-07/edpb_guidelines_202102_on_vva_v2.0_adopted_en.pdf},
}

@inproceedings{makhdoumi2014information,
  title={From the information bottleneck to the privacy funnel},
  author={Makhdoumi, Ali and Salamatian, Salman and Fawaz, Nadia and M{\'e}dard, Muriel},
  booktitle={2014 IEEE Information Theory Workshop (ITW 2014)},
  pages={501--505},
  year={2014},
  url={https://doi.org/10.1109/ITW.2014.6970882},
}

@inproceedings{neumann2019improving,
  title={Improving speech emotion recognition with unsupervised representation learning on unlabeled speech},
  author={Neumann, Michael and Vu, Ngoc Thang},
  booktitle={IEEE International Conference on Acoustics, Speech and Signal Processing (ICASSP)},
  pages={7390--7394},
  year={2019},
  url ={https://doi.org/10.1109/ICASSP.2019.8682541},
}

@article{chorowski2019unsupervised,
  title={Unsupervised speech representation learning using {WaveNet} autoencoders},
  author={Chorowski, Jan and Weiss, Ron J and Bengio, Samy and Van Den Oord, A{\"a}ron},
  journal={IEEE/ACM transactions on audio, speech, and language processing},
  volume={27},
  number={12},
  pages={2041--2053},
  year={2019},
  url={https://doi.org/10.1109/TASLP.2019.2938863}
}

@inproceedings{zhou2021vaw,
  title={{VAW-GAN} for disentanglement and recomposition of emotional elements in speech},
  author={Zhou, Kun and Sisman, Berrak and Li, Haizhou},
  booktitle={IEEE Spoken Language Technology Workshop (SLT)},
  pages={415--422},
  year={2021},
  url={https://doi.org/10.1109/SLT48900.2021.9383526},
}

@inproceedings{Niekerk2020,
  author={Benjamin van Niekerk and Leanne Nortje and Herman Kamper},
  title={{Vector-Quantized Neural Networks for Acoustic Unit Discovery in the ZeroSpeech 2020 Challenge}},
  year=2020,
  booktitle={Proc. Interspeech},
  pages={4836--4840},
  doi={10.21437/Interspeech.2020-1693},
  url={http://dx.doi.org/10.21437/Interspeech.2020-1693}
}

@article{feng2021attribute,
  title={Attribute inference attack of speech emotion recognition in federated learning settings},
  author={Feng, Tiantian and Hashemi, Hanieh and Hebbar, Rajat and Annavaram, Murali and Narayanan, Shrikanth S},
  journal={arXiv preprint arXiv:2112.13416},
  year={2021},
  url={https://arxiv.org/abs/2112.13416},
}

@inproceedings{han2020voice,
  title={Voice-indistinguishability: Protecting voiceprint in privacy-preserving speech data release},
  author={Han, Yaowei and Li, Sheng and Cao, Yang and Ma, Qiang and Yoshikawa, Masatoshi},
  booktitle={International Conference on Multimedia and Expo (ICME)},
  pages={1--6},
  year={2020},
  organization={IEEE},
  url={https://doi.org/10.1109/ICME46284.2020.9102875},
}

@inproceedings{tomashenko2022privacy,
  title={Privacy attacks for automatic speech recognition acoustic models in a federated learning framework},
  author={Tomashenko, Natalia and Mdhaffar, Salima and Tommasi, Marc and Est{\`e}ve, Yannick and Bonastre, Jean-Fran{\c{c}}ois},
  booktitle={International Conference on Acoustics, Speech and Signal Processing (ICASSP)},
  pages={6972--6976},
  year={2022},
  organization={IEEE},
  url={https://doi.org/10.1109/ICASSP43922.2022.9746541},
}

@inproceedings{Fang2019,
  author={Fuming Fang and Xin Wang and Junichi Yamagishi and Isao Echizen and Massimiliano Todisco and Nicholas Evans and Jean-Francois Bonastre},
  title={{Speaker Anonymization Using X-vector and Neural Waveform Models}},
  year=2019,
  booktitle={Proc. 10th ISCA Speech Synthesis Workshop},
  pages={155--160},
  doi={10.21437/SSW.2019-28},
  url={http://dx.doi.org/10.21437/SSW.2019-28}
}

@article{KINNUNEN201012,
title = {An overview of text-independent speaker recognition: From features to supervectors},
journal = {Speech Communication},
volume = {52},
number = {1},
pages = {12-40},
year = {2010},
issn = {0167-6393},
url = {https://doi.org/10.1016/j.specom.2009.08.009},
author = {Tomi Kinnunen and Haizhou Li}
}

@article{hansen2015speaker,
  title={Speaker recognition by machines and humans: A tutorial review},
  author={Hansen, John HL and Hasan, Taufiq},
  journal={IEEE Signal processing magazine},
  volume={32},
  number={6},
  pages={74--99},
  year={2015},
  publisher={IEEE},
  url={https://doi.org/10.1109/MSP.2015.2462851},
}

@Book{cramer2015mpc,
 title                = {Secure multiparty computation and secret sharing},
 author               = {Cramer, Ronald and Damgård, Ivan and Nielsen, J.B.},
 publisher            = {Cambridge},
 year                 = 2015,
 month                = jul,
 url = {cambridge.org/9781107043053},
}

@article{liu2021machine,
  title={When machine learning meets privacy: A survey and outlook},
  author={Liu, Bo and Ding, Ming and Shaham, Sina and Rahayu, Wenny and Farokhi, Farhad and Lin, Zihuai},
  journal={ACM Computing Surveys (CSUR)},
  volume={54},
  number={2},
  pages={1--36},
  year={2021},
  publisher={ACM New York, NY, USA},
  url={https://doi.org/10.1145/3436755},
}

@phdthesis{wang2021wake,
  title={Wake word detection and its applications},
  author={Wang, Yiming and others},
  year={2021},
  school={Johns Hopkins University},
  url={http://jhir.library.jhu.edu/handle/1774.2/64380},
}

@misc{backstrom2021edpbcomments,
   url = {https://edpb.europa.eu/sites/default/files/webform/public_consultation_reply/edpb_comments.pdf},
   title={Comments to 'Guidelines 02/2021 on Virtual Voice Assistants' of the European Data Protection Board (EDPB)},
   author = {International Speech Communication Association (ISCA)},
   editor = {Tom Bäckström and Andreas Nautsch},
   year = 2021,
   }

@article{willis2020deception,
  title={Deception by design},
  author={Willis, Lauren E},
  journal={Harvard Journal of Law \& Technology},
  volume={34},
  pages={115},
  year={2020},
  url={https://ssrn.com/abstract=3694575},
}

@inproceedings{nass1993anthropomorphism,
  title={Anthropomorphism, agency, and ethopoeia: computers as social actors},
  author={Nass, Clifford and Steuer, Jonathan and Tauber, Ellen and Reeder, Heidi},
  booktitle={INTERACT'93 and CHI'93 conference companion on Human factors in computing systems},
  pages={111--112},
  year={1993},
  url={https://doi.org/10.1145/259964.260137},
}

@article{cornelius2021acceptance,
  title={Acceptance of anthropomorphic technology: a literature review},
  author={Cornelius, Samia and Leidner, Dorothy},
  year={2021},
  url={https://doi.org/10.24251/HICSS.2021.774},
  }

@book{thompson2021machine,
  title={Machine Law, Ethics, and Morality in the Age of Artificial Intelligence},
  author={Thompson, Steven John},
  year={2021},
  publisher={IGI Global},
  doi={10.4018/978-1-7998-4894-3}
}

@book{so2021voice,
   title = "Voice content and usability",
   author = {Preston So},
   url = {https://abookapart.com/products/voice-content-and-usability},
   year = 2021,
   publisher = {A book a part},
   }

@article{iber2021auditory,
  title={Auditory augmented process monitoring for cyber physical production systems},
  author={Iber, Michael and Lechner, Patrik and Jandl, Christian and Mader, Manuel and Reichmann, Michael},
  journal={Personal and Ubiquitous Computing},
  volume={25},
  pages={691--704},
  year={2021},
  publisher={Springer},
  url={https://doi.org/10.1016/j.ijhcs.2016.06.002},
}

@article{hildebrandt2016continuous,
  title={Continuous sonification enhances adequacy of interactions in peripheral process monitoring},
  author={Hildebrandt, Tobias and Hermann, Thomas and Rinderle-Ma, Stefanie},
  journal={International Journal of Human-Computer Studies},
  volume={95},
  pages={54--65},
  year={2016},
  publisher={Elsevier},
  url={https://doi.org/10.1007/s00779-020-01394-3},
}

@techreport{ISObiometric,
type = {Standard},
key = {ISO/IEC 19795-1:2021},
month = may,
year = {2021},
title = {{ISO/IEC 19795-1:2021 Information technology -- Biometric performance testing and reporting -- Part 1: Principles and framework}},
address = {Geneva, CH},
institution = {International Organization for Standardization},
url={https://www.iso.org/standard/73515.html},
}

@article{brummer2006application,
  title={Application-independent evaluation of speaker detection},
  author={Br{\"u}mmer, Niko and Du Preez, Johan},
  journal={Computer Speech \& Language},
  volume={20},
  number={2-3},
  pages={230--275},
  year={2006},
  publisher={Elsevier},
  url={https://doi.org/10.1016/j.csl.2005.08.001},
}

@inproceedings{adler2006towards,
  title={Towards a measure of biometric information},
  author={Adler, Andy and Youmaran, Richard and Loyka, Sergey},
  booktitle={2006 Canadian conference on electrical and computer engineering},
  pages={210--213},
  year={2006},
  organization={IEEE},
  url={https://doi.org/10.1109/CCECE.2006.277447},
}

@article{mahendran2021privacy,
   author={Mahendran, Darshini and Luo, Changqing and Mcinnes, Bridget T.},
journal={IEEE Access}, 
  title={Review: Privacy-Preservation in the Context of Natural Language Processing}, 
  year={2021},
  volume={9},
  pages={147600-147612},
  doi={10.1109/ACCESS.2021.3124163}
  }

@article{schneble2018cambridge,
  title={The Cambridge Analytica affair and Internet-mediated research},
  author={Schneble, Christophe Olivier and Elger, Bernice Simone and Shaw, David},
  journal={EMBO reports},
  volume={19},
  number={8},
  pages={e46579},
  year={2018},
  url={https://doi.org/10.15252/embr.201846579},
}

@online{ovn2023,
   title={Open Voice Network},
   publisher={The Linux Foundation},
   url={https://openvoicenetwork.org},
   year=2023,
   }

@online{mydata2023,
   title={MyData Global},
   url={https://www.mydata.org},
   year=2023,
   }

@online{spsc2023,
   title={ISCA SIG "Security and Privacy in Speech Communication"},
   url={https://www.spsc-sig.org/},
   year=2023,
   publisher={International Speech Communication Association},
   }

@techreport{friedman2000informed,
   author={Batya Friedman and Edward Felten and Lynette I. Millett},
   title = "Informed consent online: A conceptual model and design principles.",
   institution={University of Washington Computer Science \& Engineering Technical Report 00–12–2 8},
   year = 2000,
   url={https://dada.cs.washington.edu/research/tr/2000/12/UW-CSE-00-12-02.pdf},
   }

@inproceedings{Teixeira2018,
  author={Francisco Teixeira and Alberto Abad and Isabel Trancoso},
  title={Patient Privacy in Paralinguistic Tasks},
  year=2018,
  booktitle={Proc. Interspeech 2018},
  pages={3428--3432},
  url={http://dx.doi.org/10.21437/Interspeech.2018-2186}
}

@article{cummings2023challenges,
  title={Challenges towards the Next Frontier in Privacy},
  author={Cummings, Rachel and Desfontaines, Damien and Evans, David and Geambasu, Roxana and Jagielski, Matthew and Huang, Yangsibo and Kairouz, Peter and Kamath, Gautam and Oh, Sewoong and Ohrimenko, Olga and others},
  journal={arXiv preprint arXiv:2304.06929},
  year={2023},
  url={https://doi.org/10.48550/arXiv.2304.06929},
}

@online{heikkila2023cambridge_ai,
   author = {Melissa Heikkilä},
   title = {A Cambridge Analytica-style scandal for AI is coming},
   url = {https://www.technologyreview.com/2023/04/25/1072177/a-cambridge-analytica-style-scandal-for-ai-is-coming/},
   journal = {MIT Technology Review},
   year = 2023,
   }

@article{beerends2013perceptual,
  title={Perceptual Objective Listening Quality Assessment ({POLQA}), The Third Generation {ITU-T} Standard for End-to-End Speech Quality Measurement Part II—Perceptual Model},
  author={Beerends, John G and Schmidmer, Christian and Berger, Jens and Obermann, Matthias and Ullmann, Raphael and Pomy, Joachim and Keyhl, Michael},
  journal={Journal of the Audio Engineering Society},
  volume={61},
  number={6},
  pages={385--402},
  year={2013},
  publisher={Audio Engineering Society},
  url={http://www.aes.org/e-lib/browse.cfm?elib=16829},
}

@inproceedings{rix2001perceptual,
  title={Perceptual evaluation of speech quality ({PESQ}) - a new method for speech quality assessment of telephone networks and codecs},
  author={Rix, Antony W and Beerends, John G and Hollier, Michael P and Hekstra, Andries P},
  booktitle={2001 IEEE international conference on acoustics, speech, and signal processing. Proceedings (Cat. No. 01CH37221)},
  volume={2},
  pages={749--752},
  year={2001},
  organization={IEEE},
  url={https://doi.org/10.1109/ICASSP.2001.941023},
}

@techreport{P800,
type = {Standard},
key = {ITU-T P.800},
month = 8,
year = 1996,
title = {{Recommendation ITU-T P.800 -- Methods for objective and subjective assessment of quality}},
address = {Geneva, CH},
institution = {International Organization for Standardization},
url={https://handle.itu.int/11.1002/1000/3638},
}

@techreport{MUSHRA,
type = {Standard},
key = {ITU-R BS.1534-3},
month = 10,
year = 2015,
title = {{Recommendation ITU-R BS.1534-3 -- Method for the subjective assessment of intermediate quality level of audio systems}},
address = {Geneva, CH},
institution = {International Organization for Standardization},
url={https://www.itu.int/rec/R-REC-BS.1534/en},
}

@phdthesis{williams2022learning,
  title={Learning disentangled speech representations},
  author={Williams, Jennifer},
  year={2022},
  school={The University of Edinburgh},
  url={http://dx.doi.org/10.7488/era/1980},
}

@inproceedings{dhiya2023privacy,
  title={Privacy and Security in the Use of Voice Assistant: An Evaluation of User Awareness and Preferences},
  author={Dhiya'Mardhiyyah, Alya and Latif, Jazlyn Jan Keyla and Tho, Cuk},
  booktitle={2023 International Conference on Information Management and Technology (ICIMTech)},
  pages={481--486},
  year={2023},
  organization={IEEE},
  url={https://doi.org/10.1109/ICIMTech59029.2023.10277724},
}

@article{franzke2021data,
  title={Data Ethics Decision Aid (DEDA): a dialogical framework for ethical inquiry of AI and data projects in the Netherlands},
  author={Franzke, Aline Shakti and Muis, Iris and Sch{\"a}fer, Mirko Tobias},
  journal={Ethics and Information Technology},
  volume={23},
  number={3},
  pages={551--567},
  year={2021},
  publisher={Springer},
  url={https://doi.org/10.1007/s10676-020-09577-5},
}

@article{leschanowsky2024review,
     title={Evaluating Privacy, Security, and Trust Perceptions in Conversational AI: A Systematic Review}, 
      author={Anna Leschanowsky and Silas Rech and Birgit Popp and Tom Bäckström},
      year={2024},
     journal = "accepted to Computers in Human Behavior",
    url={https://arxiv.org/abs/2406.09037},
}

@article{noe2022towards,
  title={Towards a unified assessment framework of speech pseudonymisation},
  author={No{\'e}, Paul-Gauthier and Nautsch, Andreas and Evans, Nicholas and Patino, Jose and Bonastre, Jean-Fran{\c{c}}ois and Tomashenko, Natalia and Matrouf, Driss},
  journal={Computer Speech \& Language},
  volume={72},
  pages={101299},
  year={2022},
  publisher={Elsevier},
  url={https://doi.org/10.1016/j.csl.2021.101299},
}

@inproceedings{shashanka2006secure,
  title={Secure sound classification: Gaussian mixture models},
  author={Shashanka, Madhusudana VS and Smaragdis, Paris},
  booktitle={2006 IEEE International Conference on Acoustics Speech and Signal Processing Proceedings},
  volume={3},
  pages={III--III},
  year={2006},
  organization={IEEE},
  url={https://doi.org/10.1109/ICASSP.2006.1660847},
}

@article{smaragdis2007framework,
  title={A framework for secure speech recognition},
  author={Smaragdis, Paris and Shashanka, Madhusudana},
  journal={IEEE Transactions on Audio, Speech, and Language Processing},
  volume={15},
  number={4},
  pages={1404--1413},
  year={2007},
  publisher={IEEE},
  url={https://doi.org/10.1109/TASL.2007.894526},
}

@inproceedings{pathak11_interspeech,
  author={Manas A. Pathak and Bhiksha Raj},
  title={{Privacy preserving speaker verification using adapted GMMs}},
  year=2011,
  booktitle={Proc. Interspeech 2011},
  pages={2405--2408},
  doi={10.21437/Interspeech.2011-626}
}

@inproceedings{parthasarathi11_interspeech,
  author={Sree Hari Krishnan Parthasarathi and Hervé Bourlard and Daniel Gatica-Perez},
  title={{LP residual features for robust, privacy-sensitive speaker diarization}},
  year=2011,
  booktitle={Proc. Interspeech 2011},
  pages={1045--1048},
  doi={10.21437/Interspeech.2011-390}
}

@inproceedings{pathak2011privacy,
  title={Privacy preserving probabilistic inference with hidden Markov models},
  author={Pathak, Manas and Rane, Shantanu and Sun, Wei and Raj, Bhiksha},
  booktitle={IEEE International Conference on Acoustics, Speech and Signal Processing (ICASSP)},
  pages={5868--5871},
  year={2011},
  organization={IEEE},
url={https://doi.org/10.1109/ICASSP.2011.5947696},
}

@article{pathak2013privacy,
  title={Privacy-preserving speech processing: cryptographic and string-matching frameworks show promise},
  author={Pathak, Manas A and Raj, Bhiksha and Rane, Shantanu D and Smaragdis, Paris},
  journal={IEEE signal processing magazine},
  volume={30},
  number={2},
  pages={62--74},
  year={2013},
  publisher={IEEE},
  url={https://doi.org/10.1109/MSP.2012.2230222}
}

@inproceedings{pathak2012ppsa,
  title={Privacy-preserving speaker authentication},
  author={Pathak, Manas and Portelo, Jose and Raj, Bhiksha and Trancoso, Isabel},
  booktitle={Information Security: 15th International Conference, ISC 2012, Passau, Germany, September 19-21, 2012. Proceedings 15},
  pages={1--22},
  year={2012},
  organization={Springer},
  url={https://doi.org/10.1007/978-3-642-33383-5%5F1},
}

@inproceedings{pathak2012ppsv,
  title={Privacy-preserving speaker verification as password matching},
  author={Pathak, Manas A and Raj, Bhiksha},
  booktitle={2012 IEEE International Conference on Acoustics, Speech and Signal Processing (ICASSP)},
  pages={1849--1852},
  year={2012},
  organization={IEEE},
  url={https://doi.org/10.1109/ICASSP.2012.6288262},
}

@ARTICLE{parthasarathi2013wordless,
  author={Parthasarathi, Sree Hari Krishnan and Bourlard, Hervé and Gatica-Perez, Daniel},
  journal={IEEE Transactions on Audio, Speech, and Language Processing}, 
  title={Wordless Sounds: Robust Speaker Diarization Using Privacy-Preserving Audio Representations}, 
  year={2013},
  volume={21},
  number={1},
  pages={85-98},
  doi={10.1109/TASL.2012.2215588}}

@INPROCEEDINGS{hendriks2013ppdistributed,
  author={Hendriks, Richard C. and Erkin, Zekeriya and Gerkmann, Timo},
  booktitle={IEEE International Conference on Acoustics, Speech and Signal Processing}, 
  title={Privacy-preserving distributed speech enhancement for wireless sensor networks by processing in the encrypted domain}, 
  year={2013},
  volume={},
  number={},
  pages={7005-7009},
  doi={10.1109/ICASSP.2013.6639020}}

@phdthesis{portelo2015privacy,
  title={Privacy-preserving frameworks for speech mining},
  author={Port{\^e}lo, Jos{\'e} Miguel Ladeira},
  year={2015},
  school={Ph. D. Dissertation. Universidade de Lisboa},
  url={http://cvis.cs.cmu.edu/cvis/docs/JosePorteloThesis.pdf},
}

@book{menezes2018handbook,
  title={Handbook of applied cryptography},
  author={Menezes, Alfred J and Van Oorschot, Paul C and Vanstone, Scott A},
  year={2018},
  publisher={CRC press},
  url={https://doi.org/10.1201/9780429466335},
}

@article{deutch1973resolution,
  title={The resolution of conflict},
  author={Deutch, Morton},
  journal={New Haven, London},
  year={1973},
  url={https://www.jstor.org/stable/j.ctt1dszxst},
}

@article{simpson2007psychological,
  title={Psychological foundations of trust},
  author={Simpson, Jeffry A},
  journal={Current directions in psychological science},
  volume={16},
  number={5},
  pages={264--268},
  year={2007},
  publisher={SAGE Publications Sage CA: Los Angeles, CA},
  url={https://doi.org/10.1111/j.1467-8721.2007.00517.x},
}

@article{ometov2018multi,
  title={Multi-factor authentication: A survey},
  author={Ometov, Aleksandr and Bezzateev, Sergey and M{\"a}kitalo, Niko and Andreev, Sergey and Mikkonen, Tommi and Koucheryavy, Yevgeni},
  journal={Cryptography},
  volume={2},
  number={1},
  pages={1},
  year={2018},
  publisher={MDPI},
  url={https://doi.org/10.3390/cryptography2010001},
}

@ARTICLE{rohit2024asr,
  author={Prabhavalkar, Rohit and Hori, Takaaki and Sainath, Tara N. and Schlüter, Ralf and Watanabe, Shinji},
  journal={IEEE/ACM Transactions on Audio, Speech, and Language Processing}, 
  title={End-to-End Speech Recognition: A Survey}, 
  year={2024},
  volume={32},
  pages={325-351},
  url={https://doi.org/10.1109/TASLP.2023.3328283}}

@book{vincent2018audio,
  title={Audio source separation and speech enhancement},
  author={Vincent, Emmanuel and Virtanen, Tuomas and Gannot, Sharon},
  year={2018},
  publisher={John Wiley \& Sons},
  url={https://doi.org/10.1002/9781119279860},
}

@inproceedings{chinen2020visqol,
  title={ViSQOL v3: An open source production ready objective speech and audio metric},
  author={Chinen, Michael and Lim, Felicia SC and Skoglund, Jan and Gureev, Nikita and O'Gorman, Feargus and Hines, Andrew},
  booktitle={2020 twelfth international conference on quality of multimedia experience (QoMEX)},
  pages={1--6},
  year={2020},
  organization={IEEE},
  url={https://doi.org/10.1109/QoMEX48832.2020.9123150},
}

@inproceedings{kairouz2015composition,
  title={The composition theorem for differential privacy},
  author={Kairouz, Peter and Oh, Sewoong and Viswanath, Pramod},
  booktitle={International conference on machine learning},
  pages={1376--1385},
  year={2015},
  organization={PMLR},
  url={https://doi.org/10.1109/TIT.2017.2685505},
}

@inproceedings{balsa2022cryptography,
  title={Cryptography, Trust and Privacy: It's Complicated},
  author={Balsa, Ero and Nissenbaum, Helen and Park, Sunoo},
  booktitle={Proceedings of the 2022 Symposium on Computer Science and Law},
  pages={167--179},
  year={2022},
  url={https://doi.org/10.1145/3511265.3550443},
}

@article{chialva2018conditionals,
  title={Conditionals in homomorphic encryption and machine learning applications},
  author={Chialva, Diego and Dooms, Ann},
  journal={arXiv preprint arXiv:1810.12380},
  year={2018},
  url={https://doi.org/10.48550/arXiv.1810.12380},
}

@inproceedings{teixeira2019privacy,
  title={Privacy-preserving paralinguistic tasks},
  author={Teixeira, Francisco and Abad, Alberto and Trancoso, Isabel},
  booktitle={ICASSP 2019-2019 IEEE International Conference on Acoustics, Speech and Signal Processing (ICASSP)},
  pages={6575--6579},
  year={2019},
  organization={IEEE},
  url={https://doi.org/10.1109/ICASSP.2019.8683595},
}

@inproceedings{yao1986generate,
  title={How to generate and exchange secrets},
  author={Yao, Andrew Chi-Chih},
  booktitle={27th annual symposium on foundations of computer science (Sfcs 1986)},
  pages={162--167},
  year={1986},
  organization={IEEE},
  url={https://doi.org/10.1109/SFCS.1986.25},
}

@article{portelo2015logsum,
  title={Logsum using garbled circuits},
  author={Port{\^e}lo, Jos{\'e} and Raj, Bhiksha and Trancoso, Isabel},
  journal={Plos one},
  volume={10},
  number={3},
  pages={e0122236},
  year={2015},
  url={https://doi.org/10.1371/journal.pone.0122236},
}

@inproceedings{portelo2014privacy,
  title={Privacy-preserving speaker verification using garbled GMMs},
  author={Port{\^e}lo, Jos{\'e} and Raj, Bhiksha and Abad, Alberto and Trancoso, Isabel},
  booktitle={2014 22nd European Signal Processing Conference (EUSIPCO)},
  pages={2070--2074},
  year={2014},
  organization={IEEE},
  url={https://ieeexplore.ieee.org/abstract/document/6952754},
}

@inproceedings{portelo2015bprivacy,
  title={Privacy-preserving query-by-example speech search},
  author={Port{\^e}lo, Jos{\'e} and Abad, Alberto and Raj, Bhiksha and Trancoso, Isabel},
  booktitle={2015 IEEE International Conference on Acoustics, Speech and Signal Processing (ICASSP)},
  pages={1797--1801},
  year={2015},
  organization={IEEE},
  url={https://doi.org/10.1109/ICASSP.2015.7178280},
}

@article{rane2013privacy,
  title={Privacy-preserving nearest neighbor methods: Comparing signals without revealing them},
  author={Rane, Shantanu and Boufounos, Petros T},
  journal={IEEE Signal Processing Magazine},
  volume={30},
  number={2},
  pages={18--28},
  year={2013},
  publisher={IEEE},
  url={https://doi.org/10.1109/MSP.2012.2230221},
}

@inproceedings{jimenez2015secure,
  title={Secure modular hashing},
  author={Jim{\'e}nez, Abelino and Raj, Bhiksha and Portelo, Jose and Trancoso, Isabel},
  booktitle={2015 IEEE international workshop on information forensics and security (WIFS)},
  pages={1--6},
  year={2015},
  organization={IEEE},
  url={https://doi.org/10.1109/WIFS.2015.7368567},
}

@inproceedings{jimenez2017two,
  title={A two factor transformation for speaker verification through $\ell_1$ comparison},
  author={Jim{\'e}nez, Abelino and Raj, Bhiksha},
  booktitle={2017 IEEE Workshop on Information Forensics and Security (WIFS)},
  pages={1--6},
  year={2017},
  organization={IEEE},
  url={https://doi.org/10.1109/WIFS.2017.8267661},
}

@article{yoneyama1983audio,
  title={The audio spotlight: An application of nonlinear interaction of sound waves to a new type of loudspeaker design},
  author={Yoneyama, Masahide and Fujimoto, Jun-ichiroh and Kawamo, Yu and Sasabe, Shoichi},
  journal={The Journal of the Acoustical Society of America},
  volume={73},
  number={5},
  pages={1532--1536},
  year={1983},
  publisher={Acoustical Society of America},
  url={https://doi.org/10.1121/1.389414},
}

@article{gardner1998_3daudio,
  title={3-D Audio Using Loudspeakers},
  author={Gardner, William G},
  volume={444},
  year={1998},
  publisher={Springer Science \& Business Media},
  url={https://link.springer.com/book/9780792381563},
}

@article{el2022differential,
  title={Differential privacy for deep and federated learning: A survey},
  author={El Ouadrhiri, Ahmed and Abdelhadi, Ahmed},
  journal={IEEE access},
  volume={10},
  pages={22359--22380},
  year={2022},
  publisher={IEEE},
  url={https://doi.org/10.1109/ACCESS.2022.3151670},
}

@inproceedings{portelo2014bprivacy,
  title={Privacy-preserving speaker verification using secure binary embeddings},
  author={Port{\^e}lo, Jos{\'e} and Raj, Bhiksha and Abad, Alberto and Trancoso, Isabel},
  booktitle={2014 37th International convention on information and communication technology, electronics and microelectronics (MIPRO)},
  pages={1268--1272},
  year={2014},
  organization={IEEE},
  url={https://doi.org/10.1109/MIPRO.2014.6859762},
}

@book{pulkki2018parametric,
  title={Parametric time-frequency domain spatial audio},
  author={Pulkki, Ville and Delikaris-Manias, Symeon and Politis, Archontis},
  year={2018},
  publisher={Wiley Online Library},
  url={https://doi.org/10.1002/9781119252634},
}

@book{pulkki2015communication,
  title={Communication acoustics: an introduction to speech, audio and psychoacoustics},
  author={Pulkki, Ville and Karjalainen, Matti},
  year={2015},
  publisher={John Wiley \& Sons},
  url={https://doi.org/10.1002/9781119825449},
}

@article{hardin1996trustworthiness,
  title={Trustworthiness},
  author={Hardin, Russell},
  journal={Ethics},
  volume={107},
  number={1},
  pages={26--42},
  year={1996},
  publisher={University of Chicago Press},
  url={https://doi.org/10.1086/233695},
}

@misc{nist2021SRE,
  author = {Omid Sadjadi and Craig Greenberg and Elliot Singer and Lisa Mason and Douglas Reynolds},
  title = {NIST 2021 Speaker Recognition Evaluation Plan},
  year = {2021},
  month = {7},
  publisher = {NIST SRE},
  url = {https://tsapps.nist.gov/publication/get_pdf.cfm?pub_id=932697},
  language = {en},
}

@article{natgunanathan2016protection,
  title={Protection of privacy in biometric data},
  author={Natgunanathan, Iynkaran and Mehmood, Abid and Xiang, Yong and Beliakov, Gleb and Yearwood, John},
  journal={IEEE access},
  volume={4},
  pages={880--892},
  year={2016},
  publisher={IEEE},
  url={https://doi.org/10.1109/ACCESS.2016.2535120},
}

@inproceedings{dwork2006differential,
  title={Differential privacy},
  author={Dwork, Cynthia},
  booktitle={International colloquium on automata, languages, and programming},
  pages={1--12},
  year={2006},
  organization={Springer},
  url={https://doi.org/10.1007/11787006_1},
}

@article{sweeney2002k,
  title={k-anonymity: A model for protecting privacy},
  author={Sweeney, Latanya},
  journal={International journal of uncertainty, fuzziness and knowledge-based systems},
  volume={10},
  number={05},
  pages={557--570},
  year={2002},
  publisher={World Scientific},
  url={https://doi.org/10.1142/S0218488502001648},
}

@techreport{ISOIEC224745_2022,
  type = "Standard",
  key = {ISO/IEC 24745:2011},
  year = 2022,
  title = "ISO/IEC 24745:2011 Information security, cybersecurity and privacy protection -- Biometric information protection",
  volume = 2022,
  address = {Geneva, Switzerland},
  institution = {International Organization for Standardization (ISO)},
  url={https://www.iso.org/obp/ui/en/#iso:std:iso-iec:24745:ed-2:v1:en},
}

@INPROCEEDINGS{sun2024privacyaware,
  author={Sun, Lunan and Guo, Caili and Chen, Mingzhe and Yang, Yang},
  booktitle={ICASSP 2024 - 2024 IEEE International Conference on Acoustics, Speech and Signal Processing (ICASSP)}, 
  title={Privacy-Aware Joint Source-Channel Coding For Image Transmission Based On Disentangled Information Bottleneck}, 
  year={2024},
  pages={9016-9020},
  doi={10.1109/ICASSP48485.2024.10447157}}

@misc{razeghi2024deep,
      title={Deep Privacy Funnel Model: From a Discriminative to a Generative Approach with an Application to Face Recognition}, 
      author={Behrooz Razeghi and Parsa Rahimi and Sébastien Marcel},
      year={2024},
      url={https://doi.org/10.48550/arXiv.2404.02696},
}

@inproceedings{vali2024porcupine,
    title     = {Privacy {PORCUPINE}: Anonymization of Speaker Attributes Using Occurrence Normalization for Space-Filling Vector Quantization},
  author    = {Mohammad Hassan Vali and Tom Bäckström},
  year      = {2024},
  booktitle = {Interspeech},
  url       = {https://doi.org/10.21437/Interspeech.2024-117},
  issn      = {2958-1796},
}

@inproceedings{hong2004privacy,
  title={Privacy risk models for designing privacy-sensitive ubiquitous computing systems},
  author={Hong, Jason I and Ng, Jennifer D and Lederer, Scott and Landay, James A},
  booktitle={Proceedings of the 5th conference on Designing interactive systems: processes, practices, methods, and techniques},
  pages={91--100},
  year={2004},
  url={https://doi.org/10.1145/1013115.1013129},
}

@INPROCEEDINGS{Kalloniatis2009,
  author={Kalloniatis, Christos and Kavakli, Evangelia and Gritzalis, Stefanos},
  booktitle={2009 13th Panhellenic Conference on Informatics}, 
  title={Methods for Designing Privacy Aware Information Systems: A Review}, 
  year={2009},
  pages={185-194},
  doi={10.1109/PCI.2009.45}}

@article{van2011privacy,
  title={Privacy by Design: an alternative to existing practice in safeguarding privacy},
  author={van Lieshout, Marc and Kool, Linda and van Schoonhoven, Bas and de Jonge, Marjan},
  journal={info},
  volume={13},
  number={6},
  pages={55--68},
  year={2011},
  publisher={Emerald Group Publishing Limited},
  url={https://doi.org/10.1108/14636691111174261},
}

@article{iachello2007end,
  title={End-user privacy in human--computer interaction},
  author={Iachello, Giovanni and Hong, Jason and others},
  journal={Foundations and Trends{\textregistered} in Human--Computer Interaction},
  volume={1},
  number={1},
  pages={1--137},
  year={2007},
  publisher={Now Publishers, Inc.},
  url={http://dx.doi.org/10.1561/1100000004},
}

@article{sivaraman2018smart,
  title={Smart IoT devices in the home: Security and privacy implications},
  author={Sivaraman, Vijay and Gharakheili, Hassan Habibi and Fernandes, Clinton and Clark, Narelle and Karliychuk, Tanya},
  journal={IEEE Technology and Society Magazine},
  volume={37},
  number={2},
  pages={71--79},
  year={2018},
  publisher={IEEE},
  url={https://doi.org/10.1109/MTS.2018.2826079},
}

@phdthesis{vestman2020methods,
  title={Methods for fast, robust, and secure speaker recognition},
  author={Vestman, Ville},
  year={2020},
  school={It{\"a}-Suomen yliopisto},
  url={http://urn.fi/URN:ISBN:978-952-61-3484-0},
}

@article{vestman2020voice,
  title={Voice mimicry attacks assisted by automatic speaker verification},
  author={Vestman, Ville and Kinnunen, Tomi and Hautam{\"a}ki, Rosa Gonz{\'a}lez and Sahidullah, Md},
  journal={Computer Speech \& Language},
  volume={59},
  pages={36--54},
  year={2020},
  publisher={Elsevier},
  url={https://doi.org/10.1016/j.csl.2019.05.005},
}

@article{roberts2007biometric,
  title={Biometric attack vectors and defences},
  author={Roberts, Chris},
  journal={Computers \& Security},
  volume={26},
  number={1},
  pages={14--25},
  year={2007},
  publisher={Elsevier},
  url={https://doi.org/10.1016/j.cose.2006.12.008},
}

@inproceedings{rahman24_spsc,
  title     = {Scenario of Use Scheme: Threat Modelling for Speaker Privacy Protection in the Medical Domain},
  author    = {Mehtab Ur Rahman and Martha Larson and Louis ten Bosch and Cristian Tejedor-García},
  year      = {2024},
  booktitle = {4th Symposium on Security and Privacy in Speech Communication},
  pages     = {21--25},
  url={https://doi.org/10.21437/SPSC.2024-4},
}

@book{singh2019profiling,
  title={Profiling humans from their voice},
  author={Singh, Rita},
  volume={41},
  year={2019},
  publisher={Springer},
  url={https://doi.org/10.1007/978-981-13-8403-5}
}

@article{acosta2022survey,
  title={A survey on privacy issues and solutions for Voice-controlled Digital Assistants},
  author={Acosta, Luca Hern{\'a}ndez and Reinhardt, Delphine},
  journal={Pervasive and Mobile Computing},
  volume={80},
  pages={101523},
  year={2022},
  publisher={Elsevier},
  url={https://doi.org/10.1016/j.pmcj.2021.101523}
}

@misc{singh2025humanvoiceunique,
      title={Human Voice is Unique}, 
      author={Rita Singh and Bhiksha Raj},
      year={2025},
      eprint={2506.18182},
      archivePrefix={arXiv},
      primaryClass={cs.SD},
      url={https://arxiv.org/abs/2506.18182}, 
}

@incollection{kroger2019privacy,
  title={Privacy implications of voice and speech analysis--information disclosure by inference},
  author={Kr{\"o}ger, Jacob Leon and Lutz, Otto Hans-Martin and Raschke, Philip},
  booktitle={IFIP International Summer School on Privacy and Identity Management},
  pages={242--258},
  year={2019},
  publisher={Springer},
  url={https://doi.org/10.1007/978-3-030-42504-3_16},
}

@article{kroger2022personal,
  title={Personal information inference from voice recordings: User awareness and privacy concerns},
  author={Kr{\"o}ger, Jacob Leon and Gellrich, Leon and Pape, Sebastian and Brause, Saba Rebecca and Ullrich, Stefan},
  journal={Proceedings on Privacy Enhancing Technologies},
  year={2022},
  url={https://doi.org/10.2478/popets-2022-0002},
}

@article{nissenbaum2004privacy,
  title={Privacy as contextual integrity},
  author={Nissenbaum, Helen},
  journal={Wash. L. Rev.},
  volume={79},
  pages={119},
  year={2004},
  publisher={HeinOnline},
  url={https://digitalcommons.law.uw.edu/wlr/vol79/iss1/10 },
}

@inproceedings{ganta2008composition,
  title={Composition attacks and auxiliary information in data privacy},
  author={Ganta, Srivatsava Ranjit and Kasiviswanathan, Shiva Prasad and Smith, Adam},
  booktitle={Proceedings of the 14th ACM SIGKDD international conference on Knowledge discovery and data mining},
  pages={265--273},
  year={2008},
  url={https://doi.org/10.1145/1401890.1401926},
}

@article{gong2018attribute,
  title={Attribute inference attacks in online social networks},
  author={Gong, Neil Zhenqiang and Liu, Bin},
  journal={ACM Transactions on Privacy and Security (TOPS)},
  volume={21},
  number={1},
  pages={1--30},
  year={2018},
  publisher={ACM New York, NY, USA},
  url={https://doi.org/10.1145/3154793},
}

@article{CASEY20038,
title = {Determining Intent — Opportunistic vs Targeted Attacks},
journal = {Computer Fraud \& Security},
volume = {2003},
number = {4},
pages = {8-11},
year = {2003},
issn = {1361-3723},
doi = {https://doi.org/10.1016/S1361-3723(03)04010-7},
author = {Eoghan Casey}
}

@article{le2023privacy,
  title={Privacy-preserving federated learning with malicious clients and honest-but-curious servers},
  author={Le, Junqing and Zhang, Di and Lei, Xinyu and Jiao, Long and Zeng, Kai and Liao, Xiaofeng},
  journal={IEEE Transactions on Information Forensics and Security},
  volume={18},
  pages={4329--4344},
  year={2023},
  publisher={IEEE},
  url={https://doi.org/10.1109/TIFS.2023.3295949},
}

@article{THEISEN201894,
title = {Attack surface definitions: A systematic literature review},
journal = {Information and Software Technology},
volume = {104},
pages = {94-103},
year = {2018},
issn = {0950-5849},
doi = {https://doi.org/10.1016/j.infsof.2018.07.008},
author = {Christopher Theisen and Nuthan Munaiah and Mahran Al-Zyoud and Jeffrey C. Carver and Andrew Meneely and Laurie Williams}
}

@inproceedings{backstrom2025beyonduser,
  title={Beyond User-centric: {Modelling} Privacy and Fairness Effects of Speech Interfaces on Community- and Society-Levels},
  author={Tom Bäckström and Fedor Vitiugin},
  booktitle = {accepted to 3rd Symposium on Security and Privacy in Speech Communication},
  year = 2025,
  publisher = {ISCA},
}

@article{salah2025privacy,
  title={Privacy-Aware Monitoring for Assisted Living},
  author={Salah, Albert Ali and Colonna, Liane and Florez-Revuelta, Francisco},
  publisher={Springer},
  year = 2025,
  url={https://doi.org/10.1007/978-3-031-84158-3},
}

@article{backstrom2025privacy,
  title={Privacy preservation in audio and video},
  author={B{\"a}ckstr{\"o}m, Tom and Ravi, Siddharth and Florez-Revuelta, Francisco},
  journal={Privacy-Aware Monitoring for Assisted Living},
  pages={77},
  year={2025},
  publisher={Springer},
  url={https://doi.org/10.1007/978-3-031-84158-3},
}

@article{li2023comprehensive,
  title={A comprehensive survey on design and application of autoencoder in deep learning},
  author={Li, Pengzhi and Pei, Yan and Li, Jianqiang},
  journal={Applied Soft Computing},
  volume={138},
  pages={110176},
  year={2023},
  publisher={Elsevier},
  url={https://doi.org/10.1016/j.asoc.2023.110176},
}

@inproceedings{raj2019probing,
  title={Probing the information encoded in x-vectors},
  author={Raj, Desh and Snyder, David and Povey, Daniel and Khudanpur, Sanjeev},
  booktitle={2019 IEEE Automatic Speech Recognition and Understanding Workshop (ASRU)},
  pages={726--733},
  year={2019},
  organization={IEEE},
  url={https://doi.org/10.1109/ASRU46091.2019.9003979},
}

@article{backstrom2025privacydisclosure,
  title={Privacy Disclosure of Similarity Rank in Speech and Language Processing},
  author={B{\"a}ckstr{\"o}m, Tom and Vali, Mohammad Hassan and Nguyen, My and Rech, Silas},
  journal={arXiv preprint arXiv:2508.05250},
  year={2025},
  url={https://doi.org/10.48550/arXiv.2508.05250},
}

@inproceedings{NEURIPS2023_e8b0c97b,
 author = {Mao, Yuhao and M\"{u}ller, Mark and Fischer, Marc and Vechev, Martin},
 booktitle = {Advances in Neural Information Processing Systems},
 editor = {A. Oh and T. Naumann and A. Globerson and K. Saenko and M. Hardt and S. Levine},
 pages = {73422--73440},
 publisher = {Curran Associates, Inc.},
 title = {Connecting Certified and Adversarial Training},
 url = {https://proceedings.neurips.cc/paper_files/paper/2023/file/e8b0c97b34fdaf58b2f48f8cca85e76a-Paper-Conference.pdf},
 volume = {36},
 year = {2023}
}

@article{chillotti2020tfhe,
  title={TFHE: fast fully homomorphic encryption over the torus},
  author={Chillotti, Ilaria and Gama, Nicolas and Georgieva, Mariya and Izabach{\`e}ne, Malika},
  journal={Journal of Cryptology},
  volume={33},
  number={1},
  pages={34--91},
  year={2020},
  publisher={Springer},
  url={https://doi.org/10.1007/s00145-019-09319-x},
}

@phdthesis{vestman2020phd,
    author = {Ville Vestman},
    title = "Methods for Fast , Robust, and Secure Speaker Recognition",
    school = {University of Eastern Finland},
    year = 2020,
    url={http://urn.fi/URN:ISBN:978-952-61-3484-0},
}

@article{schaar2010privacy,
  title={Privacy by design},
  author={Schaar, Peter},
  journal={Identity in the information society},
  volume={3},
  number={2},
  pages={267--274},
  year={2010},
  publisher={Springer},
  url={https://doi.org/10.1007/s12394-010-0055-x},
}

@phdthesis{nespoli2025phd,
    author = "Francesco Nespoli",
    title = {Voice Conversion and Text-To-Speech for Privacy Protection Applications},
    school = {Imperial College London},
    year = 2025,
}

@inproceedings{meyer2025usecasesvoiceanonymization,
      title={Use Cases for Voice Anonymization}, 
      author={Sarina Meyer and Ngoc Thang Vu},
      year={2025},
      booktitle = {5th Symposium on Security and Privacy in Speech Communication},
      url={https://arxiv.org/abs/2508.06356}, 
}

@inproceedings{vauquier25_interspeech,
  title     = {{Legally validated evaluation framework for voice anonymization}},
  author    = {{Nathalie Vauquier and Brij Mohan Lal Srivastava and Seyed Ahmad Hosseini and Emmanuel Vincent}},
  year      = 2025,
  booktitle = {{Interspeech}},
  pages     = {{3229--3233}},
  url       = {https://doi.org/10.21437/Interspeech.2025-1699},
  issn      = {{2958-1796}},
}

@article{cohen2020towards,
  title={Towards formalizing the GDPR’s notion of singling out},
  author={Cohen, Aloni and Nissim, Kobbi},
  journal={Proceedings of the National Academy of Sciences},
  volume={117},
  number={15},
  pages={8344--8352},
  year={2020},
  publisher={National Academy of Sciences},
  url={https://doi.org/10.1073/pnas.1914598117},
}

%

\begin{IEEEbiography}[{\includegraphics[width=1in,height=1.25in,clip,keepaspectratio]{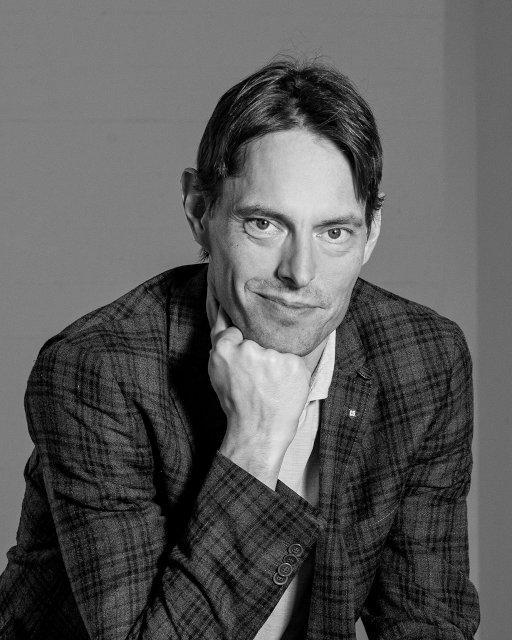}}]{Tom Bäckström}
D.Sc. (tech.) is an associate professor at Aalto University, Finland (2019-). He obtained his Master's and Doctoral degrees at Helsinki University of Technology (the predecessor of Aalto) in 2001 and 2004, respectively. During his time at International Audio Laboratories Erlangen, Germany (2008-2016), he made contributions to several international speech and audio coding standards such as MPEG USAC and 3GPP EVS, and became a professor (W2) at Friedrich-Alexander University Erlangen-Nürnberg (FAU) (2012-2016). Before his current position, he was a professor practice at Aalto University (2016-2019). He is the President (2025--2027) and Board Member (2023--2027) of the International Speech Communication Association (ISCA). He was the initiator and chair of ISCA SIG ``Security and Privacy in Speech Communication'' (2019-2022). His current research interests include privacy, trust, coding, enhancement, and transmission of speech as well as machine learning.
\end{IEEEbiography}






 
\balance

\end{document}